\documentclass{emulateapj} 
\usepackage{apjfonts}
\usepackage{booktabs}
\usepackage{multirow}
\usepackage{xcolor}
\usepackage{soul}

\def\cm{cm$^{$-$3}$}
\def\ea{{\em{et al. }}}

\newcommand{\Lsun}{\hbox{$L_\odot$}}

\def\cm3{cm$^{-3}$}

\def\beq{\begin{equation}}
\def\eeq{\end{equation}}
\def\lesssim{\mathrel{\hbox{\rlap{\hbox{\lower4pt\hbox{$\sim$}}}\hbox{$<$}}}}
\def\gtrsim{\mathrel{\hbox{\rlap{\hbox{\lower4pt\hbox{$\sim$}}}\hbox{$>$}}}}

\def\aj{AJ}
\def\pasp{PASP}
\def\apj{ApJ}

\def\apjs{ApJS}
\def\apjl{ApJL}
\def\aap{A\&A}

\def\mnras{MNRAS}
\def\nat{Nature}
\def\na{New~Astron.}

\def\v1d{{\sc v1d}}

\newcommand{\mum}{\ifmmode{\rm \mu m}\else{$\mu$m}\fi}

\shorttitle{$\eta$~Carinae IR - Sub-millimeter Emission}

\begin{document}

\title{$\eta$ Carinae's Dusty Homunculus Nebula from Near-Infrared to Submillimeter Wavelengths:  Mass, Composition, and Evidence for Fading Opacity}

\author{Patrick W. Morris\altaffilmark{1}, Theodore~R.~Gull\altaffilmark{2}, D. John Hillier\altaffilmark{3}, M.~J.~Barlow\altaffilmark{4},  Pierre Royer\altaffilmark{5}, \\ Krister Nielsen\altaffilmark{6}, John Black\altaffilmark{7}, and Bruce Swinyard\altaffilmark{8,9} 
}

\email{pmorris@ipac.caltech.edu}

\altaffiltext{1}{California Institute of Technology, IPAC, M/C 100$-$22, Pasadena, CA 91125.}
\altaffiltext{2}{NASA Goddard Space Flight Center, Code 667, Greenbelt, MD 20771.}
\altaffiltext{3}{Department of Physics \& Astronomy, University of Pittsburgh, 3941 O'Hara Street, Pittsburgh, PA 15260.}
\altaffiltext{4}{Department of Physics \& Astronomy, University College London, Gower St, London WC1E 6BT, UK.}
\altaffiltext{5}{Katholieke Universiteit Leuven, Institute of Astronomy, Celestijnenlaan 200 D, BE 3001 Leuven, Belgium.}
\altaffiltext{6}{Department of Physics, IACS, Catholic University of America, Washington, DC 20064.}
\altaffiltext{7}{Department of Earth \& Space Sciences, Chalmers University of Technology, Onsala Space Observatory, SE-43992 Onsala, Sweden.}
\altaffiltext{8}{Space Science \& Technology Department, Rutherford Appleton Laboratory, Chilton, Didcot, Oxon, UK.}
\altaffiltext{9}{Deceased.}

\begin{abstract}
 
Infrared observations of the dusty, massive Homunculus Nebula around the luminous blue variable $\eta$ Carinae are crucial to characterize the mass-loss history and help constrain the mechanisms leading to the Great Eruption.  We present the 2.4 - 670 $\mu$m spectral energy distribution,  constructed from legacy ISO observations and new spectroscopy obtained with the {\em{Herschel Space Observatory}}. Using radiative transfer modeling, we find that the two best-fit dust models  yield compositions which are consistent with CNO-processed material, with iron, pyroxene and other metal-rich silicates, corundum, and magnesium-iron sulfide in common.  Spherical corundum grains are supported by the good match to a narrow 20.2 $\mu$m feature.  Our preferred model contains nitrides AlN and Si$_3$N$_4$ in low abundances. Dust masses range from 0.25 to 0.44 $M_\odot$ but $M_{\rm{tot}} \ge$ 45 $M_\odot$ in both cases due to an expected high Fe gas-to-dust ratio.  The bulk of dust is within a 5$''$ $\times$  7$''$ central region.  An additional compact feature is detected at 390 $\mu$m.  We obtain $L_{\rm{IR}}$ = 2.96 $\times$ 10$^6$ $L_\odot$, a 25\% decline from an average of mid-IR photometric levels observed in 1971-1977.  This indicates a reduction in circumstellar extinction in conjunction with an increase in visual brightness, allowing 25-40\% of optical and UV radiation to escape from the central source.   We also present an analysis of $^{12}$CO and $^{13}$CO $J = 5-4$ through $9-8$ lines, showing that the abundances are consistent with expectations for CNO-processed material.  The [$^{12}$C~{\sc{ii}}]  line is detected in absorption, which we suspect originates in foreground material at very low excitation temperatures.   

\end{abstract}


\section{Introduction}

Massive stars evolving from the core hydrogen-burning main sequence may experience one or more episodes of extreme mass loss, reaching several 10$^{-3} M_\odot$ yr$^{-1}$ over years to decades, often accompanied by large variations in brightness, $\Delta m_V \geq$ 2.5 mag, and excursions in effective temperatures $T_{\rm{eff}}$ by up to several thousand degrees kelvin.  Stars with these characteristic signatures of instability are the luminous blue variables (LBVs; Conti 1984; Humphreys \& Davdison 1994), representing a short-lived (a few 10$^3$ - 10$^4$ years) phase of evolution between massive O-type stars and Wolf-Rayet (W-R) stars.  The mechanisms leading to such extreme behavior, and in particular whether they are a consequence of the ``natural'' evolution of a single massive star or must be the result of interactions between companions in a closed binary or multiple-star system, have been matters of intense theoretical and observational attention since the class was recognized among a very small number of stars in the Milky Way, of which $\eta$~Carinae, AG Carinae, and P Cygni are the most well known.

\begin{deluxetable*}{l r r r c}
  \tablewidth{0pc}
  \tablecaption{ISO and {\it{Herschel}} Observation Summary}.

 \tablehead{\colhead{ObsID}\tablenotemark{a} & 
 \colhead{Instrument Mode} & 
 \colhead{Obs. Date} & 
 \colhead{Phase\tablenotemark{b}} & 
 \colhead{Spectral Range} \\
 \colhead{} & 
 \colhead{} &
 \colhead{} &
 \colhead{} &
 \colhead{(THz)} 
 } 
 
 \startdata  

07100250 & SWS01-4 & 27 Jan 1996 & 9.6604  & 6.64 - 127.12 ($2.4-45.2$ $\mu$m) \\ 
07900923 & LWS01 & 04 Feb 1996 & 9.6642  &  1.52 - 6.67 ($45.0-197.0$ $\mu$m) \\   
1342180565 & HifiPointModeFastDBS & 22 Jul 2009 & 12.0952  & 0.894   \\  
1342180566 & HifiPointModeFastDBS & 22 Jul 2009 & 12.0952  & 0.894   \\ 
1342181167 & HifiPointModeFastDBS & 02 Aug 2009 & 12.1005  & 1.464   \\ 
1342181170 & HifiSScanModeFastDBS & 02 Aug 2009 & 12.1005 & $0.553-0.633$ \\
1342201814\tablenotemark{c} & HifiPointModeFastDBS & 01 Aug 2010 & ... & 1.896 \\
1342201817\tablenotemark{d} & HifiPointModeFastDBS & 01 Aug 2010 & ... & 1.896 \\
1342228699\tablenotemark{e} & SpireSpectroPoint  & 16 Sep 2011 & 12.4838  & $0.450-1.550$   \\ 
1342228700\tablenotemark{f} & SpireSpectroPoint  & 16 Sep 2011 & 12.4838  & $0.450-1.550$   \\ 
1342232979 & HifiPointModeFastDBS & 25 Nov 2011 & 12.5183  & 0.498   \\ 
1342232980 & HifiPointModeDBS & 25 Nov 2011 & 12.5183  & 0.513   \\ 
1342232982 & HifiSScanModeFastDBS &  25 Nov 2011 & 12.5183  & $0.804-0.853$ \\
1342232983 & HifiPointModeDBS & 25 Nov 2011 & 12.5183  & 0.816   \\ 
1342232984 & HifiPointModeDBS & 25 Nov 2011 & 12.5183  &0. 841   \\ 
1342234009 & HifiPointModeFastDBS & 13 Dec 2011 & 12.5274  & 1.465    \\ 
1342234010 & HifiPointModeFastDBS & 13 Dec 2011 & 12.5274  & 1.468   \\ 
1342234012 & HifiPointModeFastDBS & 13 Dec 2011 & 12.5274  & 1.897   \\ 
1342234013 & HifiPointModeFastDBS & 13 Dec 2011 & 12.5274  & 1.884  \\ 
1342235769 & HifiSScanModeDBS & 29 Dec 2011 & 12.5340 & $0.861-0.953$ \\
1342235782 & HifiPointModeDBS & 30 Dec 2011 & 12.5355  & 1.210  \\ 
1342235809 & HifiSScanModeDBS & 30 Dec 2011 & 12.5355  & $0.631-0.700$ \\
1342235829 & HifiPointModeFastDBS & 31 Dec 2011 & 12.5360  & 0.968   \\ 
1342235830 & HifiPointModeDBS & 31 Dec 2011 & 12.5360  & 1.051   \\ 
1342235831 & HifiSScanModeFastDBS & 31 Dec 2011 & 12.5360  & $0.950-1.060$ \\
1342238648 & HifiPointModeDBS & 03 Feb 2012 & 12.5531  & 0.621   \\ 
1342238649 & HifiPointModeDBS & 03 Feb 2012 & 12.5531  & 0.565   \\ 
1342247037 & HifiPointModeDBS & 15 Jun 2012 & 12.6187  & 0.578   \\ 
1342247088 & HifiPointModeFastDBS & 16 Jun 2012 & 12.6192  & 1.666   \\ 
1342250485 & HifiPointModeFastDBS & 01 Sep 2012 & 12.6573  & 1.773   \\ 
1342251009 & HifiPointModeDBS & 12 Sep 2012 & 12.6629  & 0.836   \\ 
1342258207 & HifiPointModeDBS & 26 Dec 2012 & 12.7148  & 0.545   \\ 
1342258208 & HifiPointModeDBS & 26 Dec 2012 & 12.7148  & 0.537   \\ 
1342258209 & HifiPointModeDBS & 26 Dec 2012 & 12.7148  & 0.525   \\ 
1342262772 & HifiPointModeFastDBS & 01 Feb 2013 & 12.7332  & 1.766   \\ 
1342263046 & HifiPointModeFastDBS & 06 Feb 2013 & 12.7352  & 1.884   \\ 

\tablenotetext{a}{The two 1996 observations refer to SWS and LWS on board ISO; all remaining observations were obtained from the HIFI and SPIRE instruments on the {\it{Herschel Space Observatory}}.}
\tablenotetext{b}{Column values refer to Damineli et al.  (2008a), who define phase 0 of epoch 11 to the He{\sc{i}} $\lambda$6678 narrow component disappearance on JD 2,452,819.8 with a period of 2022.7 $\pm$ 1.7 days.}
\tablenotetext{c}{Carina North:  10$^{\rm{h}}$ 43$^{\rm{m}}$ 35.01$^{\rm{s}}$  $-$59$^\circ$ 34$'$ 04$''$.59 (J2000).}
\tablenotetext{d}{Carina South:  10$^{\rm{h}}$ 45$^{\rm{m}}$ 11.46$^{\rm{s}}$  $-$59$^\circ$ 47$'$ 35$''$.42 (J2000).}
\tablenotetext{e}{Bright photometric mode.}
\tablenotetext{f}{Nominal photometric mode.}

\enddata

\label{obs_t}

\end{deluxetable*}

Clues to the history of the mass ejection phase(s) are harbored in the circumstellar nebulae observed around many of the known LBVs,  exhibiting a range of masses, from $\approx$ 0.1 M$_\odot$ for P Cygni (Smith \& Hartigan 2006), to $\approx$ 5 $M_\odot$ for R71 in the Large Magellanic Cloud (Voors et al. 1999; Morris et al. 2008; Guha Niyogi et al. 2015), to at least 15 $M_\odot$ in $\eta$~Carinae's Homunculus Nebula (Morris \ea 1999, hereafter MWB99) and to $\approx$ 20 $M_\odot$ in AG Carinae (Voors \ea 2000).  While most present-day LBVs are observed in stellar wind ionization states of B-type stars or hotter, some mass ejection events may have occurred over longer timescales in a previous cooler phase based on the observed properties of the nebular dust and kinematic ages (e.g., Voors et al. 1999, 2000).   

The eruptions that formed the Homunculus Nebula around $\eta$~Carinae, on the other hand, are well recorded as a major event that occurred in 1838-1843 and created the bipolar lobes, followed by several lesser events in the 1890s that probably formed the ragged equatorial ``skirt'' (Smith \& Gehrz 1998; see also the Hubble Space Telescope images published by Morse et al. 1998), and then periods of rapid brightening in the 1940s through 1951. A detailed record of $\eta$~Carinae's historical light curve is given by Frew (2004) and Smith \& Frew (2011).  The central source is by now established to be a binary system with a highly eccentric orbit ($e$ = 0.9) and a period of 5.52 yr as first deduced by Damineli (1996) according to the disappearance of the He~{\sc{i}} 6678 {\rm{\AA}} emission line near periastron (defining Phase 0).  Pittard \& Corcoran (2002) have shown the period to be 5.54 yr based on X-ray emission variability, and Teodoro et al. (2016) have demonstrated through analysis of the He~{\sc{ii}} $\lambda$4686 emission line that the true periastron occurs $\sim$4 days after the disappearance of the He~{\sc{i}} line.   The present-day primary (erupting) star has recently been modeled as a merger between two massive stars (Portegies Zwart \& van den Heuvel 2016) that lead to the Great Eruption ---   a scenario that has been proposed for LBVs and superluminous supernovae in general (Justham et al. 2014).   

Both present-day stars are difficult to observe spectroscopically, due to strong obscuration of radiation at visual wavelengths by the circumstellar dust.   As summarized by Hillier (2008), broad emission lines of H~{\sc{i}}, He~{\sc{i}}, and Fe~{\sc{ii}} primarily arise from the central source, with further similarities to the spectrum of the extreme P Cygni star HDE~316285, while narrow emission lines are now known to originate from small, slow-moving gas/dust condensations offset from the primary star by approximately 0$''$.1 and are believed to be located in the equatorial plane.  The condensations have been named ``Weigelt blobs'' after their discovery by means of optical speckle interferometry by Weigelt \& Ebersberger (1986).  Weigelt \& Kraus (2012) and Teodoro et al. (2017) provide in-depth reviews and analysis of these structures. It is this combination of broad emission from the stars and narrow emission from the Weigelt blobs that gives rise to the classic $\eta$~Carinae spectrum. There are arguments that suggest that circumstellar dust may have been modifying the spectrum of the central source (dominated by the primary) --- a situation that is changing as the central source has increased its brightness significantly relative to the Weigelt blobs and the rest of the Homunculus (e.g., Hillier 2008; Gull et al. 2009; Mehner et al. 2012).

The Homunculus Nebula itself is observed to be enriched with N and deficient in C and O (Davidson, Walborn, \& Gull 1982; Hillier et al. 2001; Humphreys \& Martin 2012) and possibly Si (Verner, Bruhweiler \& Gull 2005), consistent with the abundance pattern for a rotating massive star in transition from CNO cycle core H burning to triple-$\alpha$-process He burning (Ekstr\"om et al. 2012). It is unparalleled as a source of infrared (IR) radiation, accounting for the bulk of $\eta$~Carinae's luminosity:  it is the brightest IR source in the sky outside of Mars and Jupiter at opposition, $F_\nu$(25 $\mu$m) $>$ 70,000 Jy, making the object difficult to observe with increasingly sensitive IR instrumentation (it could not be observed with any instruments on the {\em{Spitzer Space Telescope}}, for example), and has an estimated IR luminosity $L_{\rm{IR}} > 10^{6} L_\odot$ (e.g, Morris et al. 1999).  The IR spectral energy distribution (SED) spanning 2.4 - 200 $\mu$m was observed in 1996 and 1997 near the peak of the cyclical radio outburst (Duncan et al. 1995; Duncan, White \& Lim 1997; Duncan \& White 2003) with the Short Wavelength Spectrometer (SWS) and Long Wavelength Spectrometer (LWS) on board the Infrared Space Observatory (ISO) and presented by Morris et al. (1999, hereafter MWB99), revealing a very strong far IR thermal continuum and significant amounts of cool dust, $T_{\rm{d}} \approx 100$ K, consequently raising the estimated mass of the Homunculus to over 15$_\odot$ (assuming a canonical gas-to-dust ratio $f_{gd} = 100$), or at least 6$\times$ higher than previously estimated.    

The analysis of the ISO observations by MWB99 was limited in two key aspects:  (1) it employed only a simple thermal and compositional model for the dust, similar to previous fits to ground-based photometric measurements of the Homunculus (e.g., Gehrz et al. 1973; Robinson et al. 1973; Cox et al. 1995); and (2) the SWS and LWS apertures were large compared to the nebula's 10$'' \times 18''$ angular size, so that the main thermal components of the dust could not be mapped.  The extent of background contamination was also uncertain.  Nonetheless, MWB99 suggested that most of the cool dust  is located in the central region rather than the lobes based on mid-IR imaging and constraints from the ISO aperture sizes, orientation, and beam profiles, possibly within a compact structure or the ``disrupted'' equatorial torus structure apparent in 18 $\mu$m imaging, extending some 5$''$ (equatorial) $\times$ 3$''$ (polar) in size.  Smith et al. (2003) pointed out that the compact torus geometry would be too small to emit such large amounts of radiation at $\eta$~Carinae's distance, but they erroneously asserted that all apertures fully encompassed the nebula (this was true only of the LWS aperture) as an argument that all of the cool dust instead resides in the bipolar lobes.  MWB99 did not argue that no dust was formed in the lobes, only that the mid-IR imaging and semi-compactness of the emitting source constrained by the ISO aperture calibrations inferred the bulk of the cool emission in the central region.  Material is at least absent where mass loss has been inhibited at stellar latitudes $\geq$ 85$^\circ$, as shown by the presence of ``holes'' in the polar caps by Teodoro et al. (2008).  

This brings us to an important change regarding the ISO data: the LWS data presented by MWB99 were presented in their best-known but not final state of photometric calibrations and nonlinearity corrections (e.g., glitches and transients; Lloyd 2003a) that were still being improved and tested with experience from the full mission database and subsequently applied in standard pipeline products from the ISO post-mission archive since 2001.  The differences to these data are not negligible; they will be discussed further below.  Smith et al. (2003) applied a similar three-temperature-component modified blackbody model to rederive dust temperatures and masses, but they curiously resorted to a low-resolution tracing of the ISO spectrum presented by MWB99 rather than exploiting the latest products from the ISO archive.   They also  fit {\rm{through}} the solid-state features, which were avoided by MWB99, recognizing that these features are unlikely to be representative of amorphous astronomical silicates with respect to grain emissivities.  The fitting by Smith et al. (2003) led to overestimates of $T_{\rm{d}}$ for the warm and cool dust (see also Brooks et al. 2005), but nonetheless confirmed the MWB99 dust mass estimates to within the quoted uncertainties.  Unfortunately, Gaczkowski et al. (2013) relied on the fit by Smith et al. (2003) to make a comparison with recent {\em{Herschel}} imaging observations of the Carina Nebula and consequently arrived at an incorrect conclusion about possible dynamical changes in dust production in the Homunculus.  In any event, the representation of the IR SED as modified blackbodies to derive dust temperatures and masses assuming astronomical silicates gives only a rough picture of the characteristics of the dusty nebula and low-end estimates of the dust mass.

In this paper we present results of detailed radiative transfer modeling of the full thermal IR SED of $\eta$~Carinae's unique Homunculus Nebula.  We combine the published ISO spectroscopy (using the updated LWS calibrations mentioned above) with unpublished observations taken with spectrometers on-board {\em{Herschel Space Observatory}}.  The {\em{Herschel}} observations presented here are part of an Open Time program (P.I. T. Gull) for exploring the far-IR and submillimeter atomic and molecular gas content of the Homunculus (T. Gull et al., in preparation).  Data obtained with the Fourier Transform Spectrometer (FTS) of the Spectral and Photometric Imaging Receiver (SPIRE) give us continuous coverage of the continuum to 670 $\mu$m and further allow us to address background corrections and the emitting source size at the long wavelengths. Our dust models are examined for their consistency with the CNO abundances of the nebula.  We support this with an analysis of rotational lines of CO emission lines observed with the Heterodyne Instrument for the Far Infrared (HIFI) in order to  derive CO/H$_2 $ and $^{12}$C/$^{13}$C abundance ratios and compare these to ground-based results by Loinard et al. (2012). 

In the next section we summarize the set of observations and their data processing.  In that section we quantify the background levels and inferred thermal source size from the SPIRE observations and examine the combined set of HIFI and SPIRE observations for indications of far IR continuum variability.  In Section~\ref{sec:dustmodeling} we present our dust models, and in Section~\ref{sec:COabundance} we present the analysis of CO emission lines observed with HIFI are presented.  We discuss the implications of our results and compare to previous studies in the Discussion section, and we conclude with future prospects in Sections~6 and 7.        

\section{Observations and Data Processing}

Observations presented here are summarized in Table~\ref{obs_t}.   The phase column values refer to Damineli \ea (2008a), who define Phase 0.0 of Epoch 11 to the He~{\sc{i}} 6678 ${\rm{\AA}}$ narrow component disappearance on JD 2452819.8 with a period of 2022.7 $\pm$ 1.7 days, or 5.54 yr.  Phase 0 corresponds to periastron in the stellar binary hypothesis (Damineli, Conti \& Lopes 1997; Hillier \ea 2001; Damineli \ea 2008a,b).

\subsection{ISO Spectroscopy}

\subsubsection{SWS 2.4 - 45.2 $\mu$m}\label{sws}

The ISO spectroscopic observations used here have been first published by MWB99, who did not provide data reduction details because of journal brevity constraints.   The observations over the 2.4 - 45.2 $\mu$m range were obtained on 1996 January 27 with the SWS (de Graauw \ea 1996) using the Astronomical Observing Template (AOT) S01 grating scan at the slowest scan rate (speed 4), providing a spectral resolution $\lambda/\Delta\lambda$ in the range of 800 - 2000.  Two complete scans were performed, in forward and reverse over the full grating ranges arranged by the four optically independent detector modules, each with their own throughput aperture, and individual sub-bands corresponding to spectral order separation. 

\begin{figure}
 \begin{center}
 \includegraphics[width=8cm]{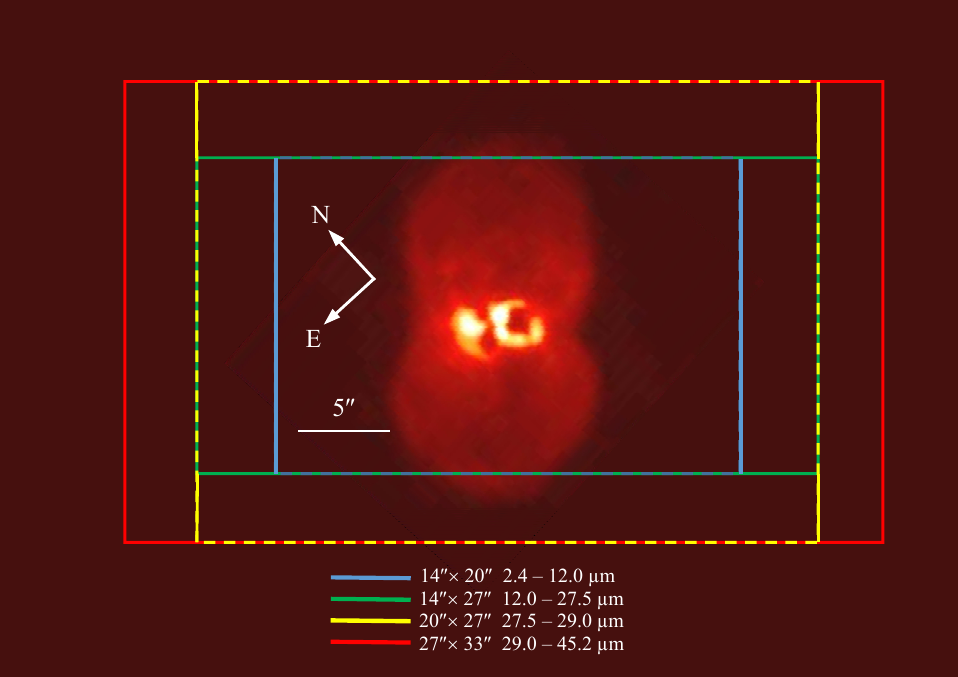}
 \end{center}
 \caption{Schematic layout of the ISO SWS apertures on $\eta$~Carinae during 1996/1997 observing.  The background image of the Homunculus is the TIMMI 18 $\mu$m observation from MWB99.  The aperture sizes of the LWS covers the $45.0-197$ $\mu$m  range from 66$''$ to 86$''$ across (Lloyd 2003b). 
   \label{apertures}}
\end{figure}

\begin{figure*}
 \begin{center}
 \includegraphics[width=8.3cm]{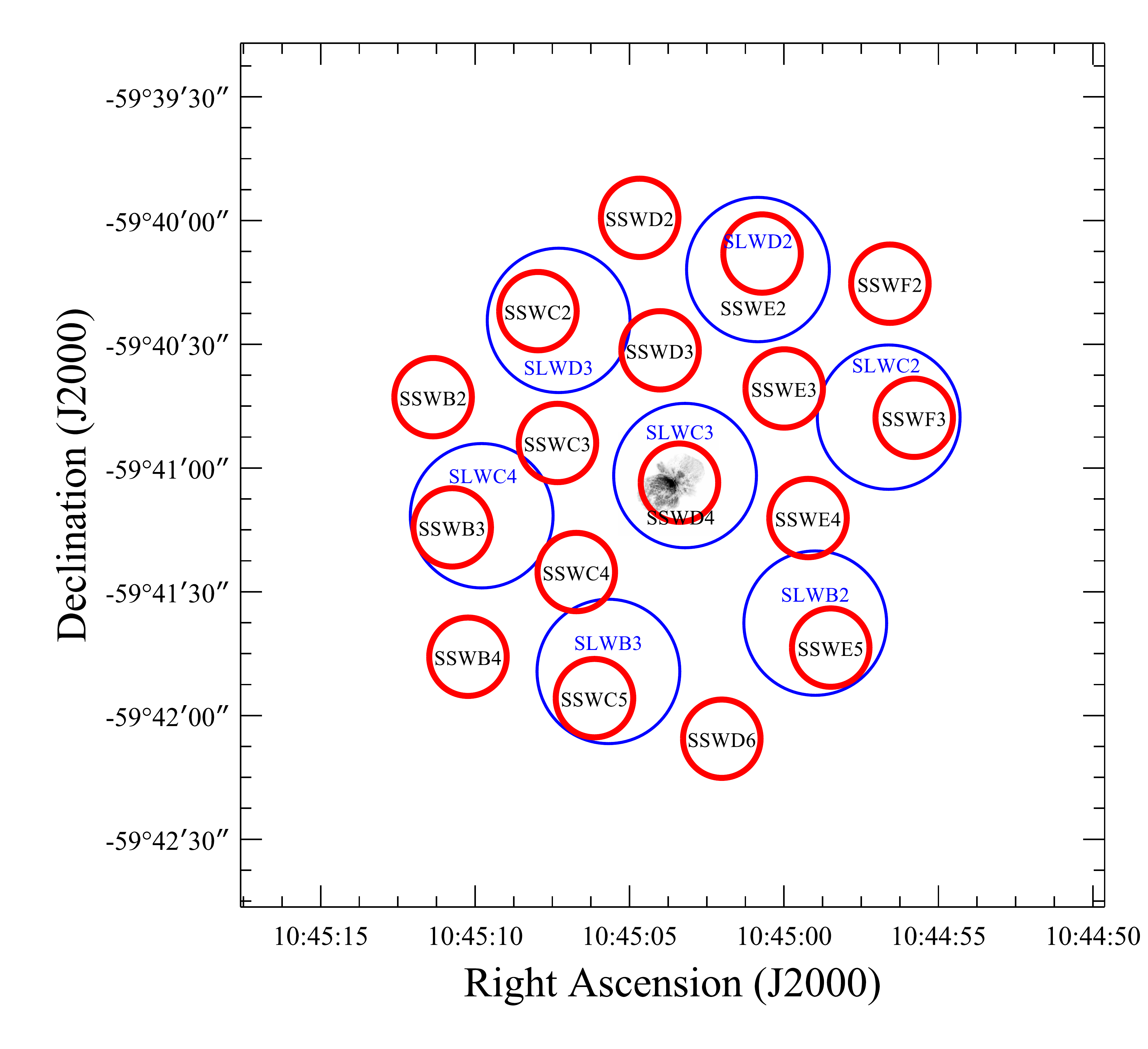}  \includegraphics[width=9.5cm]{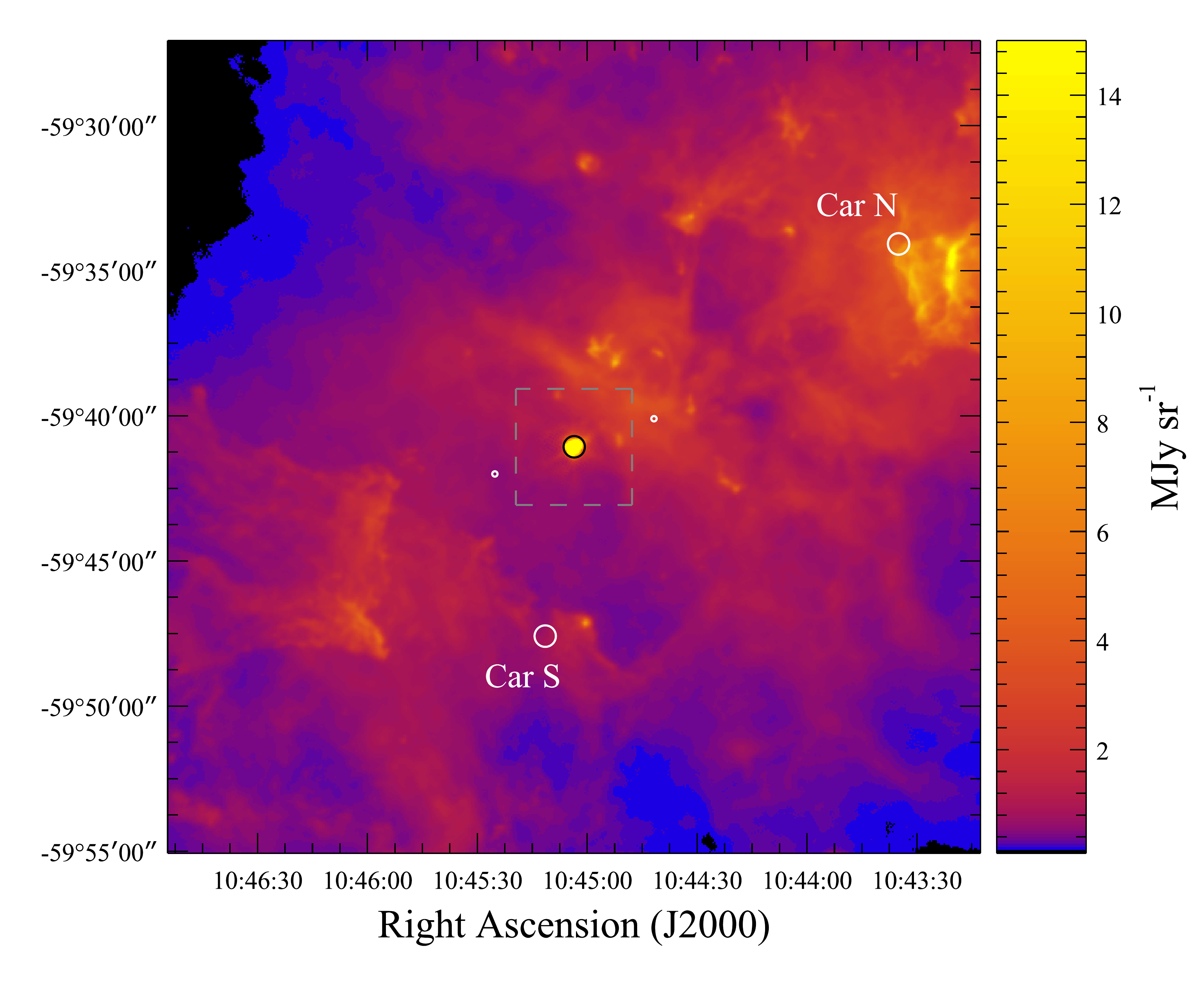}
 \end{center}
 \caption{Left:  footprint of the unvignetted SPIRE detectors as oriented during observations of $\eta$~Carinae on 2011 September 16, with the Homunculus shown to scale from optical HST imaging (Morse et al. 1998).  The SSW detectors (red) cover $194-313$ $\mu$m with beam sizes $16''-22''$ (FWHM), the SLW detectors (blue) cover $303-671$ $\mu$m with $28''-43''$ beams.  Right: positions of HIFI observations analyzed in this study, overlaid on a reference PACS 160 $\mu$m channel scan map (obsid 1342211616) centered on $\eta$~Carinae, on a logarithmic stretch and clipped at 15 MJy sr$^{-1}$ to highlight the cloud structure.     The Carina N position is $\approx$3$'$.0 to the west of HD93129,  an OIf+OV binary and the visually brightest stellar object in Trumpler 16 (Sota et al. 2014).  The two small circles to the SE and NW of $\eta$~Carinae are the beam switch positions for the C$^+$ observations discussed in Sec.~\ref{carbonisotope}, plotted at the beam size of 11$''$.2 HPBW.  The dashed gray box indicates the SPIRE field of view shown on the left.  
   \label{spireoffdets}}
\end{figure*} 

Basic data processing was carried out with the ISO off-line pipeline (OLP) software version 7.0 in the IDL-based SWS Interactive Analysis package IA3\footnote{IA3 is a joint development of the SWS consortium. Contributing institutes have been SRON, MPE, KUL, and the ESA Astrophysics Division.}. The standard pipeline produces photometrically and wavelength-calibrated spectral segments for each of the 12 pixels in the four detector bands on a grating scan and spectral order basis.   Dark currents were inspected for nonlinear behavior that can arise from a cosmic-ray hit on any element of the detector array or amplification electronics, and sky signal was initially masked where discontinuities were noticed above 3 $\sigma$ of the RMS noise level. The mean flux level within each resolution bin was then estimated using the stable detector output, and the masked detector signal was reincorporated by fitting the regions between the jumps with a first- or second-order polynomial, and rectifying the output to the mean flux. This method results in improvements of 10 - 20$\%$ in continuum signal-to-noise ratios (S/Ns). 

Special care was required for the photometric calibrations, due to the high-input signal from $\eta$~Carinae.  The slower scan speed of the S01-4 AOT results in twice the dwell time per resolution element than the three faster speeds (3 s),  and some partial saturation occurred in the 19.5 - 29.0 $\mu$m range of Band 3, where the signal from $\eta$~Carinae has peaked. However, there are also twice as many readouts (48) along each integrated signal ramp, allowing for a statistically better fit of the unsaturated portions during linearization of the ramps to a calibrated photometric scale.  In the 26.5 - 29.0 $\mu$m range of the second scan, $\sim$60\% of the signal ramps are too saturated for reliable slope fitting, due to the accumulated memory (un-reset by voltage biasing) of the preceding scanning.  These affected data were removed, and we assign a higher absolute photometric uncertainty to the remaining data, 30\%, compared to 23$\%$ in the 12.0-26.5 $\mu$m range of the rest of Band 3, 25$\%$ over 29.5 - 45.2 $\mu$m (Band 4), 8$\%$ over 2.4-4.1 $\mu$m (Band 1), and 15\% over 4.1 - 12.0 $\mu$m (Band 2).   These are formal uncertainties at the key wavelengths in each sub-band (see Table~1 in de Graauw \ea 1996).  The flux calibration at 3.5 $\mu$m (in Band 1) agrees well inside the 8$\%$ formal uncertainty with ground-based measurements by Harvey \ea (1978), and L-band photometry by Ghosh \ea (1988).   The L-band photometry varies less than 0.1 mag over a 20 yr period through 1994.  Our signal nonlinearity corrections were tested qualitatively on S01-4 observations of red hyper-giant NML~Cyg, the brightest of the SWS secondary stellar calibrators, and Mars, which was observed at opposition in 1997 July (and peaked slightly higher in flux density at 25 $\mu$m during this time compared to $\eta$~Carinae).  The ramp linearization and memory corrections resulted in good continuity between bands at the relative and absolute levels.

During observing, the minor axes of the four SWS apertures corresponding to the cross-dispersed grating axes were oriented at a position angle of 49$^{\circ}$ from celestial north, nearly  parallel with the major axis of the Homunculus Nebula (schematically shown in Figure~\ref{apertures}).  The orientation is retrievable from the spacecraft pointing history files (IIPH) and is a serendipitous result of the roll angle of the telescope during the periods of $\eta$~Carinae's limited visibility to ISO, constrained by Sun/Earth/Moon/Jupiter avoidances.  The apertures for Bands 1 and 2 have the same projected widths, covering 14$''$ of the projected $\sim18''.5$ infrared nebula.  Since the triangular photometric spatial response beam shapes are consistent (Beintema, Salama \& Lorente 2003), we do not expect a discontinuity between these two corresponding wavelength ranges regardless of source extent on the sky.  
The Homunculus is fully contained in the Band 3E and 4 apertures, which also have flattened cross-dispersed beam profiles.  Here we would expect a sizable sudden increase in continuum fluxes starting at $\lambda > 27.5 \mu$m (Band 3E) and 29.0 $\mu$m (Band 4) if emission is extended over these wavelengths, due to the change in aperture size and beam shape differences.  This effect is quite obvious in many extended nebulae observed with SWS, such as NGC~7027 (e.g., Bernard-Salas et al. 2001) and the roughly 20$''$ diameter ring nebula around AG Carinae with significant IR emission away from the central star (Voors et al. 2000).  Such effects are not observed in the calibrated $\eta$~Carinae spectrum, however, leading us to conclude that the primary source of continuum emission at $\lambda < 30 \mu$m is no more than 5$''$ across along the equatorial axis and 7$''$ along the polar axis.   These outside dimensions are in excellent agreement with the earliest measurements of $\eta$~Carinae's angular size over the $4.3-20$ $\mu$m range by Westphal \& Neugebauer (1969), Gehrz et al. (1973), and Aitken \& Jones (1973), and with the roughly 5$''$ $\times$ 3$''$ volume containing the two apparent loops and broken torus-like structure that dominates in mid-IR imaging---physical features that have been numerically modeled by Portegies Zwart \& van den Heuvel (2016) as  results of a stellar merger.  

\subsubsection{LWS $45.0-196.8$ $\mu$m}\label{lws}

Observations of $\eta$~Carinae using the LWS (Clegg et al. 1996) were obtained with the L01 AOT, which is similar to the SWS S01 grating scan AOT, and taken in the same visibility window.  The LWS has aperture diameters of 66$'' - 86''$, and (like the SWS) exhibited peaked or triangular beam profiles (Lloyd 2003b). The data were first reduced in the LWS Interactive Analysis (LIA) package,\footnote{LIA is a joint development of the ISO-LWS Instrument Team at Rutherford Appleton Laboratories (RAL, UK ---  the PI Institute) and the Infrared Processing and Analysis Center (IPAC/Caltech, USA).}, version 8.1, with calibration files based on the final OLP software version 10.1. Dark current subtraction and absolute responsivity corrections for bright sources were done interactively, following procedures similar to those described above for the SWS data. Specifically we treated glitches and memory tails (Lloyd 2003a), and averaged multiple scans using standard clipping and arithmetic mean routines, and then resampled the spectra with bin sizes preserving the native wavelength sampling ($\lambda / \Delta\lambda$ = 150 - 300).

The LWS01 detector is known to have been the least photometrically stable of the 10 detectors, and since this is where the instrument has seen the highest-input source signal (3.0 - 4.0 $\times$ 10$^3$ Jy), we regard the relative shape of the data over the 43 - 50.4 $\mu$m range to be less certain than in the calibrated signal output of the remaining detectors.  Similarly, since the SWS Band 4 Ge:Be detectors had the highest glitch rates and the longest hysteresis time constants, it is very likely that the slight ``upturn'' at the very end of the SWS spectrum 42 - 45 $\mu$m is due to residual memory effects not corrected by our procedures.  The SWS-to-LWS overlap range 42 - 50 $\mu$m is therefore ignored for broadband spectral features and effectively regarded in terms of an average flux of 3.55 $\times$ 10$^3$ Jy at 46 $\mu$m. 

The final calibrated LWS spectrum compares favorably to the ``highly processed data products'' (HPDPs) available in the ISO archive\footnote{ https://www.cosmos.esa.int/documents/1145917/1145942/technote34.html}.  The corresponding HPDP spectra have somewhat larger discontinuities between detector bands in no systematic direction, possibly a result of the autonomous (noninteractive) routines used in the processing, and it is not uncommon to find these LWS data in the literature with offsets that have been applied ad hoc using the most stable LWS04 detector as a reference.  Nonetheless,  agreement of the spectral fragments from each detector is  within 5\% at nearly all wavelengths.   The LWS spectrum presented by MWB99 which was based on OLP 6.0 processing, on the other hand, has systematically higher fluxes by $\approx$30-35\% at nearly all wavelengths.  This will naturally have an effect on derived temperatures and dust masses, as well as the energetics of the nebula (cf. MWB99 and Smith et al. 2003).  We show the net difference below, but it is a moot point for this paper since we abandon the simplistic three-component modified blackbody approach to analyze the dust continuum.     

\subsection{Herschel Spectroscopy}

The spectroscopic observations presented here are the portions relevant to our study obtained with SPIRE and HIFI in the OT~1 program {\em{The Homunculus: Clues to Massive Ejection from the Most Massive Stars}} (T. Gull, P.I.).   Range spectroscopy with the Photodetector Array Camera and Spectrometer (PACS) instrument was also obtained toward and around the Homunculus and will be presented by T. Gull et al. (in preparation).  

\subsubsection{SPIRE}\label{SpireObsSec}

\begin{figure*}
 \begin{center}
 \includegraphics[width=15cm]{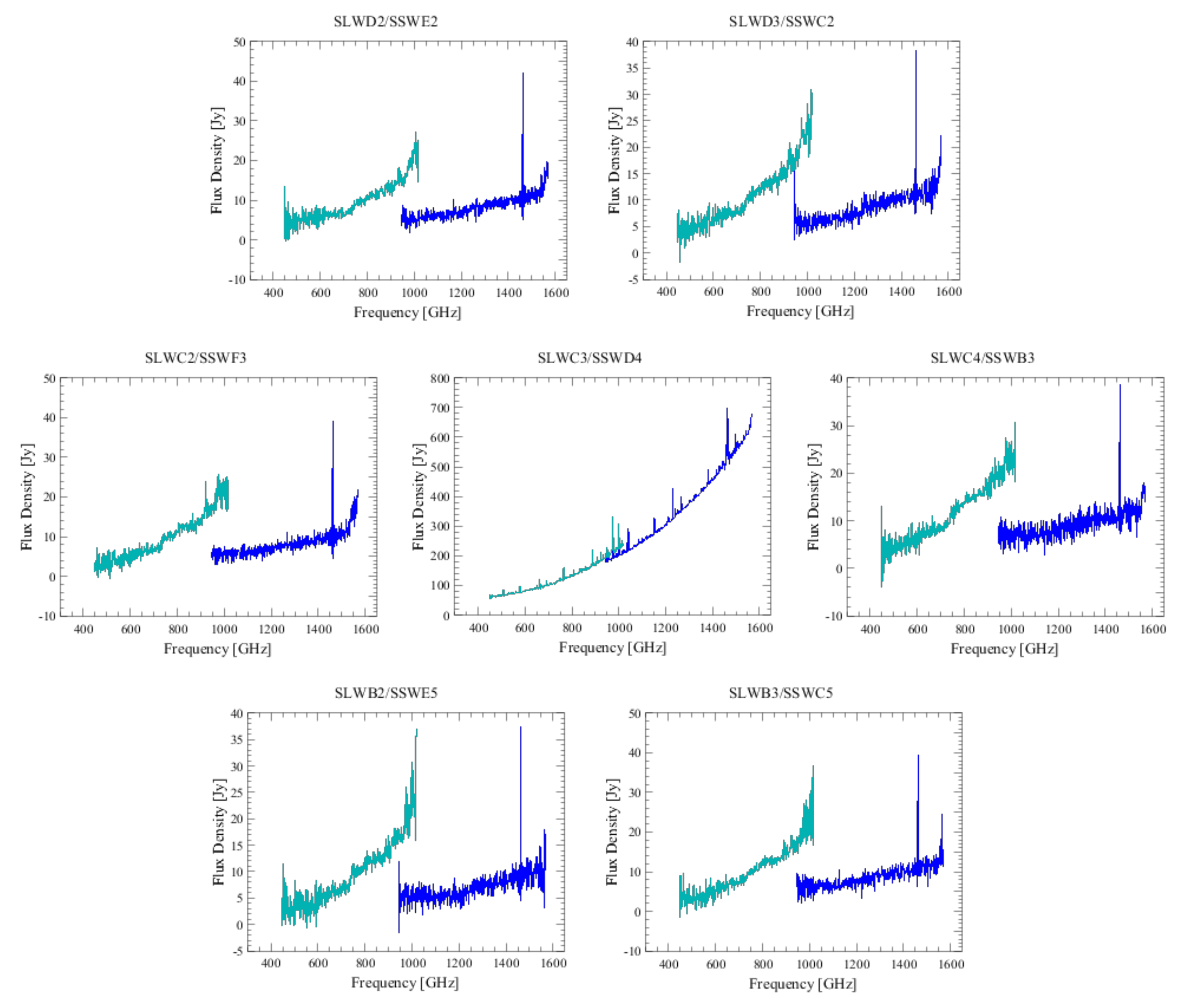}
 \end{center}
 \caption{Spectral output from the co-aligned detectors shown in Fig.~\ref{spireoffdets}.   Output from the short-wavelength (SSW) detectors are shown in dark blue, and from the long-wavelength (SLW) detectors in cyan.
   \label{spirecoalignedDets}}
\end{figure*}

The SPIRE observations were taken with its FTS (Griffin \ea 2010) on 2011 September 16 in two separate observations on the same date near apastron (see Table~\ref{obs_t}), using the nominal and bright detector modes.   The nominal mode is intended for source fluxes up to around 500 Jy over any portion of the available 194 - 671 $\mu$m (447 - 1544 GHz) range and provides maximum sensitivity for faint spectral features using optimal voltage biasing of the bolometer detector arrays.   The bright mode operates at 3 to 4 times lower detector sensitivity with much higher biasing and an out-of-phase analog signal amplifier/demodulator.  The SED of $\eta$~Carinae over {\it{Herschel}} wavelengths is on the Rayleigh-Jeans tail, with flux densities from several tens to around 700 Jy.   It has been shown by Lu \ea (2014) that the photometric calibrations of the bright and nominal modes generally agree to within 2\%, including these $\eta$~Carinae observations that were included in their analysis, when using latest pipeline processing software and calibration tables (discussed below).   Given this agreement, and lacking any noticeable saturation effects or differences between the two observations in the strengths of the spectral features, we use the bright-mode spectrum out of consideration for the higher fluxes at the short-wavelength end.  

The SPIRE FTS employs two bolometer detector arrays, corresponding to the long-wavelength or SLW module covering 303 - 671 $\mu$m (446.7-989.4 GHz) and the short-wavelength or SSW module covering 194 - 313 $\mu$m (959.3-1544 GHz).  The data have been processed in  the final version 14.1 of the Standard Product Generation (SPG) pipeline at the {\it{Herschel}} Science Centre and analyzed in the {\it{Herschel}} Interactive Processing Environment (HIPE; Ott 2010). The FTS absolute and relative photometric calibration is based primarily on Uranus.  Detailed analysis of the flux uncertainties shows that the absolute fluxes at the levels of $\eta$~Carinae are uncertain by up to 6\%. (Swinyard et al 2014; Hopwood et al. 2015).   

The beam widths of the FTS vary nonlinearly with wavelength, between 16$''$ amd 22$''$ FWHM for the SSW section and from 28$''$ to  43$''$ in the SLW section.  There is a  $\approx$40\% difference in beam size in the overlap region between the two sections, 22$''$ in the SSW band versus 37$''$ in the SLW band at 310 $\mu$m, resulting in discontinuities in flux levels between the two sections.  There are two reasons for this:  (1) extended background emission around the Homunculus is present and can be characterized and corrected in the off-axis detectors (see Figures~\ref{spireoffdets} and \ref{spirecoalignedDets}), and (2) the Homunculus is neither a point source nor fully extended, whereas the SPIRE spectral data products are provided to users for both limiting cases.   SPIRE data processing tools in HIPE include background subtraction using the unvignetted co-aligned off-axis detectors and an interactive semi-extended source correction tool, described by Wu \ea (2013).  Details on these last steps are given in Section~\ref{extend}.

\begin{figure}
 \begin{center}
 \includegraphics[width=9cm]{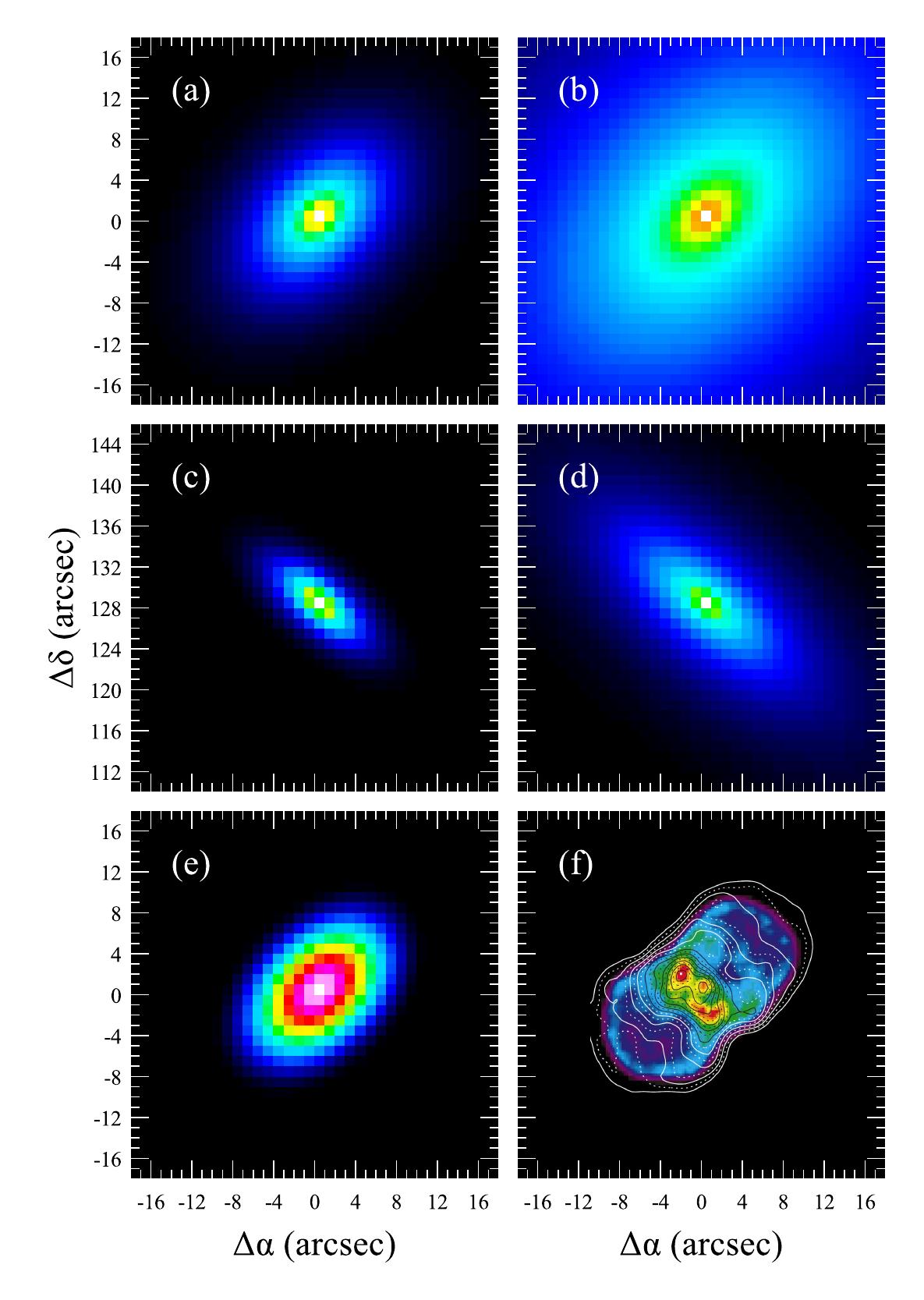}
 \end{center}
 \caption{Input source profiles for extended source corrections to the SPIRE SW and SL sections for  (a) a Sersic model with $n$ = 1.5 and scale diameter of 5$''$.0 and rotation angle of 55$^\circ$; (b)  a Sersic  with $n$=2, scale diameter of 10$''$.0, and rotation angle 55$^\circ$;  (c) a Sersic  with $n$ = 1.25, scale diameter of 5$''$.0 and rotation of 140$^\circ$;(d) a Sersic with $n$ = 2.0, scale diameter of 10$''$.0, rotation angle 140$^\circ$; and (e) a Gaussian profile with FWHM = 10$''$.0 and rotation angle 55$^\circ$.   A comparison TIMMI image of 17 $\mu$m emission with 10 $\mu$m contours is shown in panel (f), from MWB99).  
   \label{sectprofiles}}
\end{figure}

\subsubsection{HIFI}\label{HifiObsSec}

High spectral resolution observations using the HIFI instrument (de Graauw \ea 2010; Roelfsema \ea 2012) have been acquired over a range of $\eta$~Carinae's orbital phase, spanning the {\it{Herschel}} mission beginning in 2009 July and August and then over 2011 September through 2013 February.  The 2009 observations were taken near periastron in calibration time during the commissioning and performance verification phase of HIFI.  All remaining observations are from our OT1 program OT1\_tgull\_3 and were acquired mostly near apastron through $\Delta\Phi \simeq$ 0.74; see Table~\ref{obs_t}.  The Point AOT observations were acquired for spectral line measurements but can also be exploited for their calibrated continuum levels at different frequencies to witness possible variations in flux levels of the far IR continuum.  The observations of the $^{12}$CO and $^{13}$CO rotational transitions were taken with the Spectral Scan AOT for all but the $^{13}$CO $J=5-4$ line, which was obtained in a Point AOT observation.      

All Point and Spectral Scan data presented here employed the Dual Beam Switch (DBS) reference scheme, which follows an alternating chop-nod pattern with the internal mirror of HIFI and the telescope between the target and the sky to correct for standing waves, backgrounds, and baseline drift.  The reference positions are always on a fixed throw distance of 3$'$ on either side of the central on-target position, set mechanically by the chopper mirror and the position angle of the telescope.  Fast chopping is used at frequencies where Allan stability times are short or system temperatures are high, pertaining mainly to the Hot Electron Bolometer (HEB) mixer bands 6 and 7, and at certain frequencies in the Semiconductor-Insulator-Semiconductor (SIS) mixer bands 3 and 4 using diplexers to provide coupling between the signal from the sky and the local oscillators (LOs).  

 HIFI operates as a double-sideband (DSB) receiver in which sky frequencies above and below the LO frequency, at $\nu_{\rm{LO}} + \nu_{\rm{IF}}$ and $\nu_{\rm{LO}} - \nu_{\rm{IF}}$, are simultaneously detected, meaning that the continuum is a co-addition of two nearby ranges mirrored around the tuned LO frequency over an intermediate-frequency (IF) bandpass.  In the SIS Bands 1 through 5 (480 - 1250 GHz) the IF bandpass is 4.0 GHz, while in the HEB Bands 6 and 7  (1410 - 1910 GHz) the IF bandpass is 2.4 GHz.  HIFI has two spectrometer systems:   the digital autocorrelation high-resolution spectrometer (HRS) and the acousto-optical wide band spectrograph (WBS).   Here we present only the WBS data which have a resolution of $\simeq$1 km s$^{-1}$.  

Observations have been processed in the SPG version 14.1,  producing calibrated 1D spectra on an antenna temperature $T_{\rm{A}}$ scale.  The processing is done on scans from each of the independent mixer polarizations referred to as H (horizontal) and V (vertical), which can be averaged together interactively in HIPE, taking into account that the H and V beams are not perfectly co-aligned (see Table 4 in Roelfsema \ea 2012), and converted to a main-beam temperature $T_{\rm{mb}}$ scale.  The spectral scans all use a frequency redundancy of 4 or higher, and have been deconvolved to a single-sideband (SSB) scale in HIPE.  Since we use products from the version 14.1 pipeline, the data quality flags have all been manually inspected and updated for artifacts as described in Section 5.3 of the HIFI Handbook,\footnote{\label{handbook}http://herschel.esac.esa.int/twiki/bin/view/Public/HifiCalibrationWeb} resulting in optimal sideband deconvolution.  To a very good approximation the beams are Gaussian, with widths (FWHM) from 11$''$.1 at 1910 GHz to 44$''$.2 at 480 GHz. 

Three HIFI Fast DBS observations of the C$^+$ line at $\nu_0$ = 1900.537 GHz listed in Table~\ref{obs_t} have been processed to avoid self-chopping effects from extended line emission contaminating the reference sky positions.  The OFF reference subtraction has been skipped in the HIFI pipeline for obsids  1342201814, 1342201817, and 1342234012, and the nonsinusoidal electrical standing waves and sinusoidal optical (LO path-related) standing waves have been treated interactively in HIPE.       

\subsection{Background and Semi-extended source corrections}\label{extend}

As shown in Figure~\ref{spirecoalignedDets}, the background levels around the Homunculus are easily detected in the SPIRE off-axis detectors.   Background subtraction is straightforward using the prescription described by Wu et al. (2013) and available as an interactive script in HIPE.  The script allows the user to select (or de-select) any of the off-axis detectors for the correction; we adhere to the standard set of the unvignetted, co-aligned detectors.  The uncertainties from this procedure are estimated to be a fraction of 1 Jy (Hopwood et al. 2015) assuming a smooth background, i.e.,  iinsignificant compared to the high flux levels of the Homunculus.  

After background subtraction, the remaining flux difference between the SSW and SLW sections is very small, $\approx$ 6 Jy in the overlap region, or $\approx$ 3\% of the average flux at 306 $\mu$m.  In other words, the SPIRE spectrum indicates that this feature causing the jump is very small compared to the beam size, as these spectra have been calibrated as point-like in the SPIRE pipeline. We can treat the remaining discontinuity as arising from semi-extended emission, as follows. 

\begin{figure}
 \begin{center}
 \includegraphics[width=8cm]{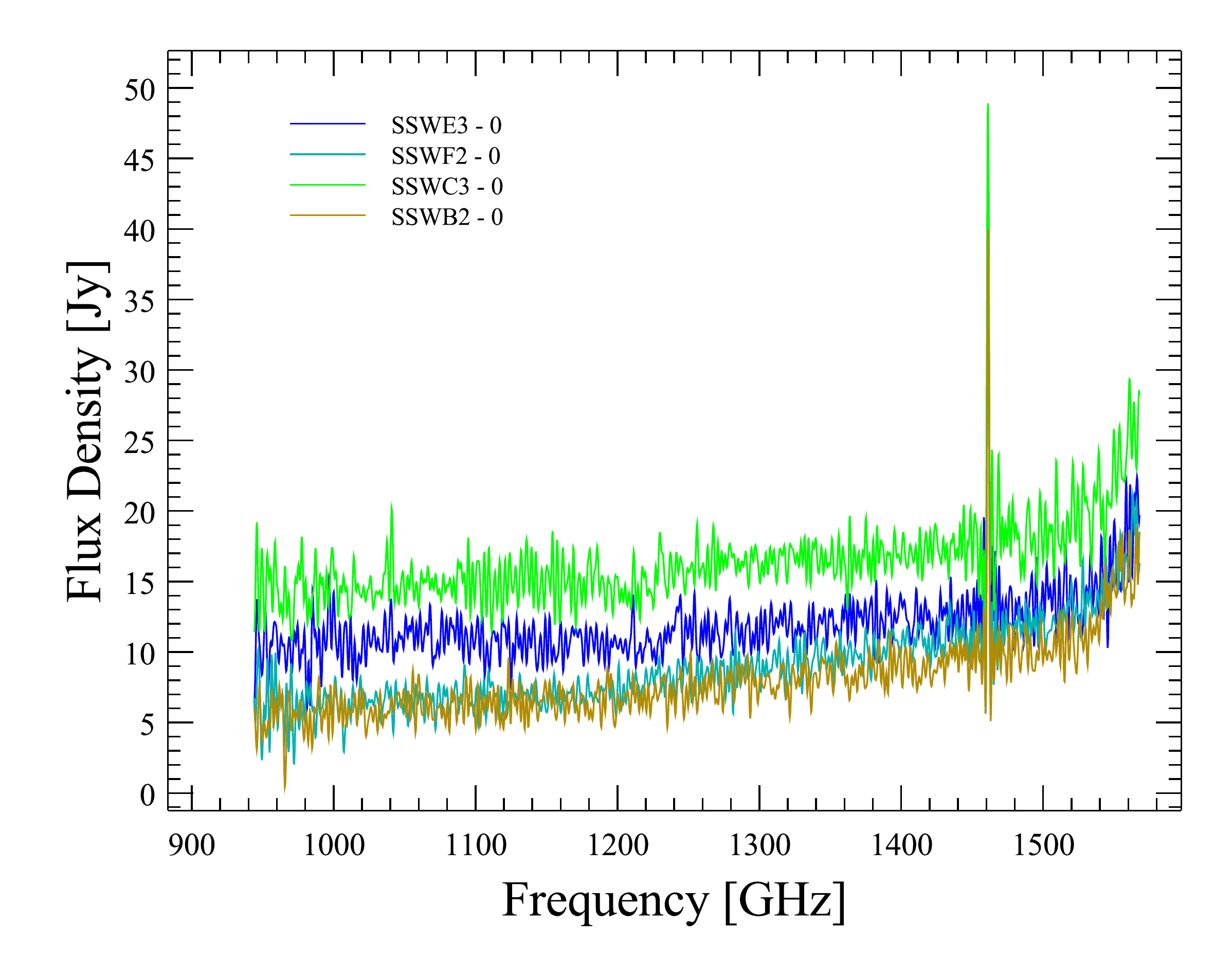}
 \end{center}
 \caption{Spectral output from SSW detectors interior to the co-aligned SSW and SLW detectors shown in Fig.~\ref{spireoffdets}, as indicated in the legend.  
   \label{spireInsideDets}}
\end{figure}     

The extended source flux density $F_{\rm{ext}}$ in Jy as formulated by Wu \ea (2013) is

\begin{equation}
F_{ext}(\nu) = F_p(\nu) \; \eta_c (\nu,\Omega_{source})  {\Omega_{source}\over{ \eta_f (\nu,\Omega_{source}) \; \Omega_{beam}(\nu)}},    
\end{equation}

\noindent where $F_p$ is the point-source flux density, $\eta_f$ is the signal coupling efficiency between source of angular size $\Omega_{source}$ and the telescope (i.e., the forward efficiency), and $\eta_c$ is an empirical correction efficiency to take into account how the beam model couples to the ``true'' source distribution and thus depends on a geometric source distribution model. $\Omega_{beam}(\nu)$ is the size of the beam at frequency $\nu$.

Three representative 2D source distribution models are available with the extended source correction algorithm, namely, a flat top hat model, a Gaussian distribution function, and a Sersic profile that is often applied to galaxies and stellar nebulae with low-brightness central regions and more extended emission away from the core.  The Sersic profile is given by $I(r) = A {\rm{exp}}[-(r/r0)^{1/n}]$ where $A$ is a normalization factor such that the area of the profile is 1.0, $r$ is the distance from the center of the profile, $r_0$ is the scale radius, and $n$ is the Sersic index. A small value of $n$ corresponds to compact sources ($n$ = 0.5 for a point source), and larger values for more extended emission ($n$ = 10.0 for a fully extended source).   Input parameters that define the source shape for all three source distribution models in the algorithm are the diameter $\Omega_{source}$, eccentricity $e$, rotation (or position angle), $x$ and $y$ detector coordinate offsets from the central pointing (needed for a suspected mis-pointing of the telescope or an off-center emission region of interest), and the Sersic index $n$ when applicable.   With the diameter set to an initial guess and remaining parameters fixed, the algorithm can perform a search to find the optimum diameter that effectively closes the gap to a mean flux ratio of 1.00 in the overlap region between SSW and SLW spectra.  The final spectrum from each model is nearly identical ($<$ 0.5\% differences at all wavelengths) regardless of input initial guesses, since it is the beam shapes that drive the corrections at the final estimated size for each corresponding model self-consistently.

\begin{deluxetable}{l l l r r}
  \tablewidth{0pc}
  \tablecaption{975 GHz source angular diameters.\tablenotemark{a}}

 \tablehead{
 \multicolumn{2}{c}{Source Model} & 
 \colhead{Rot. Angle} & 
 \colhead{$e$} & 
 \colhead{$\Omega_{s}$} \\
 \colhead{} &
\colhead{} & 
 \colhead{(deg)} & 
 \colhead{} & 
 \colhead{(arcsec)} 
 } 
 
 \startdata  

\multirow{4   }{*}{Sersic} & \hspace{-2em} $n$ = 1.5 & 55.0 & 0.75 & 0.875   \\ 
                                     & \hspace{-2em} $n$ = 2.0 & 55.0 & 0.75 & 0.35   \\
                                     & \hspace{-2em} $n$ = 1.25   & 140.0 & 0.90 & 1.75  \\ 
                                     & \hspace{-2em} $n$ = 2.0   & 140.0 & 0.90 & 0.55  \\ 
\hline \\
\multirow{2}{*}{Gaussian} & &  55.0 & 0.75 & 4.75  \\ 
                                        & & 140.0 &0.90 & 5.25   \\ 

\enddata
\tablenotetext{a}{Angular diameters $\Omega_s$ refer to best-fit major-axis diameters of the emitting source that provides optimum calibration of the background-subtracted SPIRE spectra.} 

\label{sect}

\end{deluxetable}

We have applied all three source distribution models to the SPIRE spectra of $\eta$~Carinae, adopting an initial conservative guess of 10$''$.0 for the source size and setting the eccentricities $e$ and rotation angles as  shown in Figure~\ref{sectprofiles} and listed in Table~\ref{sect} to represent simplified geometries for emission from the lobes and the equatorial region.  For the  Sersic model, $n$ = 2.0 confines 80\% of the power within an ellipse with major axis $a = 8''$.   When the initial guess is more than a factor of two from the optimized diameter, we apply the algorithm again using the result of the first iteration with smaller $\Delta\Omega_{source}$ bin sizes, converging to a final optimum value.   Again, to a very high degree of accuracy the corrected fluxes are independent of the assumed input model even if that model is not a perfect representation of the source, as long as a consistent  source size is used.

The results given in Table~\ref{sect} demonstrate that the largest extent of the emitting source at 306 $\mu$m is 5$''$, as fitted by the Gaussian profile on both the polar and equatorial axes. If we regard the Sersic model values to be more realistic, then the source is confined to significantly smaller sizes.

Before reaching a final conclusion on the indicated source size, consideration must be made about the background correction;  we must acknowledge that the smooth background assumption is not completely valid.  As can be seen in Figure~\ref{spireoffdets} there are several SSW detectors closer to the Homunculus (within $\approx$30$''$ of the core region) that we have not used in the corrections, lacking co-aligned SLW detectors. The inner SSW detector output indicates higher levels of background emission compared to the exterior co-aligned detectors, by a few Jy; see Figure~\ref{spireInsideDets}, where we have plotted spectra from two pairs of detectors oriented radially from the core region,  showing a clear radial drop-off of the fluxes.  An ad hoc application of inferred background levels that would fill the beams at corresponding SLW wavelengths completely closes the gap in the on-source spectrum, removing all indications of continuum source extension in the point-calibrated spectrum of $\eta$~Carinae.    Rather than commit to the conclusion that the far IR emission is escaping from a point-like region, the indications are for a $\sim$309 $\mu$m source size in the range of 2$''-5''$. This is {\it{not}} implied to be the size of the cool thermal emission region, regardless of the assumed source geometries, but only a feature that leads to the residual  discontinuity in flux at this wavelength.  We will show below that it is most likely associated with a structure observed with higher angular resolution at longer wavelengths.   

There are several important implications for these results concerning the radial decrease of background levels from the Homunculus:

\begin{enumerate}
\item The HIFI fluxes that are corrected for the backgrounds at positions 3$'$ on either side of the nebula are likely to be undercorrected.  However, we examine the continuum levels only for relative changes due to variability (Sec.~\ref{variability}).
\item The higher backgrounds within 30$''$ of $\eta$~Carinae could indicate lower-density material that has been ejected, or is still being ejected, along the lines proposed by Gomez et al. (2010) from ground-based submillimeter observations. 
\item While we have no indications of background contamination in the SWS observations (which would otherwise show jumps in the spectrum where the aperture sizes change toward the longer wavelengths; see Sec.~\ref{sws}), most certainly the LWS data are affected in a way that must be taken into account in the dust models.   
\end{enumerate}

\begin{figure}
 \begin{center}
 \includegraphics[width=8cm]{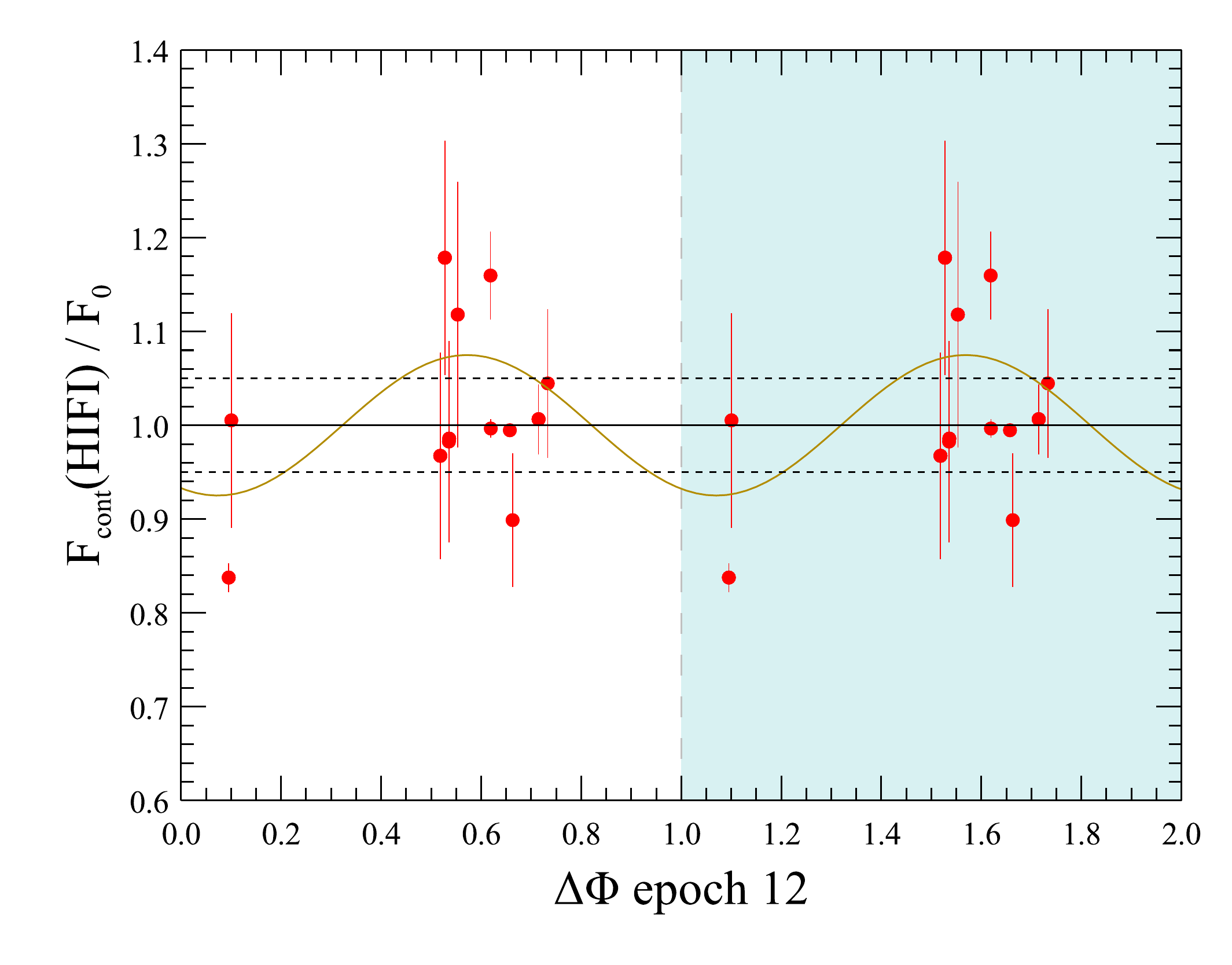}
 \end{center}
 \caption{Continuum levels extracted from HIFI observations of $\eta$~Carinae during epoch 12.  Data points in the shaded region at $\Delta \Phi > 1.0$ are repeated  for fitting the sinusoidal function, shown by the orange curve.  The flux normalization $F_0$ to the SPIRE FTS spectrum is described in Sec.~\ref{variability}.
\label{hificontinuumphase}}
\end{figure}   

\subsection{Far-IR continuum variability}\label{variability}

 Variability at near IR wavelengths has been studied by Whitelock et al. (2004) in the $JHKL$ bands during $2000-2004$, and addressed by Chesneau et al. (2005) in $JHKL'$ and narrowband 3.74 $\mu$m and 4.05 $\mu$m filters at various locations across the core region and the nebula, with the objective of confronting differences between their measurements of the correlated flux from the SED of the primary star and the CMFGEN non-LTE model atmosphere by Hillier et al. (2001).   Chesneau et al. (2005) have emphasized a number of challenges to establishing direct comparisons with the model atmosphere, including stellar wind asymmetries and intrinsic variability of the primary star, and influences of the dust-enshrouded companion star as a phase-dependent source of ionization.    Gomez et al. (2010) have also studied variability at 870 $\mu$m over an extended region around $\eta$~Carinae with the Atacama Pathfinder Experiment (APEX) telescope shortly after a radio maximum near apastron (cf. Duncan \& White 2003), and found indications of significant continuum variability compared with previous measurements, attributing the variation to the ionized stellar wind rather than dust emission.     Since we have demonstrated that the bulk of the cool dust observed with SPIRE must be concentrated around the central region, temporal measurements of the far IR continuum levels may further elucidate the importance of variability versus wind asymmetries.

\begin{figure*}
 \begin{center}
 \includegraphics[width=16cm]{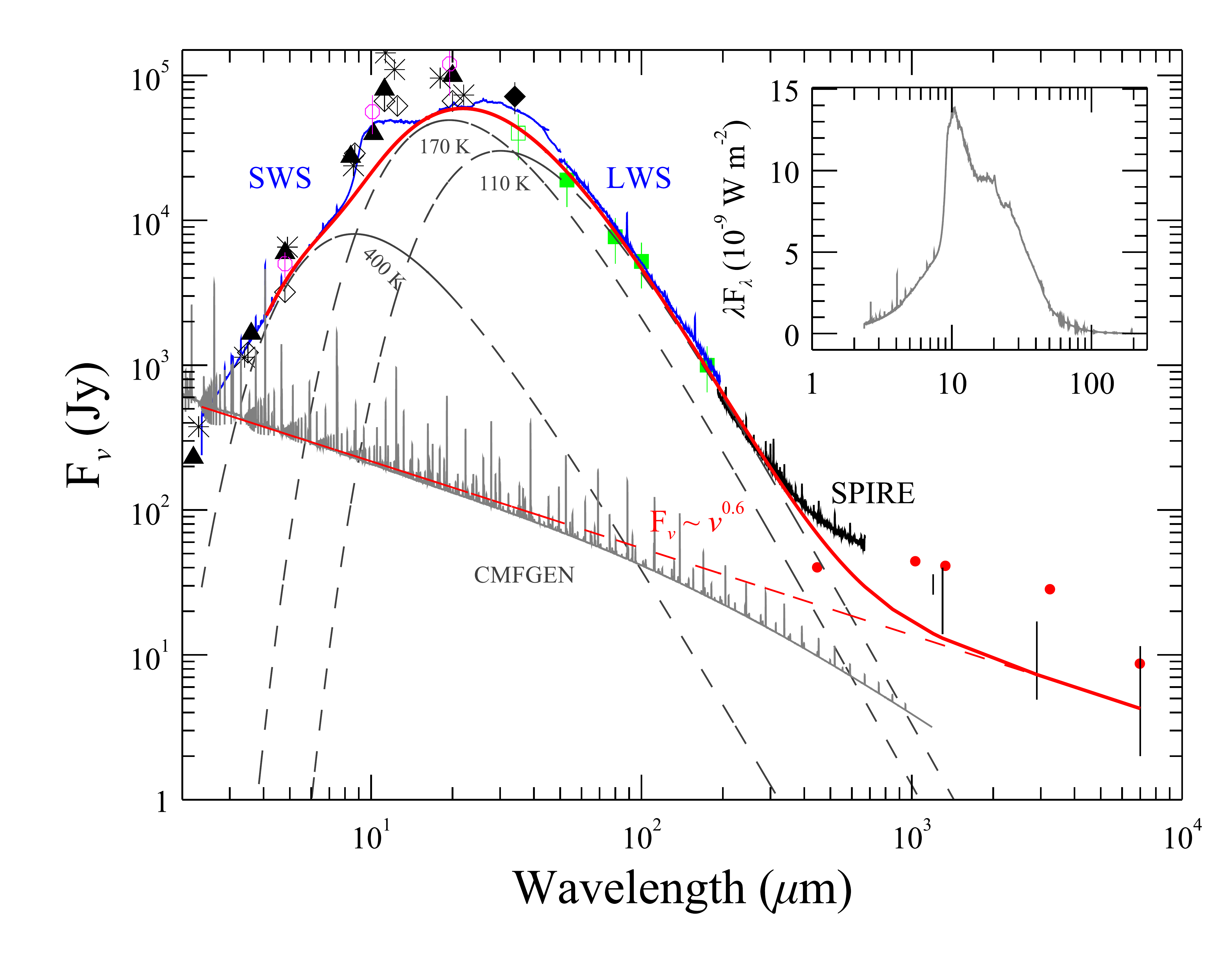}
 \end{center}
 \caption{Combined ISO and {\em{Herschel}} SED of $\eta$~Carinae's Homunculus Nebula, including published ground-based IR observations. The triangles at $2.0-20$ $\mu$m are from Robinson et al. (1973), stars at $2.3-22$ $\mu$m are from Gehrz et al. (1973), diamonds at $3.5-34.0$ $\mu$m are from Sutton et al. (1974), magenta circles at $4.8-19.5$ $\mu$m are from Westphal \& Neugebauer (1969), green squares at $35-175$ $\mu$m are from Harvey et al. (1978), red circles at $450-7000$ $\mu$m are ALMA observations from Abraham et al. (2014), and vertical black lines are compiled from Brooks et al. (2005) and Gomez et al. (2010).  The height of the black lines indicates the range of variability from those studies.  The filled symbols at $2.0-175$ $\mu$m are those data points selected for SED fitting by Cox et al. (1995).   The first ALMA point at 450 $\mu$m was taken at an angular resolution of 0$''$.45.  A CMFGEN model atmosphere for the primary star is shown in light gray (see text for a description), along with modified blackbodies to approximate the thermal SED following MWB99.  The red dashed line shows a reference $F_\nu \sim \nu^{0.6}$ power law for an optically thick free-free wind of the primary star.  The solid red line is the sum of the blackbodies and power law.  The inset shows the SED on a $\lambda_{\mu{\rm{m}}}$ {\em{vs.}} $\lambda$F$_\lambda$ scale to emphasize the enormous influence of the 10 $\mu$m band on the IR luminosity. 
\label{etafullsed} }
\end{figure*}

We examine our HIFI observations for indications of variability at multiple frequencies over 3/4 of an orbital phase in epoch 12 (see Table~\ref{obs_t}), adopting the SPIRE SED taken at $\Delta \Phi$ = 0.4828 as a reference.  As a heterodyne instrument, HIFI is designed for line stability rather than high-accuracy broad-band or continuum measurements, limited primarily by standing wave residuals compared to the baselines for line sources with zero continuum.   We estimate uncertainties of 5\% in setting the SPIRE reference levels and compute the errors on the HIFI continuum measurements from the RMS noise in the data and the standard deviations between measurements taken on the same day  --- there are always at least two for the H and V spectra originating from the optically independent mixer assemblies.  

We  proceed first by converting the HIFI data from the antenna temperature scale to flux densities in order to utilize the SPIRE SED as the reference at corresponding frequencies. In the framework of the intensity calibration of HIFI, the correction algorithm has the form 

\begin{equation}
 F_{ext}(\nu) = T_A(\nu) \; {2\; k\over{(A_{geom} \;  \eta_A) \; K(x)}}
\end{equation}

\noindent where $K(x) = [1 - {\rm{exp}}(-x^2)]/x^2$ is the source distribution function of $x \propto (\Omega_{source} / \Omega_{beam})$, with $\Omega_{beam} \sim \lambda \; \sqrt{1/(\eta_A \;  A_{\rm{geom}})}$, the effective area of the telescope $A_{\rm{geom}}$ = 8.45 m$^2$, and with $T_A$, $\eta_A$, and $\eta_B$ as above in Sec~{\ref{HifiObsSec}}.  The values of $\eta_A$ and $\eta_B$ at each observed wavelength $\lambda$ are derived from beam calibration observations (Roelfsema et al. 2012) and stored in calibration look-up tables with each set of observation products.$^{13}$    The HIFI conversion algorithm used here assumes a Gaussian source profile. However, we are not actually using this algorithm to estimate a source size; we simply adopt an equivalent source size of 5$''$ consistent with the semi-extended source corrections to the SPIRE data and put HIFI fluxes on a Jy scale.

 In order to utilize the SPIRE spectrum as a reference, we next apply an offset of $-$4.3 Jy to all HIFI continuum measurements so that the HIFI data taken at $\Phi$ = 12.5183 (obsid 1342232979) match that of SPIRE at 498 GHz.  This removes a systematic offset that may be present from the different locations of the background correction measurements applied in the data processing.  It also assumes that no variability has occurred between $\Phi$ = 12.4838 to 12.5183.    
 The results are shown in Figure~\ref{hificontinuumphase}, in which the measurements are mirrored in the shaded region in order to check for periodicity.

Indications of variability at the mean frequency of 500 GHz (600 $\mu$m) are tenuous, due to the size of the error bars (neither HIFI nor the SPIRE FTS is a precision photometer) and nonuniform temporal coverage, but a simple least-squares fit of a sine wave to the data suggests a periodicity in phase with the orbital period, at a peak-to-peak amplitude of $\approx$15\% around the normalized continuum.  The indicated increase in continuum levels peaks near $\Delta \Phi$ = 0.57, following apastron passage, which might suggest  either that the dust temperatures are decreasing near the central region during apastron (thus raising the far-IR continuum) or that more dust is being destroyed than created during periastron passage, or possibly from changes in ionization in the stellar wind of the primary.  In any case, any large variations (cf. Gomez et al. 2010 at 870 $\mu$m) would be more apparent; the indications of variability at the indicated levels here should be regarded with caution and are too low to affect our analysis of the dust grains below.

\subsection{The Full IR to millimeter SED of $\eta$~Carinae}\label{fullsed}

The 2.4 $\mu$m$-$0.7 mm SED of $\eta$~Carinae is shown in Figure~\ref{etafullsed}.    We include ground-based mid-IR measurements by Robinson et al. (1973), Sutton et al. (1974), and Harvey et al. (1978), showing reasonably good agreement with the SWS and LWS continuum levels at most wavelengths.  Some  data near the peak of the SED are considerably higher, however, a factor of $\approx$1.5 between the ground-based flux points at 20 $\mu$m, for example.  Mid-IR variability between the Robinson et al. (1973) and Sutton et al. (1974) observations at this level is unlikely (but not out of the question), as they were obtained $1-2$ yr apart.  The filled symbols between 2 and 175 $\mu$m were selected by Cox et al. (1995) to fit the IR to submillimeter SED, obtaining $L_{\rm{IR}}$ = 5 $\times$ 10$^{6}$ $L_\odot$.   From our SED we obtain a lower value,  $L_{\rm{IR}}$ = 2.96 $\times$ 10$^{6}$ $L_\odot$.  The difference is clearly due to the high-end flux points they selected, especially at 11 and 20 $\mu$m.  The 10 $\mu$m band dominates the IR luminosity, as shown in the inset plot in Figure~\ref{etafullsed}: integrated over 8 to 16 $\mu$m, it makes up 75\% of the total luminosity. The submillimeter portion of the SED contributes less than 5\% to the total over the $2.0-1000$ $\mu$m baseline.  Gomez et al. (2010) obtained $L$($12-1000$ $\mu$) = 1.6 $\times 10^6 L_\odot$, while we measure a value just slightly higher, 1.7 $\times 10^6 L_\odot$ over the same range (and using the same distance of 2.3 kpc to $\eta$~Carinae; Allen \& Hillier 1993). 

 The continuity between the SWS and LWS spectra is quite good (with the caveats about residual photometric nonlinearities in the overlap range; Sec.~\ref{lws}). The obvious discontinuity between the LWS and SPIRE spectrum near 200 $\mu$m is due to the uncorrected backgrounds in the LWS aperture at these longer wavelengths.   We will take this into account in our dust modeling.  We also include points from observations compiled by Brooks et al. (2005), and at the APEX Telescope by Gomez et al. (2010).     The vertical bars follow Brooks et al. (2005) and Gomez et al. (2010) to represent the range of variability as indicated in their study.  

There is very good continuity between the SPIRE spectrum and the ALMA observations (which were obtained late in epoch 12 with beams 0$''$.45$-$2$''$.88 FWHM from short to long wavelengths, thus the lower flux point at 450 $\mu$m); both are significantly higher over most submillimeter wavelengths compared to the ranges reproduced from Brooks et al. (2005) .  Structures observed in selected H recombination lines and the excess emission peaking near 3 mm have been postulated by Abraham et al. (2014) to be due to the detection of an unresolved source of free-free emission near the core created in the 1941 ejection event that formed the ``Little Homunculus'' (Ishibashi et al. 2003), dubbing this new thermal feature a ``Baby Homunculus.''  Gull et al. (2016) have since pointed out that this feature is more likely to be an accumulated ionized source of emission, due to the periodic interacting wind structures, mapped in the 3 cm radio continuum by Duncan \& White (2003) and extended by [Fe~{\sc{ii}}] and [Fe~{\sc{iii}}] structures observed with the Space Telescope Imaging Spectrograph,  starting in mid-2009 and followed across the most recent 5.5 yr cycle.  The continuity between the SPIRE and ALMA continuum flux levels would suggest that the $\approx$2$''$ source measured from the semi-extended source correction of the SPIRE data (Sec.~\ref{extend}) is related to these inner  wind-wind structures.  
 
Interestingly, the flux brightening reported by Gomez et al. (2010) at 850 $\mu$m is based on observations with APEX/LABoCa (which has a similar beam size to the SPIRE data), taken 1.7 yr after the maximum in the radio and X-ray cycle, which occurs close to apastron (Duncan, White \& Lim 1997), whereas the SPIRE and ALMA data have also been acquired close to and just after apastron passage ($\Delta \Phi$ = 0.48 and 0.69, respectively) and are higher at 850 $\mu$m and nearly all upper ranges to the data compiled by Brooks et al. (2005) at similar wavelengths.  This could indicate even higher increases in submillimeter emission correlated with the radio cycle, and it is also in phase with the low-amplitude variability indicated (albeit tenuously) in our HIFI continuum measurements (Sec.~\ref{variability}). Variability at a level of $\approx$25\% has also been seen in the 3 mm continuum between the ALMA data by Abraham et al. (2014) and Australia Telescope Compact Array observations by Loinard et al. (2016), indicated to be lower 1.67 yr after the ALMA observations, which were taken near periastron.  

\begin{figure}
 \includegraphics[width=8.5cm]{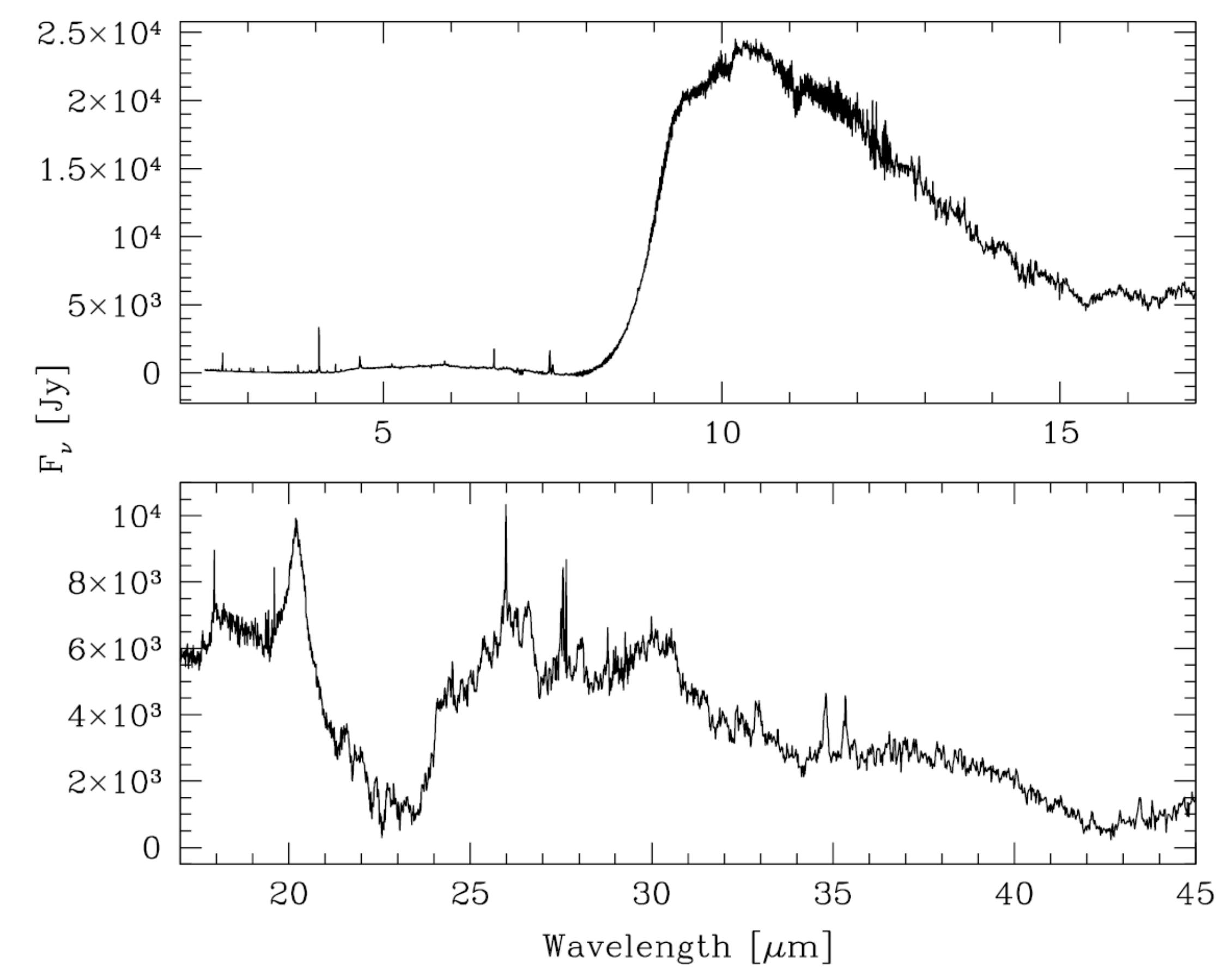}
 \caption{Solid-state dust bands over the SWS range, obtained by differencing the total spectrum with the broadband Fe dust contribution from Model A (presented in Sec.~\ref{sec:dustmodeling}).  The upturn in the $42-45.2$ $\mu$m range is unreliable as a result of possible high-input signal memory residuals here and in the LWS01 detector output (see Sec.~\ref{lws}).    
\label{swsdust} }
\end{figure}

Following MWB99, we show an updated three-component modified blackbody fit for reference.  We follow MWB99 more closely, consistently avoiding the principle solid-state dust features at 9-15 $\mu$m and 23-38 $\mu$m for the fitting.   The best-fit temperature of the cool modified blackbody remains $T_{d,{\rm{cool}}}$ =  110~K , but the total contribution of this component to the total emission is slightly lower.   The lower IR luminosity estimate begins to alleviate the problem of energy output from such a small region indicated by the ISO and {\em{Herschel}}/SPIRE beam constraints, but this model is only a coarse representation of the thermal emission in the spatially integrated spectrum.   We will address the inferred source sizes below. The total modified blackbody curve remains a difficult or misleading tool for setting a reference baseline that reveals the astonishing collection of solid-state features present in the SWS spectrum, as many of these are blended (i.e., the method tends to overfit the baseline).  This is better accomplished with the full dust model.   In Figure~\ref{swsdust}, we show the narrow dust features after subtraction of the broadband contributions from our modeling (presented below) from the total observed SED.    We point out the significant, broad emission band over the 17 - 23 $\mu$m range, including substantial emission near 18 $\mu$m.  Emission at these wavelengths was claimed (incorrectly) by Chesneau et al. (2005) {\em{not}} to be present in this same ISO spectrum as a justification for a dominant contribution by corundum (Al$_2$O$_3$) rather than silicates to the 10 $\mu$m band.  We will address the balance of dust species in our modeling below.

We also show the CMFGEN model atmosphere for the primary star (Hillier et al. 2001) computed at the time of the SPIRE observation and extrapolated to 1mm.  By comparison to the $F_\nu \sim \nu^{0.6}$ power law for an optically thick free-free wind, the CMFGEN has an obvious downturn beginning already in the mid-IR.  This is due to the recombination of H in the outer wind, under the influence of the ionizing companion star at this epoch. At most epochs, the companion will likely ionize the outer wind, leading to an increase in free-free emission, and is probably one reason for variability at these longer wavelengths. 

\section{Mass and Composition of the Dust in the Homunculus}\label{sec:dustmodeling}

\subsection{Dust model inputs and fitting}

We use the radiative transfer code DUSTY (Ivezic et al. 1999) to model the dust in the Homunculus Nebula.  This code solves the
1D radiative transfer equation for a spherically symmetric dust shell centered on a source of radiation. The inputs are an SED for the source, the physical parameters of the dust, a radial density profile, and the optical depth of the dust shell at a reference wavelength. We employ the CMFGEN model SED shown in Figure~\ref{etafullsed} for the central source. The optical properties of the dust are specified by a set of optical constants (the real and complex indices of refraction $n$ and $k$), which are derived from reflectance measurements of representative samples in the laboratory.  The code then requires a grain size distribution, density profile, optical depth, and the temperature of the dust at the inner radius of the shell.  Up to 10 sets of optical constants may be provided at a specific grain size distribution, temperature, density profile, and optical depth.    

\begin{figure}
 \begin{center}
 \includegraphics[width=8.5cm]{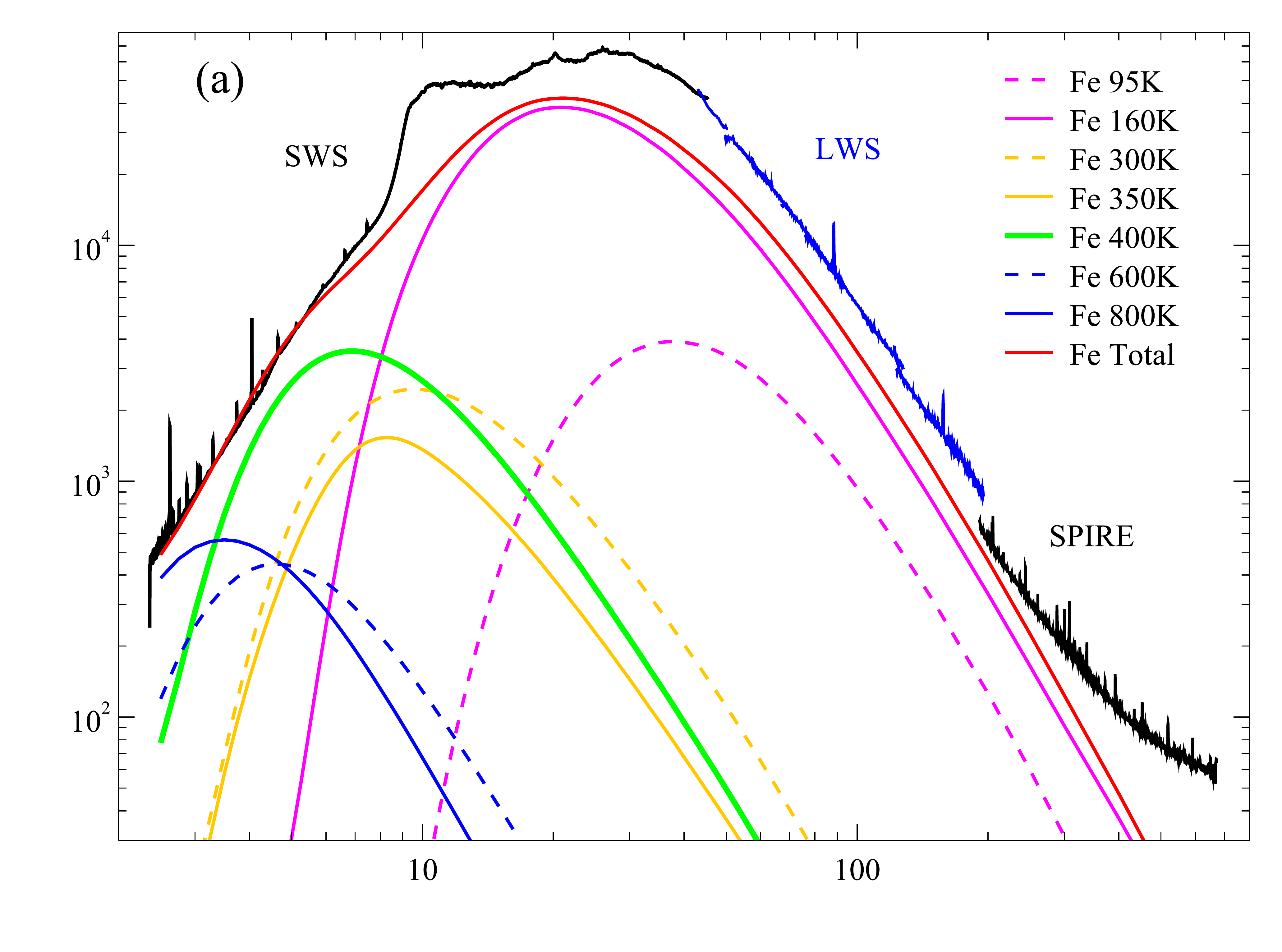}\\
 \includegraphics[width=8.5cm]{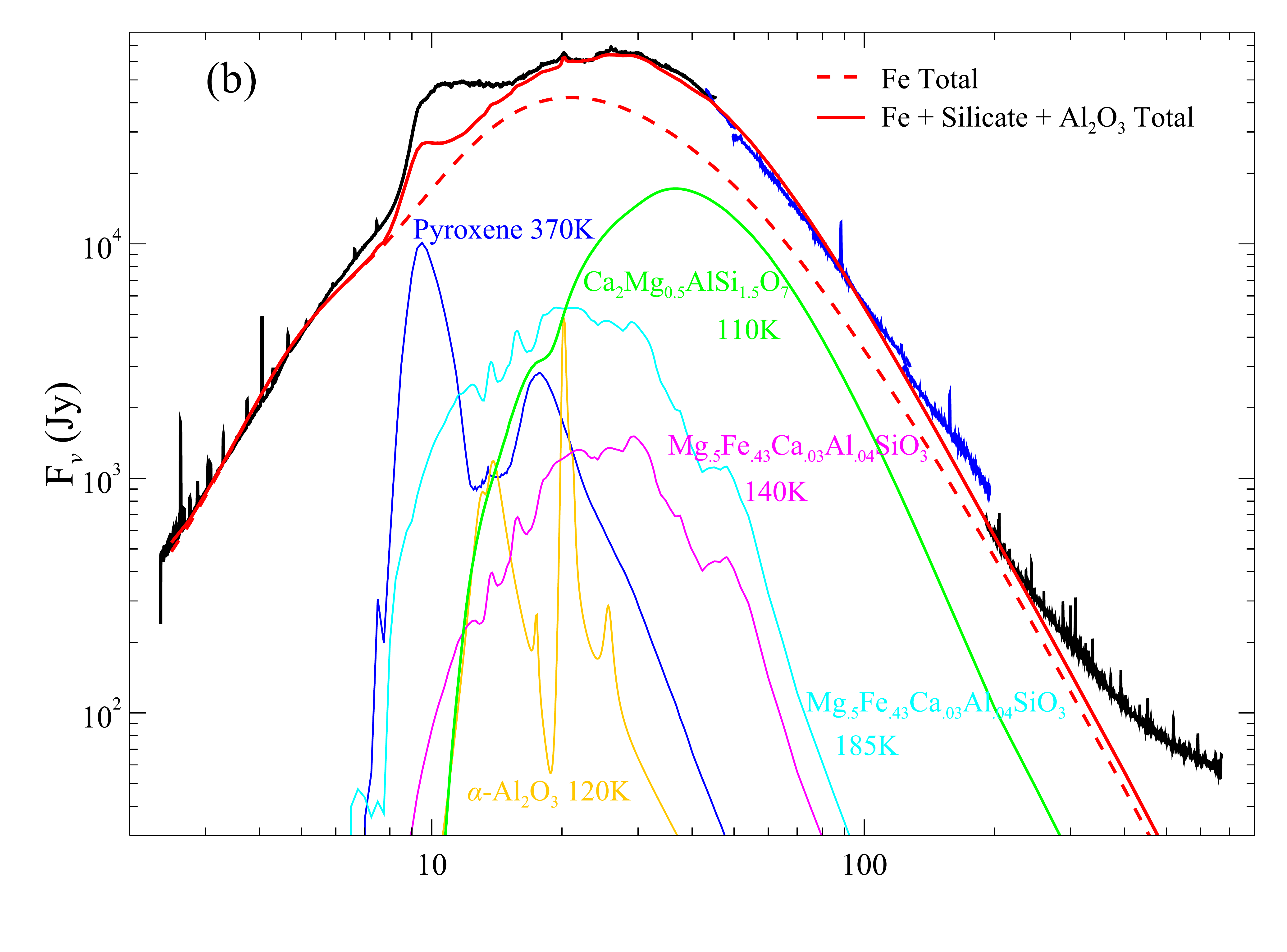}\\
 \includegraphics[width=8.5cm]{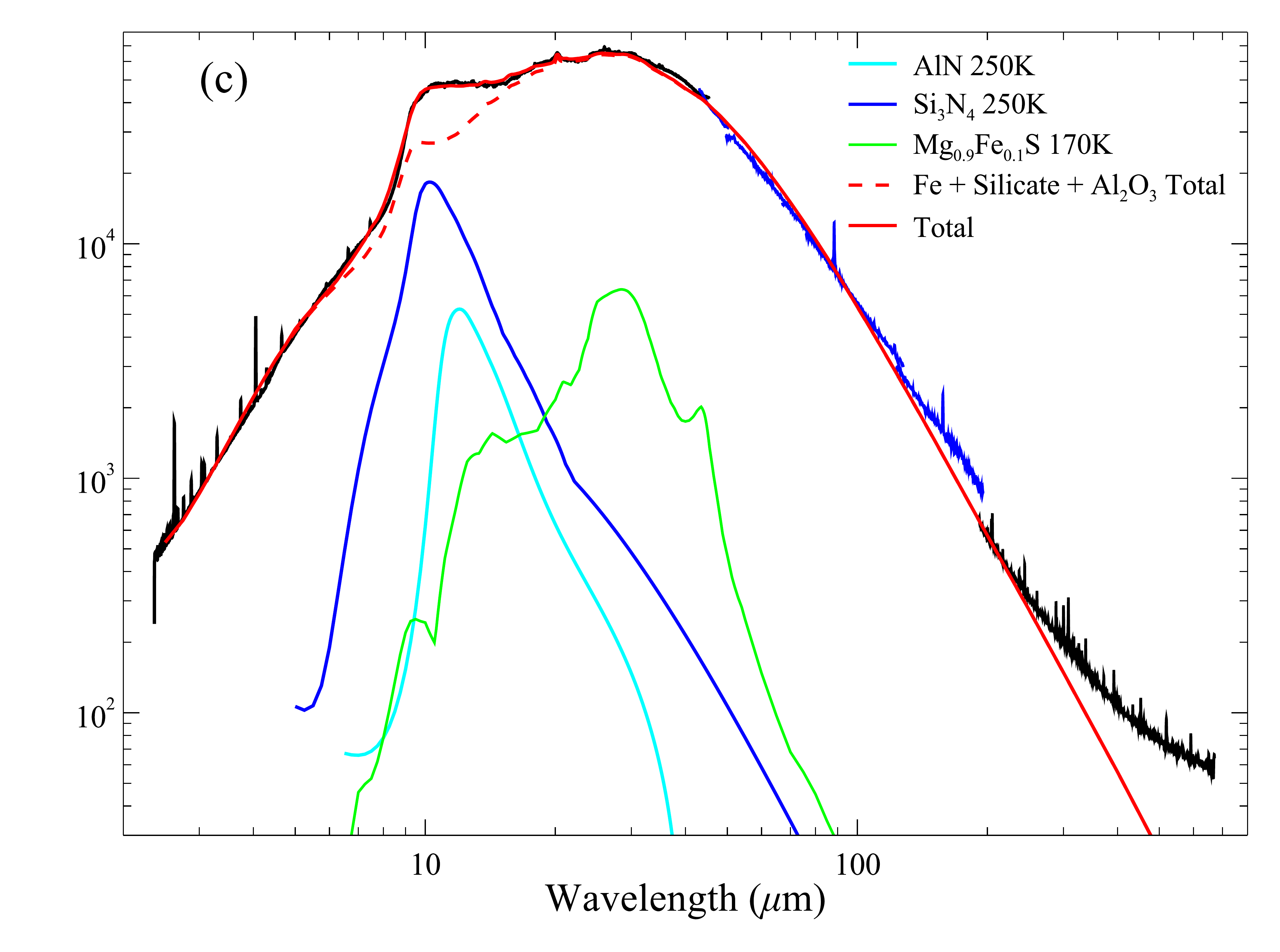}
 \end{center}
 \caption{Model A: best-fit composition dominated by pure Fe grains, metal-rich crystalline and amorphous silicates, corundum, iron-magnesium sulfite, and nitrates.  The fitting has been done simultaneously but plotted for clarity of the relative contributions of the dust constituents to the major spectral features.  Plot (a) shows the Fe grain contributions at $T_{\rm{d}}$ between 95 and 800~K; (b) shows the addition of corundum and silicates; and (c) shows the sulfite and nitride contributions.  The fitting at $\lambda > 50 \mu$m weights the LWS fluxes at their midpoint and assumes that fluxes at longer wavelengths are contaminated by background emission.  The SPIRE spectrum is corrected for background emission.     
   \label{fig:ModelA}}
\end{figure}

The spherical symmetry assumption for our spatially unresolved composite ISO spectrum uses the simplest possible geometry, with the fewest free parameters, and is justified for lack of other constraints on the actual geometry.  While we have argued that the bulk of the dust emission is concentrated in the inner $\approx$5$''$ based on the ISO aperture and SPIRE beam size constraints  (Sec.~\ref{extend}), the geometry of the dust is probably quite complex; MWB99 suggested that the coolest thermal component ($T_{\rm{d}} \approx$ 110 K) is primarily located in a configuration related to the disrupted torus-like structure apparent in 18 $\mu$m imaging (see Fig.~\ref{sectprofiles}f). Dust may also be forming in regions of interaction between the present-day stellar winds of the companion stars, and with prior ejected material.  Hence, the dust distribution does not resemble a typical red supergiant or asymptotic giant branch (AGB) star ejection shell, and we avoid the ambiguities of geometry for an initial analysis of the dust chemistry, temperatures, and masses.  Under these conditions, the input dust density profile and optical depths are somewhat arbitrary, and thus we adopt the DUSTY default $r^{-2}$ density profile and fix the relative thickness of the shell to less than 100 and $\tau_{\rm{d}}$(0.55 $\mu$m) $\leq$ 2.0.  This bounding helps reduce the propensity for a high number of degenerate solutions because of the large number of free parameters in the modeling, and will be examined as part of the uncertainties on our results (Sec.~\ref{dustmodelerrors}).

Grain sizes $a$ are represented by the so-called ``MRN'' (Mathis, Rumple \& Nordsieck 1977) power-law distribution $n(a) \propto a^{-q}$ for $a_{\rm{min}} \leq a \leq a_{\rm{max}}$.  The grains are assumed to be spherical, homogeneous, and nonaligned as given by standard Mie theory.   While other grain shape distributions can be found in the literature, namely, the Continuous Distribution of Ellipsoids (CDE; Bohren \& Huffman 1983) and the Distribution of Hollow Spheres (DHS; Min et al. 2005), both of which can alter dust band profiles in comparison to spherical grains under the same temperature and density conditions, it is clear when fitting a limited number of observations with many free input parameters that grain shape, size, temperature, and composition cannot be disentangled and must be fit in combination.  Adopting the most fundamental and least ambiguous shape possible therefore avoids overfitting the problem.  Despite these multiparameter limitations, fitting a dust spectrum across a wide wavelength baseline that covers the main optically active modes of vibration (e.g., Jaeger et al. 1998) provides some constraints, and we may indeed have support for spherical grains in at least one major dust species as will be shown below. 

The optical properties of dust species spanning the widest range of compositions possible have been assembled from the literature into over 50 tables of indices of refraction $n$ and $k$ as a function of wavelength.  The compositions include the principle families of silicates, glasses, metal oxides, pure metals, sulfites, carbonaceous grains (carbon and carbides), and nitrates.  These materials include crystalline and amorphous grain structures and may include the same species measured at different temperatures in the laboratory.   The silicates have varied degrees of metal (Fe, Mg, Ca, Al, Na) inclusions, as well as both O-rich and O-poor species, where the latter are more common in the circumstellar environments of evolved stars (Ossenkopf, Henning \& Mathis 1992).  C and O are both known to be at very low abundances overall, while N is enhanced in the Homunculus as a consequence of the CNO cycle, but abundance variations are present in the outer ejecta (Smith \& Morse 2004) and rotational transitions of the CO molecule have been detected (Loinard et al. 2012 and below); hence, we include the possibility of both extremes of carbonaceous dust and nitrates.  

Our modeling emphasizes four key spectral regions:  (1) the power law-like continuum $2.4-9$ $\mu$m, (2) the uniquely broad and strong band peaking at 10 $\mu$m, (3) the $15-30$ $\mu$m region containing a number of narrow to broad dust bands, and (4) the featureless long-wavelength region 35 to $\sim$250 $\mu$m covering the LWS data and portions of the SWS and SPIRE data.  The SPIRE spectrum used in the fitting has been corrected for background emission as described above, while the LWS data have not been corrected (they can only be inferred).  Fitting is done to all regions simultaneously, and we allow grain sizes to run logarithmically from 0.001 to 100 $\mu$m using both a power index $q$ of 3.5 (the MRN default) and 0 (fixed sizes) and with $T_{\rm{d}}$ varying between 50 and 1500 K for most species (particularly at the hot end for pure metal grains).  

The mass of dust component $i$ is calculated as

\begin{equation}
M_{{\rm{d}},i} = {D^2 F_{{\nu},i}(\lambda)\over{\kappa_{{\nu},i}(\lambda) B_\nu(\lambda,T_{{\rm{d}},i})}}
\label{eq:dustmass}
\end{equation}
where $D = 2.3$ kpc is the distance to $\eta$~Carinae, and for a dust component $i$,  $F_{{\nu},i}(\lambda)$ is the flux density,  
$\kappa_{{\nu},i}$ is the mass absorption coefficient, and $B_\nu(\lambda,T_{{\rm{d}},i})$ is the Planck function evaluated at the derived dust temperature.  The mass absorption coefficient $\kappa_\nu$ is equal to 3$Q_{\rm{abs}}$/4$\rho a$, where $\rho$ is the dust bulk density, typically adopted to equal 3.0 g cm$^{-3}$, and $Q_{\rm{abs}}$ is the grain absorption efficiency. The values of $Q_{\rm{abs}}$ are calculated from the optical constants and grain size using Mie theory as described by Bohren \& Huffman (1983).  The total mass is then calculated by summing Eq.~\ref{eq:dustmass} over all components $i$ of the fit.

\begin{figure}
 \begin{center}
 \includegraphics[width=9cm]{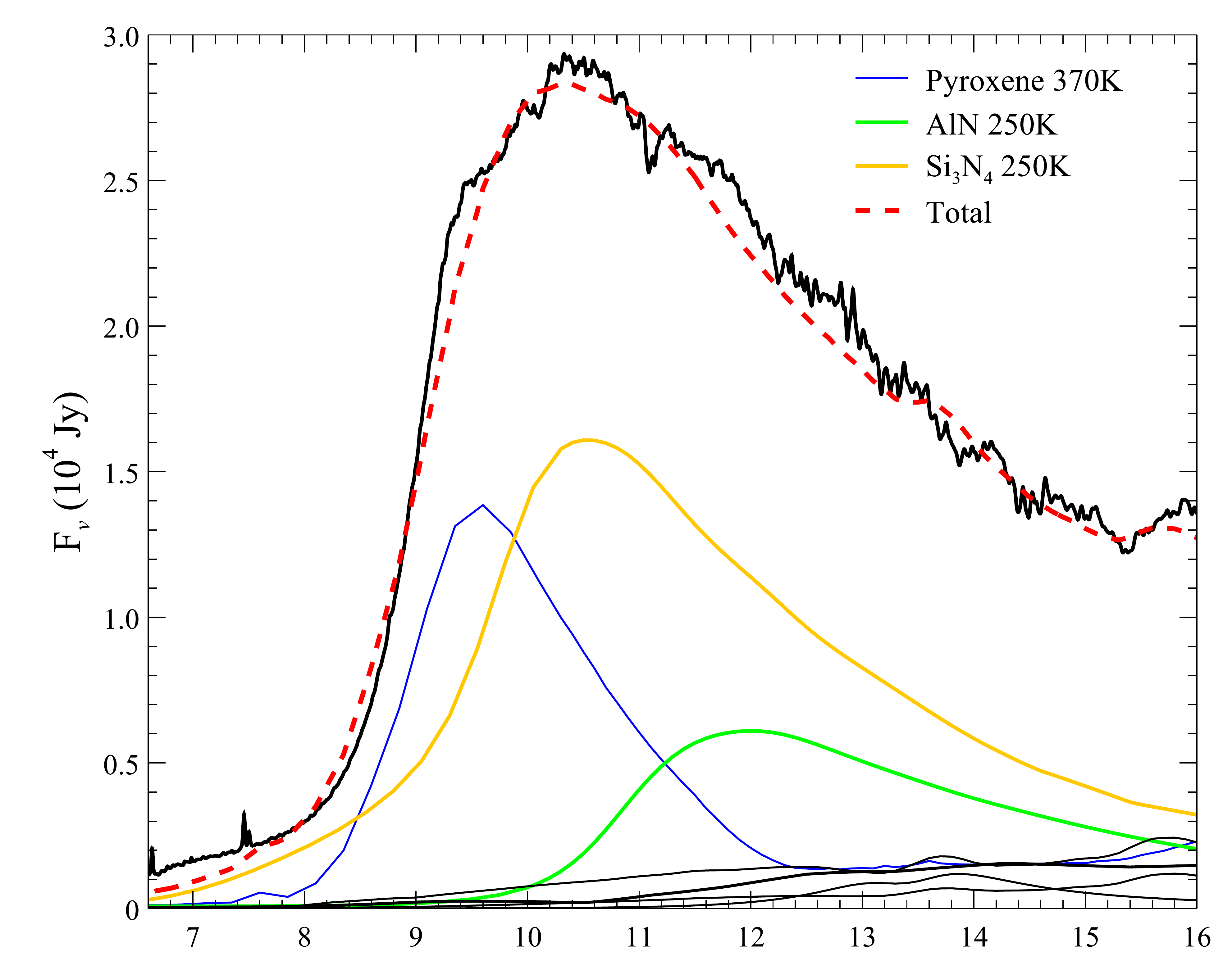} \\
 \includegraphics[width=9cm]{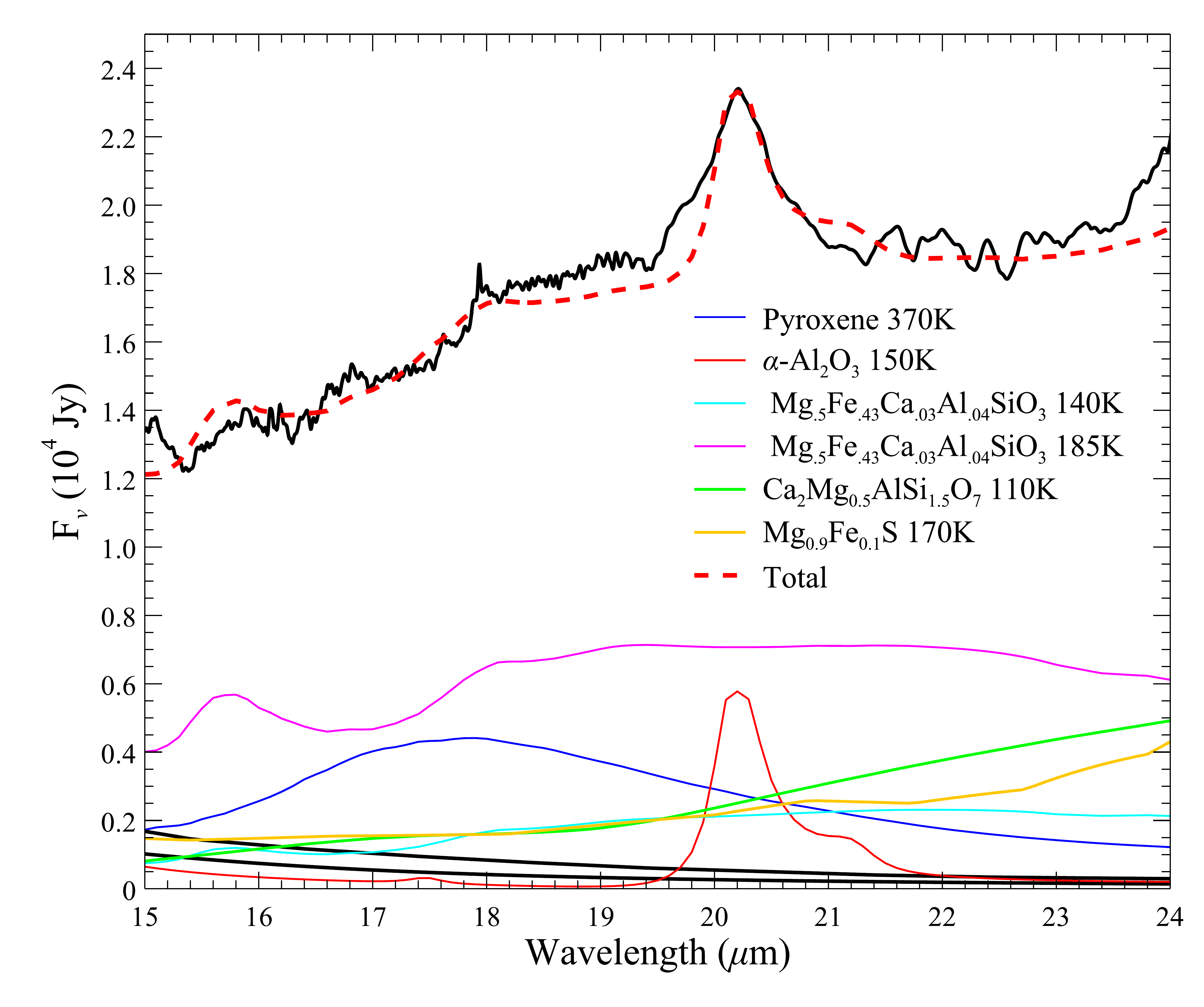} 
 \end{center}
 \caption{Magnification on the 10 and 20 $\mu$m bands, highlighting the principle contributing dust species in Model A.  The observed spectrum (black) and dust modeling have been subtracted by the total contribution from Fe grains, shown in Fig.~\ref{fig:ModelA}.   
\label{modelA1020bands} }
\end{figure}

Next we present our two sets of models emerging as best fits to the 2.4 - 250 $\mu$m dust spectrum,  representing the range of derived dust compositions, temperatures, and masses.  The results are summarized in Table~\ref{table:dustmodels}.

\subsection{Model A:  the dominance of Fe and indications of nitrides, corundum, and sulfites}\label{ModelA} 

Figure~\ref{fig:ModelA} shows the results of the fitting in which the featureless portion of the dust continuum is dominated by pure Fe grains at temperatures between 95~K and at least 800 K (panel (a)) and the aluminosilicate glass (ASG) Ca$_2$Mg$_{0.5}$AlSi$_{1.5}$O$_7$ at 120 K (panel (b)).  These all require relatively small grains, which have higher extinction efficiencies compared to large grains, and dominate the dust mass in the three lowest-temperature Fe components and the ASG; see Table~\ref{table:dustmodels}.  The Rayleigh-Jeans portion of the aggregate model is equally weighted from SWS wavelengths to the midpoint of the LWS spectrum, then avoids the uncorrected background contamination at longer LWS wavelengths, and then is weighted to unity over the 190 - 215 $\mu$m portion of the SPIRE spectrum (which is background corrected but contains the onset of the source of excess emission peaking at longer wavelengths; see Fig.~\ref{etafullsed}). The ASG at the tabulated grain size and temperature provides a similar broad but somewhat more peaked profile compared to Fe that provides a decent match to the $\sim$23-45 $\mu$m range of the observed SED, particularly in combination with Mg$_{0.9}$Fe$_{0.1}$S for the 25-30 $\mu$m plateau of emission (panel (c)), better than other species in this Fe-dominated model.

Hot Fe grains provide the best fit to the power-law-like range of the SED to $\sim$9 $\mu$m, for $T_{\rm{d}} \geq$ 300 K.  As pointed out by MWB99, the grains are likely to be distributed over a continuous range of temperatures up to $\sim$1500 K and are probably not in thermal equilibrium.  Our fitting has used temperature intervals of 50 K, and as shown in Figure~\ref{fig:ModelA},  the 300, 350, 400, 600, and 800 K components provide the measurable contributions in this wavelength range, none contributing more than $\sim$1\% to the total dust mass.   

Panels (b) and (c) in Figure~\ref{fig:ModelA} show the contributions of the remaining dust species emerging from the fitting that fills in the observed band structure.  As shown in Figure~\ref{modelA1020bands}, the fit to the strong 10 $\mu$m band with the unusually broad red wing is well fit by a simple combination of amorphous pyroxene at 350 K and two nitrides at 250 K.   All three are only minor contributors to the total dust mass, but they provide enormous amounts of scattered radiation needed to match the high luminosity ($\approx$52 $L_\odot$) of  this band.  The shape of the band at $\lambda \gtrsim$ 14.5 $\mu$m is also influenced by a similarly low concentration of the metal-rich crystalline silicate Mg$_{.5}$Fe$_{.43}$Ca$_{.03}$Al$_{.04}$SiO$_3$ (essentially an olivine or mafic silicate with Ca and Al inclusions, known in terrestrial settings as augite) at 185~K, which is a more important contributor to the 20 $\mu$m band discussed below.  The composition we derive in this model for the 10 $\mu$m band deviates significantly from the olivine and corundum mix proposed by Chesneau et al. (2005), most certainly due to the limited wavelength coverage ($7.5-13.5$ $\mu$m) in their ground-based observations and exclusion of nitrides as potential carriers in their model.  The implications for nitrides as key carriers of the 10 $\mu$m feature will be discussed in Section~\ref{sec:Discussion}.    

\begin{figure}
 \begin{center}
 \includegraphics[width=8.5cm]{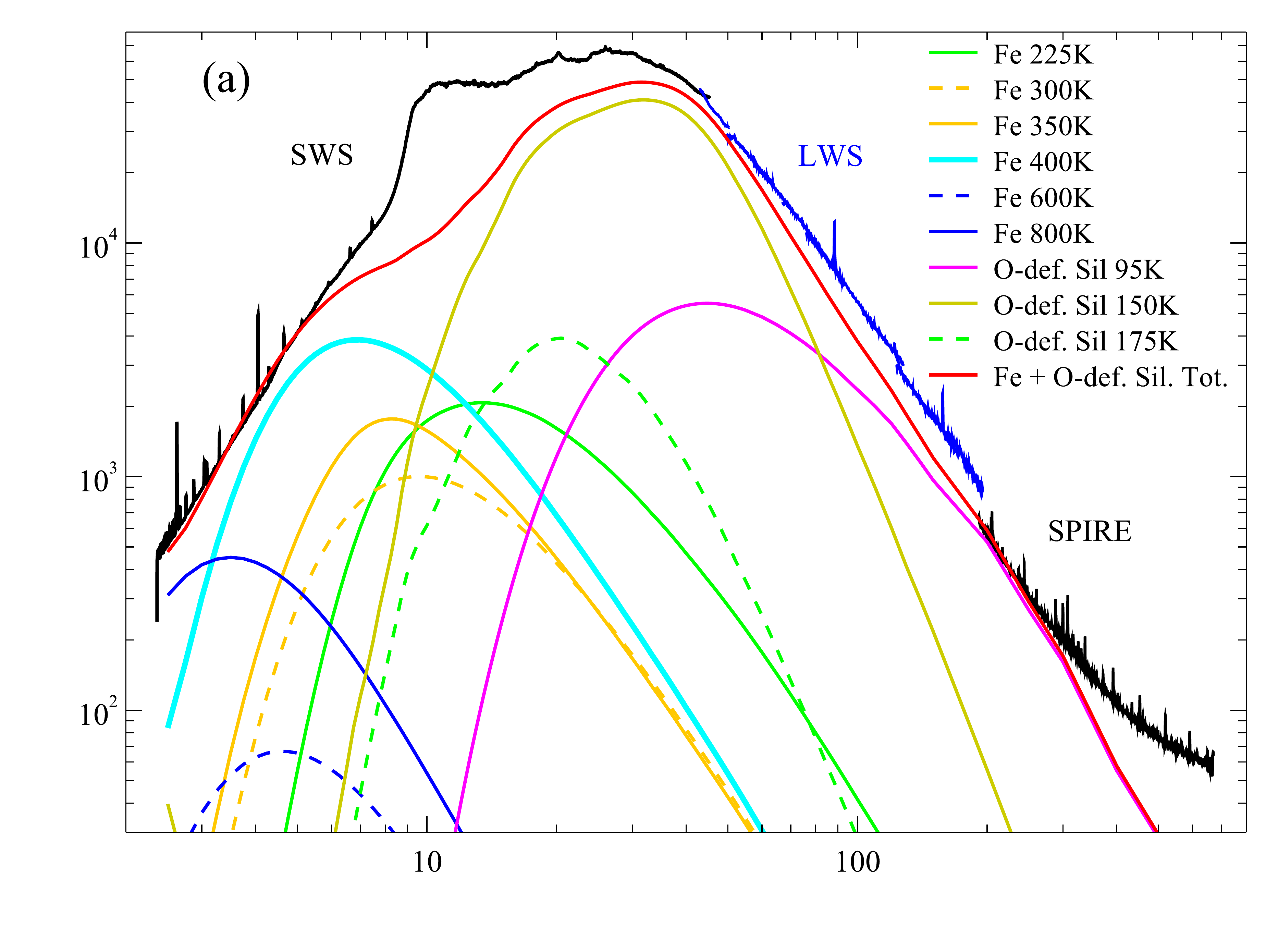}\\
 \includegraphics[width=8.5cm]{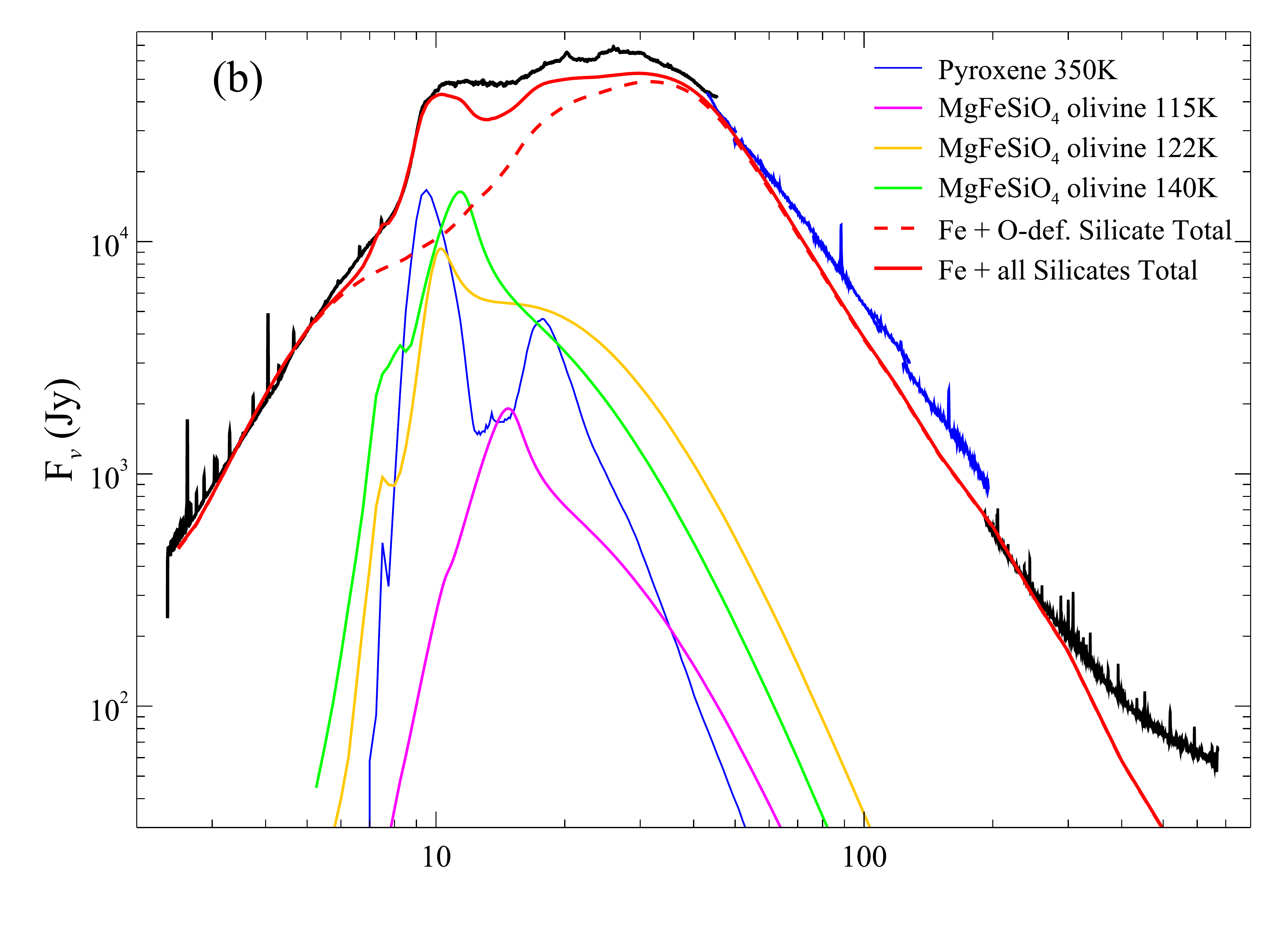}\\
 \includegraphics[width=8.5cm]{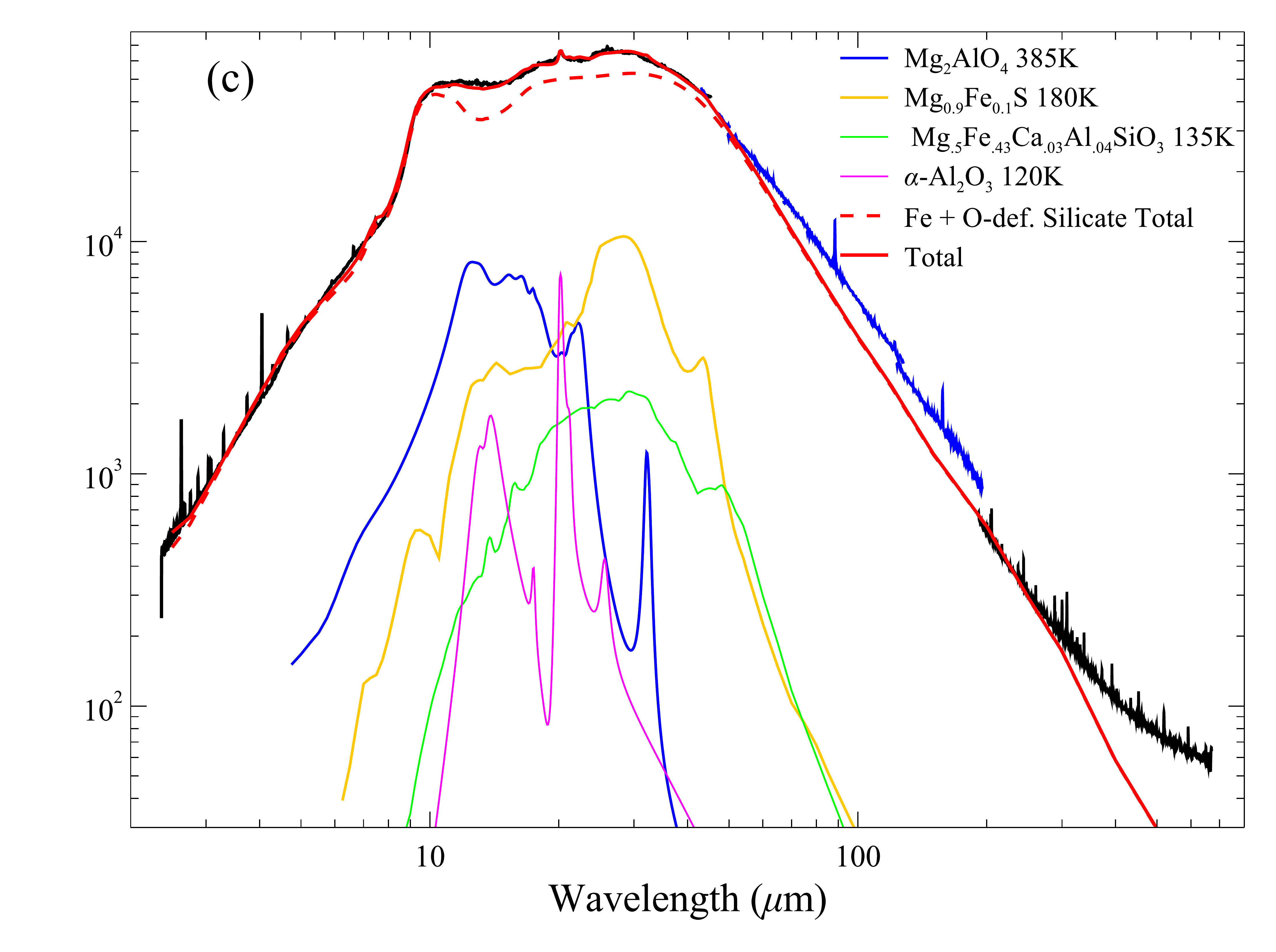}
 \end{center}
 \caption{Model B: composition is  dominated by warm and hot ($T_{\rm{d}}$ = $225- 800$ K) pure Fe grains and cool ($T_{\rm{d}}$ = $95-175$ K) O-deficient silicates making up the bulk of the emission, metal-rich crystalline and amorphous silicates,  corundum, spinel, and magnesium-iron sulfite.  Like Model A, the fitting has been done simultaneously but plotted to show the relative contributions of the dust types to the major spectral features.  Here the long-wavelength fit weights the end of the SWS range and the 200-350 $\mu$m range of the background-corrected SPIRE, shortward of the excess feature extending into the submillimeter range in ALMA observations (see Fig.~\ref{etafullsed}). This demonstrates the possible extent of background contamination over the LWS range.        
   \label{fig:ModelB}}
\end{figure}

Another interesting feature in Model A is the good match between corundum (Al$_2$O$_3$) at 120 K and the narrow Lorentzian-like profile at 20.2 $\mu$m, as well as the surrounding plateau in combination with the crystalline metal-rich silicate augite; see Figure~\ref{modelA1020bands}.  Only spherical grains of corundum calculated in Mie theory produce such narrowband structure compared to the CDE and DHS grain shapes; see Fabian et al. (2001), Mutschke et al. (2009), Imai et al. (2009), and Koike et al. (2010) for studies of grain shape effects on the optical properties of various silicate grains.  We would otherwise need to add an {\rm{ad hoc}} synthetic Lorentzian profile to account for the contribution at 20.2 $\mu$m, but this would fail to add any opacity to the SED at 16 $\mu$m, where a secondary band from corundum contributes in our model.  The 20.2 $\mu$m match gives us some confidence that spherical grains not only are the least ambiguous shape to assume but also may indeed be observationally consistent with at least some dust carriers.

 \subsection{Model B: the dominance of Fe, olivine, and O-poor silicates}\label{ModelB}  

Model B is similar to Model A in terms of a near-equal goodness of fit and uses Fe carriers at comparable grain sizes and temperatures to fit the region to $\sim$9.0 $\mu$m.  However, the remaining ``featureless'' part of the SED now requires substantial contributions from olivine and O-poor silicates at temperatures between 95 and 150 K in order to match the Rayleigh Jeans side of the composite spectrum.  The absorption efficiencies of these carrier grains are markedly lower compared to the astronomical silicates adopted in the simple modes by MWB99 and Smith et al. (2003), driving up the dust masses.

\begin{figure}
 \begin{center}
 \includegraphics[width=9cm]{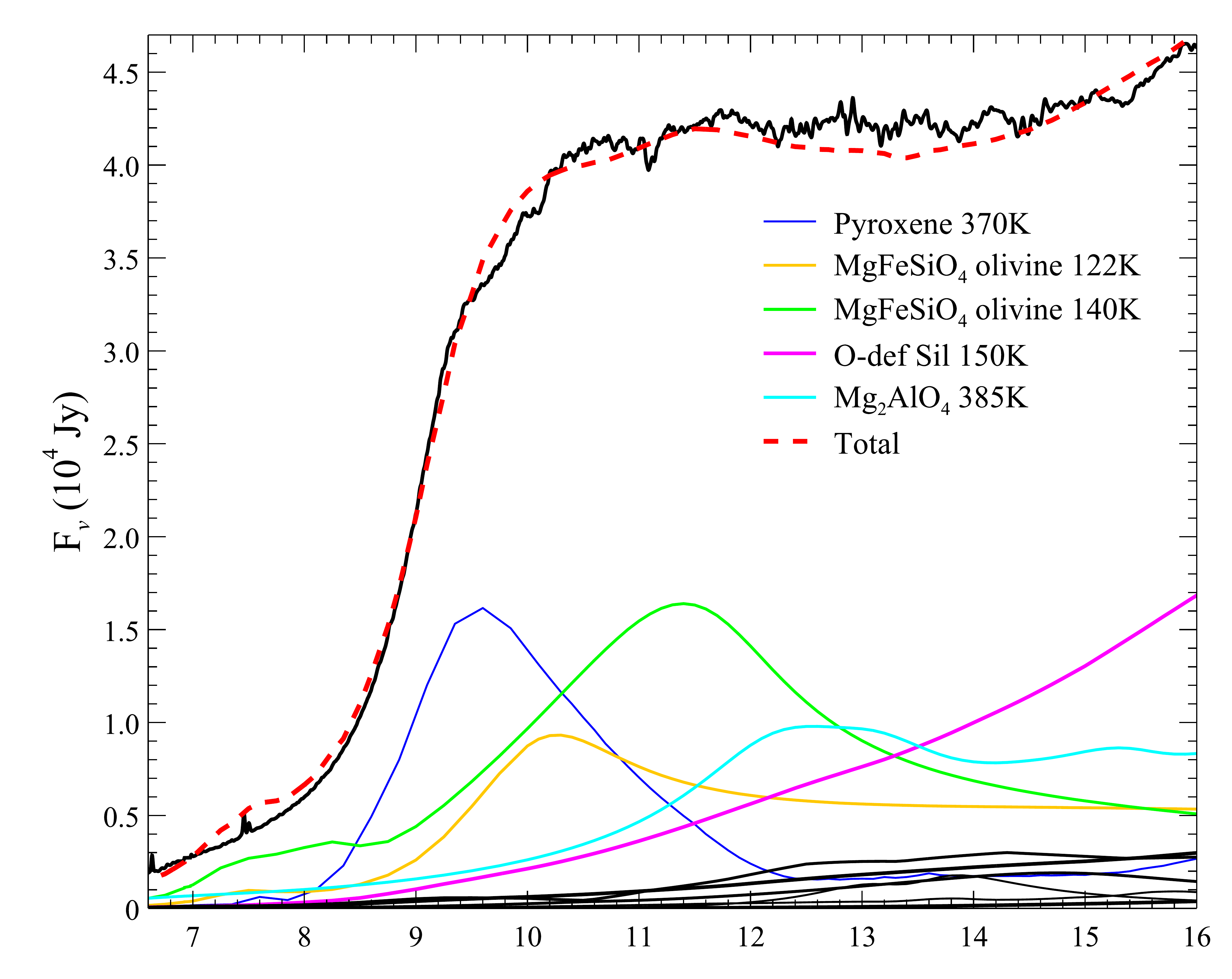} \\
 \includegraphics[width=9cm]{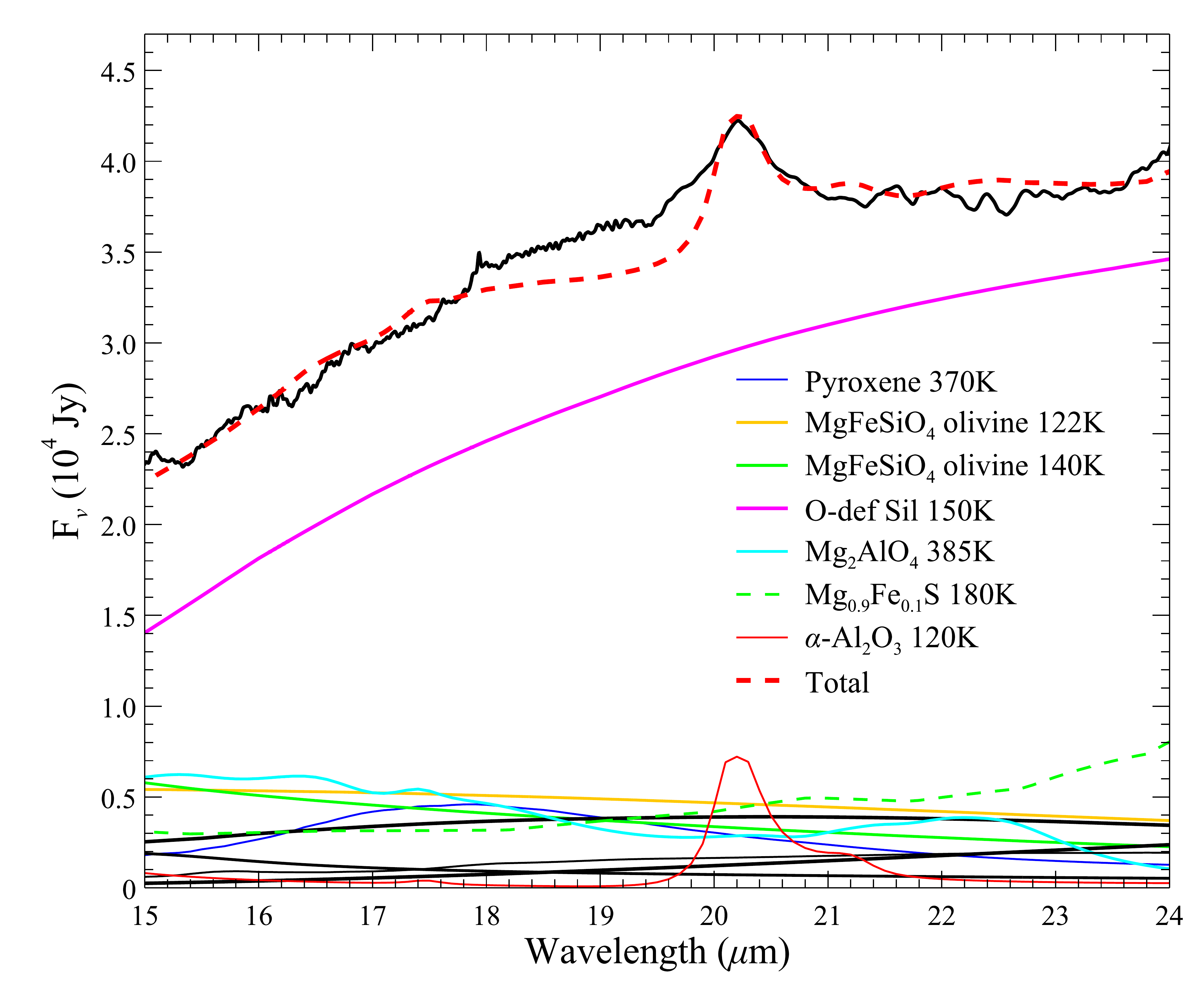} 
 \end{center}
 \caption{Magnification on the 10 and 20 $\mu$m bands, highlighting the principle contributing dust species in Model B.  The observed spectrum (black) and total dust model (red dashed line) have been subtracted by the total contribution from Fe grains, shown in Fig.~\ref{fig:ModelB}.  An additional offset of $-$2.4 $\times$ 10$^4$ Jy has been applied to  the observed spectrum and total model for clarity. 
\label{modelB1020bands} }
\end{figure}

A contribution from warm pyroxene is similarly needed to produce the blue portion of the 10 $\mu$m band, but the rest is a somewhat complicated mix of 115-140 K olivine, magnesium oxide (spinel), and a stronger contribution from Mg$_{0.9}$Fe$_{0.1}$S.  Corundum also remains the only carrier that can provide the narrow 20.2 $\mu$m feature.   Corundum has also been shown to be a component in the dust modeling of the 10 $\mu$m band from VLTI/MIDI spectra (7.5 - 13.5 $\mu$m) by Chesneau et al. (2005). In their models, however, this is present at higher temperatures ($>$ 250 K) and proportions, emitting as a broad band (a result of the CDE approximation to the grain shapes), in combination with olivine to fit the observed 10 $\mu$m feature.   The full profile in our spatially integrated spectrum exhibits a substantial red wing to $\approx$15.4 $\mu$m, where cooler corundum along with the metal-rich silicate augite does contribute in our models (including Model B below, also in combination with olivine) in minor proportions and relies on spherical grains in order to reproduce the narrow 20.2 $\mu$m band.    A close-up comparison of the main contributors in Model B to the 10 and 20 $\mu$m bands is shown in Figure~\ref{modelB1020bands}.

A main difference in how the SED was fit for Model B compared to Model A is that we left the entire LWS range unweighted.   This gives a comparative and conservative sense of the extent of possible background contamination in these uncorrected observations, which this model indicates to start at  around 60 $\mu$m.  If we force a similar match to the midpoint of the LWS range, a stronger cool contribution from either the Fe or O-poor silicate (or a combination both) would result, driving up the dust mass even further.   As it is, we already regard $M_{\rm{d}}$(A) = 0.25 $M_\odot$ to be a very high amount of dust mass, while $M_{\rm{d}}$(B) = 0.45 $M_\odot$ is astonishing.

 \subsection{Dust model uncertainties}\label{dustmodelerrors}  

The formal fitting uncertainties on both models are very similar (taking into account the difference in weighting of the LWS spectrum as just described) --- the derived dust temperatures have a maximum uncertainty of $\pm$5 K, affecting the dust masses though the predicted values of the emergent fluxes $F_{{\nu},i}(\lambda)$ and in the Planck function in Eq.~\ref{eq:dustmass} most strongly at low $T_{\rm{d}}$.  The highest-mass component in Model B, namely, the olivine at 115 K, has a conservative uncertainty range of $-$0.03 to $+$0.05 $M_\odot$.  Grain sizes $a$ are derived together with $T_{\rm{d}}$ and their uncertainties over the same range of temperature uncertainties have very little effect on derived masses because of the invariance of $a$ with $Q_{\rm{abs}} / a$ for 2$\pi \mid k \mid a  / \lambda \ll 1$.  

As previously alluded, the main source of uncertainty on the dust masses is due to the lack of constraints on the dust-emitting geometry.  For large, easily observed ejection shells around AGB and red supergiant (RSG) stars and the LBV AG Carinae, the number of shells and their inner and outer radii can be constrained with IR imaging, and appropriate density profiles can be assumed with knowledge of the outflow type.   As stressed already for $\eta$ Carinae's environment, however, we have only an inferred size of the dust-emitting region from aperture constraints.  The 12 and 18 $\mu$m imaging from MWB99 suggests that the bulk of the dust is likely to be present in the equatorial region in toroidal structures, possibly also in an inner unresolved torus or disk, and/or in pinwheel-like structures created by colliding stellar winds in the orbital plane similar to those observed in some WC binaries (e.g., Monnier et al. 2007), and to an unsettled extent in the bipolar lobes.  Thus,  in the representation of a single shell model we bounded the shell radii and optical depths to reduce some of the degeneracies of solutions from the large number of free parameters (a common issue even for well-constrained geometries and relatively simple dust chemistries), and in doing so, we applied Eq.~\ref{eq:dustmass}, which assumes that dust is optically thin and therefore yields lower limits on the values of $M_d$.

While we cannot improve on our assumptions for the dust geometry without higher angular resolution IR observations, which may then justify 2D or 3D modeling, we can give a sense of the effects on $M_d$ when shell radii and optical depth limits are relaxed for example dust species.  Using the highest-mass silicate component in Model A,  Ca$_2$Mg$_{0.5}$AlSi$_{1.5}$O$_7$, we can obtain a similar output flux distribution for a shell with inner radius $r_1$ = 1.85 $\times$ 10$^{16}$ cm, a shell thickness $\Delta r$ = 15 $\times$ $r_1$, and an optical depth at 0.55 $\mu$m of $\tau_{0.55}$ =  1.2.  The dust density has been assumed to vary as $r^{-2}$.  The dust mass can then be calculated as 

\begin{equation}
M_{{\rm{d}},i} = {16 \pi^2 a \rho \tau_{\lambda} r' \over{3 Q_{\rm{abs},\lambda}}}
\label{eq:dustmass2}
\end{equation}
\noindent where $r'$ is the radial center of mass and other parameters are the same as defined above for Eq.~\ref{eq:dustmass}.  The value of $r'$ for an $r^{-2}$ density profile is simply the average of the inner and outer radii.  Eq.~\ref{eq:dustmass2} yields $M_d$ = 0.04 $M_\odot$ in this example, which is close to that estimated from the optically thin case mainly because $\tau$ has not increased very much above unity.   Interestingly, the shell's derived angular size in this geometry is found to be 0$''$.5 and 8$''$.0 at the inner and outer radii, respectively, which is not inconsistent with the size of the emitting region inferred by aperture constraints (Sec.~\ref{sws}) where the cool dust peaks.    Higher optical depths can be offset by a thinner shell size, compared to the initial limit $\Delta r \leq$ 100 for Models A and B, but there is no straightforward way to estimate this for all species in simultaneous fitting without descending into a very high number of degenerate solutions involving all dust parameters.  With these factors taken together, the dust masses given in Table~\ref{table:dustmodels} should be regarded as lower limits.  An additional uncertainty on the gas-to-dust ratio will be discussed when we address the implied total mass of the Homunculus from each model (Section~\ref{sec:Discussion}).     

\begin{deluxetable*}{l c l l l l}
  \tablewidth{0pc}
  \tablecaption{Homunculus Dust Mass by Grain Composition}.

 \tablehead{
 \colhead{Species} & 
\colhead{Type\tablenotemark{a}} &
\colhead{Ref.\tablenotemark{b}} &
 \colhead{$T_{\rm{d}}$} & 
 \colhead{$a_{\rm{min}}$} & 
 \colhead{$M_{\rm{d}}$}  \\
 \colhead{} & 
 \colhead{} & 
 \colhead{} & 
 \colhead{(K)} &
 \colhead{($\mu$m)} &
 \colhead{($M_{\rm{\odot}}$)} 
 } 
 
 \startdata  
\multicolumn{6}{c}{Model A} \\  \hline \\ 
 Fe (pure) & C &  1 & 95 &  0.1 &  0.11  \\ 
 Fe (pure) & C &  1 & 160 & 1.0 & 0.09  \\
 Ca$_2$Mg$_{0.5}$AlSi$_{1.5}$O$_7$ & A & 6 & 110 &  0.01 & 0.03 \\
 Fe (pure) & C &  1 & 300 & 0.01 & 0.01 \\
 Fe (pure) & C &  1 & 400 & 0.01 & 2.71 $\times$ 10$^{-3}$ \\
 $\alpha$-Al$_2$O$_3$ corundum & C & 5 & 120 &  0.1 & 2.50 $\times$ 10$^{-3}$ \\
{Si$_3$N$_4$} & {C} & {8} & {250} &  {0.1} & {1.22 $\times$ 10$^{-3}$} \\ 
Mg$_{.5}$Fe$_{.43}$Ca$_{.03}$Al$_{.04}$SiO$_3$ & C & 3, 4 & 185 & 4.0 & 7.40 $\times$ 10$^{-4}$ \\
Mg$_{.5}$Fe$_{.43}$Ca$_{.03}$Al$_{.04}$SiO$_3$ & C & 3, 4 & 140 & 3.5 & 6.71 $\times$ 10$^{-4}$ \\
{Mg$_{0.9}$Fe$_{0.1}$S} & {C}  & {7} & {170} &  {3.0} & {6.10 $\times$ 10$^{-4}$} \\
''Cosmic'' pyroxene & A & 2 & 350 & 0.2 & 1.51 $\times$ 10$^{-4}$ \\
{AlN} & {C} & {8} & {250} &  {0.6} & {1.12 $\times$ 10$^{-4}$} \\
 Fe (pure) & C &  1 & 600 & 0.01 & 3.87 $\times$ 10$^{-5}$ \\
 Fe (pure) & C &  1 & 350 & 1.0 & 2.38 $\times$ 10$^{-5}$ \\
Fe (pure) & C &  1 & 800 &  0.01 & 1.01 $\times$ 10$^{-5}$ \\ \cmidrule{6-6}
& & & &   &Total \ 0.25 \\

\hline \\ 

\multicolumn{6}{c}{Model B} \\  \hline \\ 

MgFeSiO$_4$ olivine & C &10 & 115 & 1.6 & 0.19 \\
O-poor silicate & A & 9 & 95 & 70.0 & 0.11 \\
MgFeSiO$_4$ olivine & C & 10 & 122 & 1.6 & 0.06 \\
MgFeSiO$_4$ olivine & C & 10 & 140 & 1.5 & 0.05 \\
O-poor silicate & A & 9 & 150 & 4.5 & 0.02 \\
 Fe (pure) & C &  1 & 300 & 0.01 & 5.11 $\times$ 10$^{-3}$ \\
 $\alpha$-Al$_2$O$_3$ corundum & C & 5 & 120 &  0.1 & 1.46 $\times$ 10$^{-3}$ \\
O-poor silicate & A & 9 & 170 & 1.5 & 1.23 $\times$ 10$^{-3}$ \\
 Fe (pure) & C &  1 & 400 & 0.05 & 7.64 $\times$ 10$^{-4}$ \\
Fe (pure) & C &  1 & 225 & 1.0 & 6.98 $\times$ 10$^{-4}$  \\
Mg$_{0.9}$Fe$_{0.1}$S & C  & 7 & 180 &  3.0 & 6.07 $\times$ 10$^{-4}$ \\
''Cosmic'' pyroxene & A & 2 & 350 & 0.2 & 1.83 $\times$ 10$^{-4}$ \\
Mg$_2$AlO$_4$ spinel & C & 11 & 385 & 2.0 & 4.67 $\times$ 10$^{-5}$ \\
 Fe (pure) & C &  1 & 350 & 1.0 & 2.90 $\times$ 10$^{-5}$ \\
Fe (pure) & C &  1 & 800 &  0.05 & 3.09 $\times$ 10$^{-6}$ \\ 
 Fe (pure) & C &  1 & 600 & 0.05 & 2.08 $\times$ 10$^{-6}$ \\\cmidrule{6-6}
& & & &  &Total \  0.44

\tablenotetext{a}{Type refers to crystalline (C) or amorphous (A) grain structure.} 
\tablenotetext{b}{Dust optical constants references: (1) Ordal et al. (1985); (2) Henning \& Mutschke (1997); (3) Jaeger et al. (1994); (4) Dorschner et al. (1995); (5) Zeidler, Posch \& Mutschke (2013); (6) Mutschke et al. (1998); (7) Begemann et al. (1994);  (8) Kischkat et al. (2012); (9) Ossenkopf et al. (1992); (10) Fabian et al. (2000); (11) Fabian et al. (2001). }

\enddata

\label{table:dustmodels}

\end{deluxetable*}

An additional source of uncertainty in Eq.s~\ref{eq:dustmass} and \ref{eq:dustmass2} is on the bulk density of the grains, particularly the Fe grains, for which we adopted the canonical value of 3.0 g cm$^{-3}$ as for the silicates.  The bulk densities are not quoted by Ordal et al. (1985) whose optical constants we employed owing to the continuous IR wavelength coverage of the measurements.  Values of $\rho_{\rm{Fe}}$ quoted in the literature range from 2.8 g cm$^{-3}$ for iron powder and filings (which are probably oxidized) to as high as 7.9 g cm$^{-3}$ (Pollack et al. 1994).  The upper value would have the effect of more than doubling the estimated dust masses of all fitted Fe components.

Finally, the fact that two models emerge from our analysis as preferred best fits should underscore that a chemically unique breakdown of the spectral signatures in the observed SED is not possible to derive.  This may be pa,rtially a consequence of unquantifiable limitations in our library of dust analogs for representing the true dust properties, but it also tied in with the lack of spatial discrimination of the features.  In fact the indications we have that the majority of the warm and cool dust is within the central 5$''$ and not in the lobes only compounds the degeneracy problem.   Nonetheless, our models are chemically sensible for the C- and O-poor and N-rich abundance pattern of the Homunculus:  we do not detect any meaningful contributions from carbonaceous compounds (graphite, carbon grains, SiC, etc.) or minerals that are precipitated with water (e.g., montmorillonite, carbonates, etc.), while nitrides, sulfites, and O-poor and metal-rich silicates are strongly favored.  From this standpoint,  the two models should be taken together as reasonable representations; the actual chemistry may be a mix of both models, and it is clear that the Homunculus is extremely massive, $M_{\rm{d}}$ > 0.25 $M_\odot$  

The abundance pattern supporting our derived dust chemistries is not without some recent ambiguities.  C and O deficiencies in the Homunculus are consistent with core evolution for a massive star and have been used to explain the lack of detection of the CO molecule in searches at submillimeter and UV wavelengths (Cox \& Bronfman 1995; Verner et al. 2005; Nielsen et al. 2005) --- that is, until an analysis of recently detected $^{12}$CO and $^{13}$CO emission by Loinard et al. (2012), using the APEX telescope. Their study indicates that CO relative to H$_2 $ is at cosmic levels in the Homunculus, while a value for [$^{12}$C/$^{13}$C] $\approx$ 5 derived from the CO lines is consistent with isotopic enhancement of $^{13}$C for material which has been lifted (or erupted in $\eta$~Carinae's case) from a star showing surface abundances consistent with CNO processing in the core.   We address these apparent inconsistencies next.

\section{CO abundances in the Homunculus}\label{sec:COabundance} 

\begin{figure}
 \begin{center}
 \includegraphics[width=8cm]{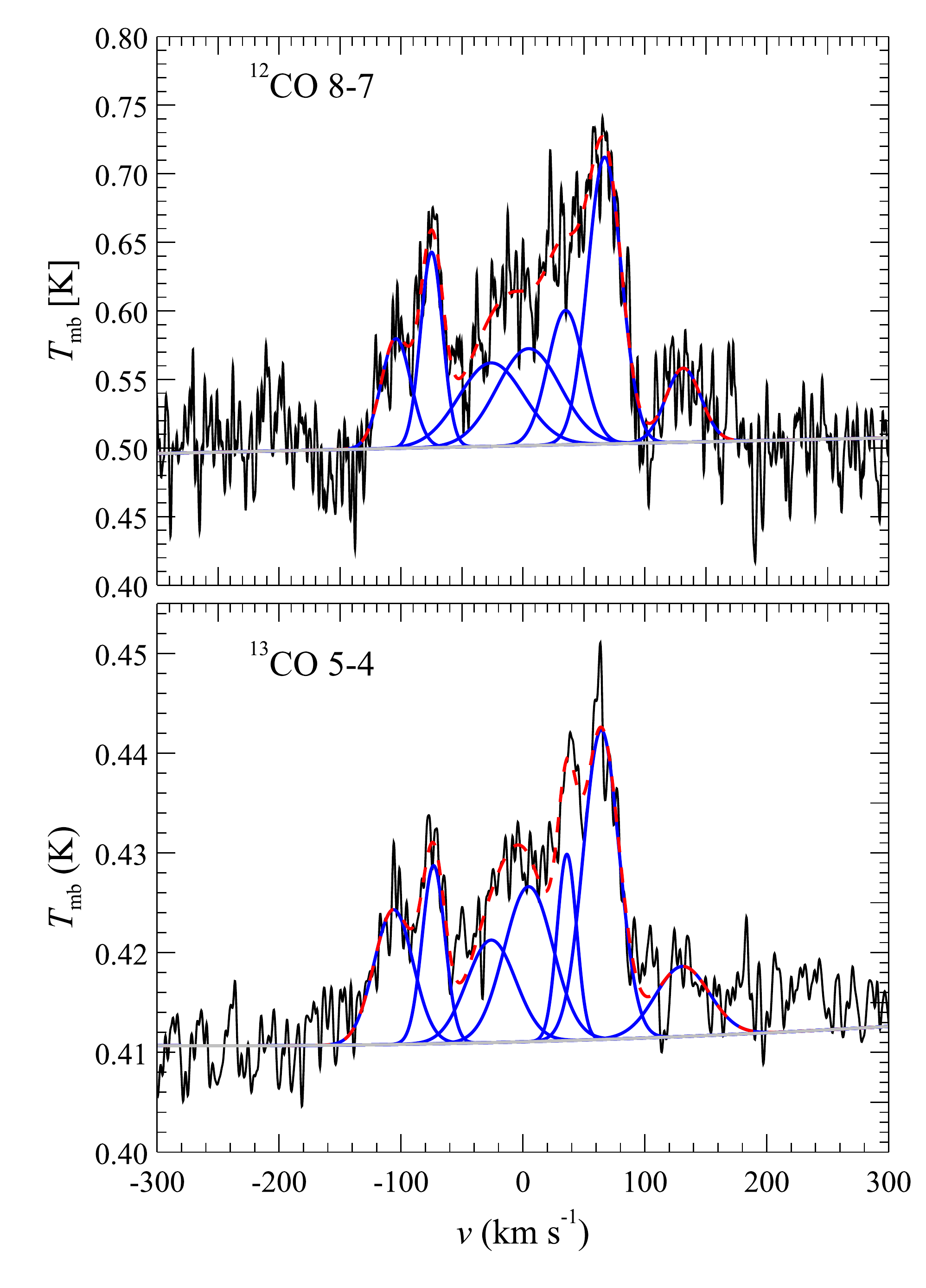}
 \end{center}
 \caption{Gaussian decomposition of the $^{12}$CO $8-7$ and  $^{13}$CO $5-4$ lines observed with HIFI. For both observed profiles the number of Gaussian components, centroids, amplitudes, and widths have been fit between $\pm$200 km s$^{-1}$ with a Levenberg-Marquardt minimization procedure, yielding  seven components (blue) in both cases with $\langle v_{\rm{lsr}} \rangle$(km s$^{-1}$) =  $-$104.1, $-$74.9,  $-$26.0, $+$4.7, $+$35.1, $+$67.0, and $+$131.5,  and respective widths $\langle \Delta v \rangle$(km s$^{-1}$) = 28.6, 21.3,  62.4, 61.2, 34.5, 32.5, and 33.9. The sum total of the fitting is shown by the red dashed lines.  
   \label{fig:COvelfits}}
\end{figure}

Following unsuccessful efforts to detect CO emission from the Homunculus by Cox \& Bronfman (1995) and Verner et al.  (2005), submillimeter rotational lines of CO in emission have been detected during separate campaigns with {\em{Herschel}}/HIFI starting in 2009, and then in 2011 with the APEX heterodyne instruments.  The APEX observations of $^{12}$CO $J = 6-5$, $4-3$, and $3-2$ have been analyzed along with $^{13}$CO $J = 6-5$ and $3-2$ and low- to intermediate-$J$  (mostly $J = 4-3$) emission from CN, HCO$^+$, HCN, H$^{13}$CN, HNC, and N$_2$H$^+$ by Loinard et al. (2012).  Their modeling of the CO lines leads them to conclude that the CO/H$_2$ abundance ratio is at cosmic levels, and not underabundant as concluded in previous observational studies.  The analysis relies critically on an assumption for the emitting source size (as we will explain further below), which was set to to $\theta_s = 1''$ based on optical depth arguments set by APEX beam constraints.  The availability of overlapping and additional transitions of CO --- up to $J$ = 9-8 and including $J$ = 5-4, which cannot be observed from the ground --- gives us the opportunity to reevaluate the CO abundance question over a higher range of upper energy levels, $\Delta E_u$ = 79 - 249 K compared to the 33 - 116 K range of the APEX observations.

The HIFI CO observations exhibit similar emission profile characteristics compared to the APEX data shown by Loinard et al. (2012):  the emission is centered near $-$4 km s$^{-1}$ and is considerably structured and asymmetric with several components of emission with a primary peak at $+$68 km s$^{-1}$ and a secondary peak at $-$75 km s$^{-1}$.  These features are unlikely to be associated with the Weigelt blobs of slow-moving nebular condensations, which have radial velocities $v_{\rm{r}} \approx -$45 km s$^{-1}$ (Gull et al. 2016 and references therein; see also Teodoro et al. 2017).  Up to five additional weaker velocity components can be resolved using Gaussian profile decomposition.  Figure~\ref{fig:COvelfits} shows two examples of the fitting to the $^{12}$CO $8-7$ and  $^{13}$CO $5-4$ profiles observed with HIFI, both highly consistent with each other.  Loinard et al. (2012) quote a value of $+$20 km s$^{-1}$ for a central velocity, but we observe a more continuous set of components in our data and the presence of weak emission at $+$130 km s$^{-1}$ and $-$100 km s$^{-1}$.  The differences are probably due to data quality:  both observations shown in Figure~\ref{fig:COvelfits} have lower RMS noise compared to the CO data shown by Loinard et al. (2012) (see their Fig. 1 for comparison).   Measures of the CO line strengths and the HIFI beam characteristics are given in Table~\ref{table:COmeasures}.

\begin{figure*}
 \begin{center}
 \includegraphics[width=16cm]{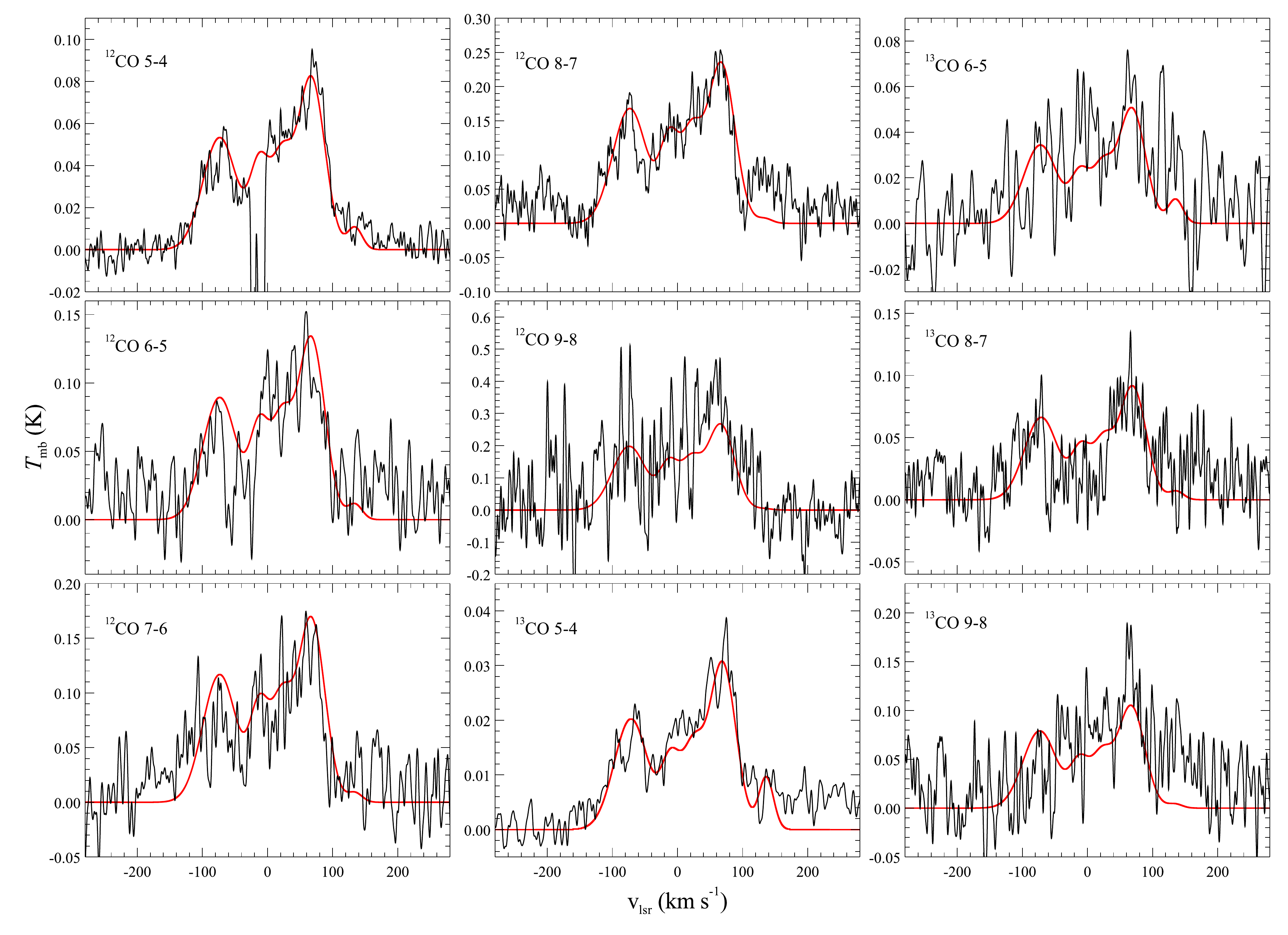}
 \end{center}
 \caption{CO spectra observed with HIFI (black) compared to synthetic spectra (red) produced by models described in Sec.~\ref{sec:COabundance}.  The synthetic profiles have been generated by fits to each of the $^{12}$CO and $^{13}$CO observations simultaneously, weighted by S/N; the final parameters are given in Table~\ref{table:COmodelParams}.  The narrow absorptions in the $^{12}$CO $J=5-4$ spectrum are due to emission at the dual beam-switched sky positions used for standing wave and background corrections.  
   \label{fig:COmodels}}
\end{figure*}

\begin{deluxetable}{l l r l l r}
  \tablewidth{0pc}
  \tablecaption{Homunculus CO line intensities, HIFI beam properties}

 \tablehead{
 \colhead{CO Transition} & 
\colhead{$E_u$} &
\colhead{$\nu_0$} &
 \colhead{$\theta_{\rm{mb}}$\tablenotemark{a}} & 
 \colhead{$\eta_{\rm{mb}}$\tablenotemark{a}} & 
 \colhead{$I_\nu$\tablenotemark{b}}  \\
 \colhead{$J_u - J_l$} & 
 \colhead{(K)} &
 \colhead{(GHz)} & 
 \colhead{(arcsec)} & 
 \colhead{} & 
 \colhead{(K km s$^{-1}$)} 
 } 
 
 \startdata  

$^{12}$CO $5-4$ &  83.0 & 576.27 & 37.1 & 0.75 & 11.0 $\pm$ 2.0 \\
$^{13}$CO $5-4$ &  79.3 &  550.93 & 38.6 & 0.75 &  4.1 $\pm$ 1.1 \\
$^{12}$CO $6-5$ & 116.2 & 691.47 & 30.5 & 0.75 & 18.0 $\pm$ 2.8\\
$^{13}$CO $6-5$ & 111.1  &  661.07 & 32.7 & 0.75 & 6.8 $\pm$ 1.2 \\
$^{12}$CO $7-6$ & 154.9 & 806.65 & 26.2 & 0.75 &  25.6 $\pm$ 3.9 \\
$^{12}$CO $8-7$ & 199.1 & 921.80 & 22.7 & 0.74 &  32.4 $\pm$ 1.8 \\
$^{13}$CO $8-7$ & 190.4 & 881.28 & 23.9 & 0.74 & 12.3 $\pm$ 2.4\\
$^{12}$CO $9-8$ & 248.9 & 1036.91 & 20.7 & 0.74 & 37.4 $\pm$ 7.2 \\
$^{13}$CO $9-8$ & 237.9 & 991.33 & 20.9 & 0.74 & 14.3 $\pm$ 4.5\\ 

\tablenotetext{a}{$\theta_{\rm{mb}}$ and $\eta_{\rm{mb}}$ are the HIFI main-beam width and main-beam efficiency, respectively, used in the conversion from antenna temperatures to the main-beam temperature $T_{\rm{mb}}$ scale.} 
\tablenotetext{b}{The line intensity $I_\nu = \int T_{\rm{mb}} dv$ integrated over the range of velocities where emission is observed.}

\enddata

\label{table:COmeasures}
\vspace{1em}
\end{deluxetable}

To model the physical properties of the CO-emitting region in the Homunculus, we use the CASSIS\footnote{CASSIS (http://cassis.irap.omp.eu/) has been developed by IRAP-UPS/CNRS.} line analysis package (Vastel et al. 2015) which computes synthetic spectra by solving radiative transfer equations with a number of adjustable input parameters (emitting source size, column density, excitation or kinetic temperature, H$_2$ density, LSR velocities, and continuum levels).  The calculations can be done in LTE or in non-LTE using the RADEX radiative transfer code (van der Tak et al. 2007).  Initially we carry out LTE predictions that can be compared with results by Loinard et al. (2012), and we perform a similar cross-check in RADEX for possible non-LTE effects.  The synthetic spectra compared to all $^{12}$CO and $^{13}$CO lines observed with HIFI are shown in Figure~\ref{fig:COmodels}, and a comparison of the best-fit parameters compared to those of Loinard et al. (2012) is given in Table~\ref{table:COmodelParams}.

The key results from modeling the HIFI CO observations are as follows:
\begin{enumerate}

\item  Over any range of column densities $N$(CO) and excitation temperatures $T_{\rm{ex}}$, we find  $\theta_s \geq 3''$.  We cannot obtain good fits to the observed profiles using a small source size $\theta_s \leq 1''$, as adopted by Loinard et al.  (2012).  Fixing  $\theta_s = 1''$ predicts intensities that are lower than observed and also creates some flat-topping of the profiles by the onset of saturation in the high-velocity components.  This effect is slightly evident on close inspection of the $v_{\rm{lsr}}$ = $-$76 km s$^{-1}$ component in the synthetic $^{12}$CO spectra shown by Loinard et al. (2012).   

\item Column densities are in the range $N$($^{12}$CO) = ($1.3-3.2$) $\times$ 10$^{17}$ cm$^{-2}$; hence, the CO/H$_2$ abundance ratio is approximately an order of magnitude lower than cosmic levels, using $N$(H$_2$) = 3.0 $\times$ 10$^{22}$ cm$^{-2}$ for the Homunculus (Smith et et. 2006).  Actually, the H$_2$ density will be $\approx$40\% higher if we assume that the 45 M$_\odot$ in the nebula is contained within a 90 arcsec$^2$ volume inferred by our results, decreasing the CO/H$_2$ abundance that much further.

\item Similarly, our results for $N$($^{13}$CO) = ($5.0-13.0$) $\times$ 10$^{16}$ cm$^{-2}$ yield $^{12}$C/$^{13}$C = 3.0.  This value far below that of the interstellar medium (ISM), 67-70 (Langer \& Penzias 1990; Milam et al. 2005), implying a relative enhancement of $^{13}$C, and is almost precisely the value of the predicted CNO equilibrium isotope ratio for massive nonrotating stars and compares favorably with predictions for rotating massive stars at the end of the W-R WN phase (e.g., Hirschi, Meynet, \& Maeder 2005; Kraus 2009; Georgy et al. 2016).  
 
\end{enumerate}     

It is not possible for us to definitively explain the differences in required source sizes needed to fit the CO lines, then leading to different CO/H$_2$ results between our study and that of Loinard et al. (2012) .  Uncertainties that can affect both studies involve line opacity effects on the estimated column densities.  An inverse proportionality between the relative peak intensities and the corresponding beam areas indicates optically thick lines, although the trend for the five $^{12}$CO lines observed with HIFI is not as strict as indicated for the three observed at APEX by Loinard et al. (2012).  Again, this may be related to a combination of weaker emission at lower $J$ and the higher baseline noise in the APEX data, or a real change in optical depth with excitation.  The CASSIS code does treat opacity effects in the interaction of spatially overlapping velocity components, treating the component with the highest $T_{\rm{ex}}$ as the innermost and using the spectrum for that component as the continuum seen by the next component and so on (E. Caux, private communication).  The code also appropriately broadens the optically thick lines.  Regardless, the extent of these effects can be accurately quantified only with future high angular resolution observations of the CO-emitting region in the Homunculus.   

Treatment of the different sources of continuum emission may affect the analysis of the higher-$J$ CO lines, which are at frequencies where the continuum arises primarily from heated dust, compared to the lower-$J$ lines at frequencies where a significant contribution to the continuum may come from free-free emission.  Our analysis of the CO lines observed with HIFI includes continuum radiation from dust only.  As seen in seen in Figure~\ref{etafullsed}, the  onset of free-free emission occurs at around 350 $\mu$m.  We estimate that the contribution of this component reaches a maximum of $\approx$25\% at the CO $J=5-4$ line frequency, falling off to zero by $J=8-7$.  It is not evident that the sources of the free-free radiation and CO emission are in the same volume, but the free-free contribution is weak, so it is probably unimportant for these lines.  This is not necessarily true for the lower-$J$ lines, if the molecular gas is exposed to the free-free photons.  In reality, the small emitting volumes of the free-free emitting region arising from wind-wind interactions near the central source and the region of molecular line emission, protected by gas-phase Fe traced by [Fe {\sc{ii}}] and [Fe {\sc{iii}}]  (Gull et al. 2016), and/or by the Fe dust indicated in our dust models, may each be physically related.    

Another contributing factor may be the somewhat  higher intensity levels measured from our HIFI CO observations compared to the APEX data.  The integrated intensity of the $^{12}$CO $J=6-5$ line (the only transition overlapping both studies) measured over the same range of LSR velocities is $\approx$30\% stronger in our dataset\footnote{The 691 GHz frequency is anyway difficult for ground-based receivers, due to low atmospheric transmission, so that quoted RMS noise error bars may not be sufficient for an accurate comparison.}  after corrections for beam sizes and efficiencies. This discrepancy is outside both sets of quoted observational uncertainties.  Variability in the lines due to changes in excitation temperatures over $\sim$3/4 of a 5.54 yr cycle on which the HIFI observations were taken  is possible (the APEX observations were taken almost exactly at apastron in the same cycle) but unlikely since we do not witness differences this large between the observed profiles and synthetic profiles from a single model with fixed parameters.   It could be argued that the S/Ns are not high enough in either datasets to be certain about this in all observed velocity components.  

The HIFI beam is larger at 691 GHz, so it is also not unreasonable to question whether there is additional warm CO emission originating from outside the Homunculus.  However, we can rule this out for the most part, since the observations are corrected for the backgrounds as part of the standing wave calibrations using the DBS mode and would require emission over the same range of velocities and relative intensities in all lines to extend just outside the Homunculus but without contaminating a $\sim$1 arcsec$^2$ area centered at the positions of the 3$'$ chop throws.  Contamination appearing as narrow absorptions due to off-source emission at both sky chop positions is indeed present in the $^{12}$CO $J=5-4$ line near $v_{\rm{lsr}}$ = 0 km s$^{-1}$ (see Fig.~\ref{fig:COmodels}),  but the contamination from this narrow background emission is clearly distinguishable as such from the rest of the profile's kinematic structure.    The CO emission over this full range of velocities and widths from outside the Homunculus has also been discounted with the smaller APEX beams by Loinard et al. (2012).

\begin{deluxetable}{l l l}
  \tablewidth{0pc}
  \tablecaption{Properties of the CO-emitting region}

 \tablehead{
 \colhead{Parameter} & 
\colhead{Loinard et al.} &
\colhead{This Study} 
 } 
 
 \startdata  

$E_u$(CO)\tablenotemark{a} (K)  & $33-116$ & $83-238$ \\  
$T_{\rm{ex}}$ (K) & $40-90$ & $145-165$ \\
$\theta_s$ (arcsec) & 1.0 & 3.0 \\ 
$N$($^{12}$CO) (10$^{17}$ cm$^{-2}$) & 65 & $1.3-3.2$ \\
$[^{12}$CO/H$_2]$\tablenotemark{b} (10$^{-5}$) & 21.7 & 0.43-1.1 \\
$[^{12}$C/$^{13}$C$]$ & 4.6 & $2.4-2.6$ \\

\tablenotetext{a}{Upper $J$ level energy ranges correspond to observed sets of CO presented by Loinard et al. (2012) and in this study.}
\tablenotetext{b}{The H$_2$ column density $N$(H$_2$) = 3.0 $\times$ 10$^{22}$ is adopted from Smith et al. (2006). }

\enddata

\label{table:COmodelParams}
\vspace{-1em}
\end{deluxetable}

\section{Discussion}\label{sec:Discussion}

\subsection{The Luminosity of $\eta$~Carinae}

A crucial parameter for understanding the $\eta$~Carinae system is its luminosity. Typically it has been assumed that this can be obtained from the infrared spectral distribution under the assumption that the majority of visible and UV light is absorbed by circumstellar dust and reradiated in the infrared. Below we examine this assumption and discuss the variability of infrared fluxes. For consistency, all previous estimates of the luminosity are scaled to a distance of 2.3\,kpc.

Early observations of $\eta$~Carinae's infrared flux were made by Neugebauer \& Westphal (1968) and Westphal \& Neugebauer (1969); their measurements are shown in Figure~\ref{etafullsed}.  The latter estimated $L = 4.5 \times 10^6$\,\Lsun\  for the energy emitted between 0.4 and 19.5\,\mum. This is a lower limit for the luminosity --  the current data indicate that $\sim$ 30\% of the energy is emitted longward of 20\,\mum. This would increase the luminosity at the time of Westphal \& Neugebauer's 1968 measurements to  $L = 6.4 \times 10^6$\,\Lsun. Later measurements made by Gehrz et al. (1973) and Robinson et al. (1973) from 1 to 20\,\mum\  are broadly consistent with those of Neugebauer \& Westphal (1968) and Westphal \& Neugebauer (1969). The fluxes were in agreement with the earlier measurements except at 3.6\,\mum. The estimated errors were 0.1 mag at 4.8, 8.4, and 11.2 \mum, but with a larger uncertainty of 30\% at 20\mum.

Cox et al. (1995) derived $L_{\rm{IR}} = 5 \times 10^6$\,\Lsun, where  $\approx$85\% of the luminosity can be attributed to emission from heated dust in two main thermal components at 210 and 430~K. They used measurements of Robinson et al. (1973) that were obtained in 1971 and 1972, 34\,\mum\ data from Sutton et al. (1974), and longer-wavelength data taken in 1977 with the Kuiper Airborne Observatory by Harvey et al. (1978). Using the data selection by Cox et al. (1995), Davidson \& Humphreys (1997)  estimated  $L = 5 \times 10^6$\,\Lsun\ after making a 10\% allowance for energy not absorbed by dust in the Homunculus, and estimated a 20\% error in $L$.

Our integration of the SED from 1 to $10^6$\,\mum\ yields a luminosity of $3.0 \times 10^6$\,\Lsun, which is significantly lower than the earlier estimates and raises several issues.  Is $\eta$~Carinae variable in the infrared? If the source is variable, is this due to intrinsic changes in luminosity, or due to changes in circumstellar extinction? What is the luminosity of $\eta$~Carinae? This is of crucial importance since estimates for the mass of the primary star are based primarily on its luminosity via the Eddington limit.

Smith et al. (1995) provide strong evidence that $\eta$~Carinae is variable at 10\,\mum.  In July 1993 they measured a 12.5\,\mum\ flux of $1.02 \times 10^{-13}$ W cm$^{2}$ \mum$^{-1}$ while 20 yr earlier they had measured a flux of  $1.63 \times 10^{-13}$ W cm$^{2}$ \mum$^{-1}$ using the same beam. By assuming that the shape of the SED had not changed, they showed that this change was consistent with a measurement of the 10\,\mum\ flux made in 1986 by  Russell et al. (1987). The 1993 flux measurement of Smith et al. (1995) compares favorably with our value of $1.14 \times 10^{-13}$ W cm$^{2}$ \mum$^{-1}$, which was obtained in 1996.

\begin{figure}
 \begin{center}
 \includegraphics[width=8cm]{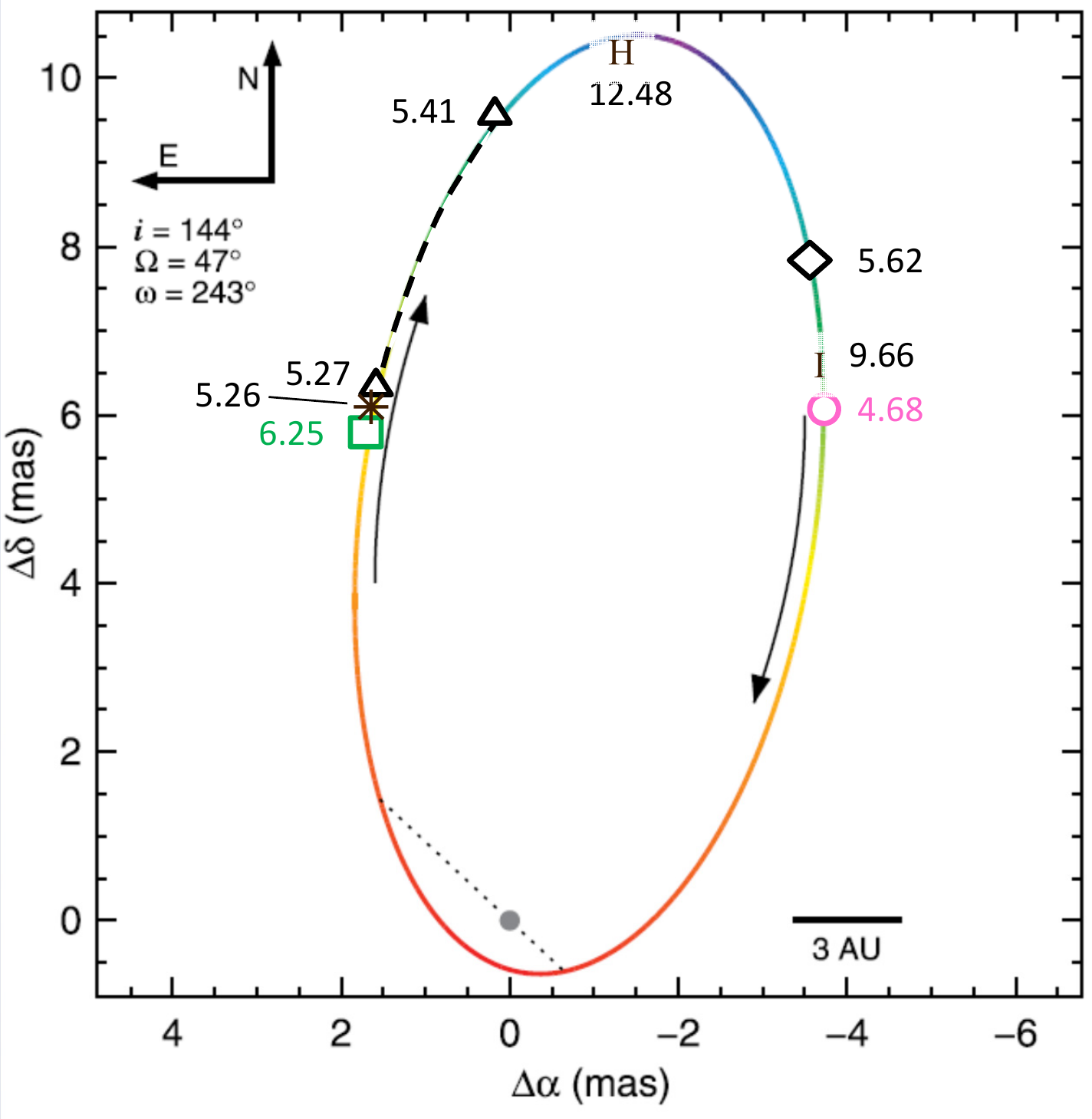}
 \end{center}
 \caption{Ground-based, ISO, and {\em{Herschel}} continuum measurements used in this study schematically overlaid onto the projected orbit of the binary system.  Symbols are the same as in Fig.~\ref{etafullsed}, 'I' and 'H' indicate the ISO/SWS and {\em{Herschel}}/SPIRE observations, respectively.  The labeled epochs  are defined in footnote (a) of Table~\ref{obs_t}.  The plot of the orbit is reproduced from Teodoro et al. (2016), by permission.  
   \label{orbit}}
\end{figure}

In the near-IR and optical, the variability of $\eta$~Carinae is well documented. In the early 1940s $\eta$~Carinae brightened very rapidly by about 1 mag (Fern{\'a}ndez-Laj{\'u}s et al. 2009), and since then it has been getting steadily brighter. Today it is about 2 mag brighter than it was in the 1970s. In addition to the overall increase in its brightness, the central source is now much brighter than in the past and is increasingly contributing a greater fraction of the observed optical flux.  In 8 yr, {\em{HST}} observations reveal that the central region became brighter by about 2 mag (Martin \& Koppelman 2004). Superimposed on the longer-term trend is the imprint of the binary with its 5.5 yr period (e.g., Whitelock et al. 2004;  van Genderen et al. 2006;  Fern{\'a}ndez-Laj{\'u}s et al. 2010; Mehner et al. 2014), as well as more random variations. Recent ground-based photometry is provided by Observatorio Astron\'omico de La Plata\footnote{http://www.etacar.fcaglp.unlp.edu.ar}.

The most likely scenario to explain the gross behavior of all observations is that  the extinction caused by circumstellar dust is declining. This can explain both the fading at infrared wavelengths and the brightening at visible wavelengths. If we assume that the bolometric luminosity of the system has remained constant, a significant amount of UV and visible flux must have been escaping from the system in 1996 (the time of the ISO measurements). Because our line of sight to the system has unusually high extinction (Hillier \& Allen 1992), it is difficult to estimate this fraction. However, under the assumption of a constant bolometric luminosity, and by comparison with the earlier measurements, at least 30\% of the flux emitted by the two stars must have been escaping in the visible/IR in 1996.  Today it could be much larger, since both the system and the central star are much brighter than in 1996.  This would be a consequence of more dust being destroyed in the shocks created in the colliding stellar winds during periastron passage than is created during the subsequent cooling phase, leading to a relative brightening at visual wavelengths after each passage.  Unfortunately, the temporal resolution of the referenced ground-based and ISO observations with respect to orbit phase is rather poor for demonstrating this, as shown in Figure~\ref{orbit}.  There is an obvious paucity of measurements from $\phi$ = 0.75 to periastron and then into the immediate cooling phase of the orbit, emphasizing the need to pursue mid-IR observations to the next periastron event in 2020 and then over the next year (presuming that the system survives) focused on the balance of dust creation and destruction in the central region and lobes.

Taking all of the above-referenced 1970s photometry into account in an average sense, we conclude that a reasonable estimate for $L$ is $4.1\times 10^6$\,\Lsun, and this is within the 20\% uncertainty placed on $L$ by Davidson \& Humphreys (1997).  This may be a lower limit --- were  we lucky in the 1970s when the first IR measurements were made? Did these measurements provide us with an accurate estimate of the bolometric flux? It is known from HST observations (and also ground-based observations) that the Weigelt blobs suffer very little circumstellar extinction. By contrast, using HST data obtained in 1998, Hillier et al. (2001) estimated that the total extinction to the star along our line of sight was close to 7 mag (but is now much less). The difference in extinction between the Weigelt blobs and the primary star is not a new phenomenon -- it must have also been present in the 1940s/1950s since narrow nebula lines, presumably arising in the Weigelt blobs, were very prominent in spectra taken at that time (e.g., Gaviola 1953).

\subsection{$\eta$~Carinae's Dust Chemistry}\label{dustchem}

The dust modeling of our combined ISO and {\em{Herschel}} spectral observations has been the main focus of this paper, and superseding the simple blackbody approach to estimating dust temperatures and masses (MWB99; Smith et al. 2003) with a full treatment of the interaction between the radiation field of the erupting star and its dusty environment covered by a wide range of potential optical properties has resulted in dust compositions that are not unexpected for the CNO-processed nebula.  Indications of some species have been made in previous studies. Corundum, for example, has been pointed out by Chesneau et al. (2005) to be one of the first species to condense starting at high temperatures ($\sim$1800 K) and high densities (e.g., Tielens 1990).    Following Mitchell \& Robinson (1978), Chesneau et al. (2005)  use warm corundum to fit the broad 10 $\mu$m feature, but in higher proportions compared to our models, mainly a consequence of the additional emission in this band unobservable from the ground beyond 13.5 $\mu$m and of our approach with fitting the full SED.   

Pure iron grains, on the other hand, may be much more difficult to condense in most astronomical environments.   New sounding rocket experiments under microgravity conditions carried out by Kimura et al. (2017) indicate that the sticking probability for the formation of Fe grains is quite small and that only a few atoms will stick per hundred thousand collisions.  This contrasts with significantly higher probabilities determined in ground-based laboratory measurements (e.g., Giesen, Kowalik \& Roth 2004) for proposed reasons such as thermal convection that can create a buffer in the nucleation site of the grains and complicate interpretations, but otherwise is not entirely understood (Kimura et al. 2017).  

Observational evidence for pure Fe dust has been shown in dust models for, e.g., the O-rich shell of the AGB star OH 127.8+0.0  (Kemper et al. 2002), in the Type IIb supernova remnant Cas A (Rho et al. 2008), and in the LBV R17 in the LMC (Guha Niyogi et al. 2014). In fact, Mitchell \& Robinson (1978) proposed Fe grains along with corundum and forsterite (Mg$_2$SiO$_4$) as the primary constituents of the 10 $\mu$m band of $\eta$~Carinae, observed at 8$-$13 $\mu$m.  Fe is thermodynamically favored in condensation models suited to LBV envelopes, $\eta$~Carinae in particular, by Gail et al. (2005).  Kemper et al. (2002) have pointed out that while metallic Fe can condense at fairly low temperatures 50 - 100 K in gas of solar composition, it condenses into FeO or FeS (cf. Jones 1990) below 700 K in O-rich environments, unless it is protected from an oxidizing environment by inclusion onto a grain.  The nucleation and survival of  Fe grains in the Homunculus are promoted by high Fe abundances (e.g. Hillier et al. 2001) and O deficiency where most of the available O (and Si) is likely to have been incorporated into silicates  and gas-phase molecules (CO, OH, HCO, etc.), which are surviving the harsh environment of the hot star radiation field (T. Gull et al, in preparation).  While some of the indicated silicates in our model fits do contain Fe inclusions, Mg atoms are thermodynamically more favorable during silicate formation (Gail \& Sedlmayr 1999).   We found no evidence of FeO, but we do see the presence of magnesium sulfide with a minority inclusion of iron (Mg$_{0.9}$Fe$_{0.1}$S).  

Detection of nitrides in astrophysical environments have also been elusive.  Si$_3$N$_4$ has been identified in meteorites and attributed to pristine presolar grains deposited into the ISM by by Type II supernovae via analysis of the elemental isotopes (Nittler et al. 1995; Lin, Gyngard \& Zinner 2010).  Nitrides have not been unambiguously detected in a circumstellar environment without challenge (e.g., Pitman, Speck \& Hofmeister 2006).  Small amounts of nitrides are predicted to occur around C-rich AGB and post-AGB stars after enhancement of N from subsolar values in the dredge-up of CNO products from the stellar interior and pulsational shocks propagating into the shells where dust will form (Fokin et al. 2001; Schirrmacher, Woitke \& Sedlmayr 2003; Pitman, Speck \& Hofmeister 2006).   Even in the N-rich environment of the Homunculus, AlN and Si$_3$N$_4$ in our preferred Model A are in minor abundances, $\approx$ 0.5\% of the total dust mass.  Their main effect is in supplying opacity to the broad 10 $\mu$m feature.     

In Section~{\ref{sec:dustmodeling}} we did not tabulate shell radii of the various dust species for the following reasons.  The thermal balance of spherical grains exposed to a stellar radiation field at temperature $T_\star$ and distance $r$ from a star with radius $R_\star$ can be derived from Kirchhoff's law, 

\begin{equation}
W \int_0^{\infty} {\overline{C_{\rm{abs}}}}(\lambda) \pi B_\lambda(T_\star) {\rm{d}}\lambda = W \int_0^{\infty} {\overline{C_{\rm{em}}}}(\lambda) \pi B_\lambda(T_d) {\rm{d}}\lambda 
\label{kirchoff}
\end{equation}

\noindent where ${\overline{C_{\rm{abs}}}}(\lambda)$ and ${\overline{C_{\rm{em}}}}(\lambda)$ are the orientation-averaged absorption and emission cross sections that are derived from the absorption and emission efficiencies $Q_{\rm{em}}$ and $Q_{\rm{abs}}$ (also rotationally averaged), and $W \equiv (R_\star / R_{\rm{d}})^2$ is the dilution factor.  If the grains are unpolarized, randomly oriented in a spherical geometry, and the radiation field is  isotropic, ${\overline{C_{\rm{abs}}}}(\lambda) \approx {\overline{C_{\rm{em}}}}(\lambda) / 4$ for blackbody radiation.  Then using Stefan-Boltzmann's law, Eq.~\ref{kirchoff} simplifies to an expression for the blackbody dust temperature as  

\begin{equation}
T_{\rm{d}}^{\rm{BB}} = T_\star W^{1/4}
\label{bbtemp}
\end{equation}

\noindent and from this one gets a simple approximation $R_{\rm{d}}$ (au) $\approx  1.72 \times 10^9 T_{\rm{d}}^{-2}$  for $\eta$~Carinae.  The approximation has been used by Smith et al. (2003) and Chesneau et al. (2005) to estimate the radial distances of the equivalent blackbody thermal components, finding a distance of 900 au (0$''$.3) for the hot 400~K component.  This estimate is useful only as a ``blackbody equivalent radius,'' since even in an environment where approximations on the geometry and isotropy of the radiation field have to be made (as we have done here), the clear non-equivalence between dust absorption and emission efficiencies for different grain compositions and sizes can lead to very different results.  These non-equivalences are taken into account in the derivation of the dust temperatures and estimated masses, but for radial distances they are numerically difficult to solve and still subject to the other approximations.  Moreover, the hottest dust, i.e., the small Fe grains above 400~K and probably reaching 1500~K to fill in the 2 - 8 $\mu$m opacity,  are unlikely to be in thermal equilibrium (e.g., Fischera 2004), so that Eq.~\ref{bbtemp} is inappropriate. That said, it is correct to point out (as done by Smith et al. 2003) that there must be significant shielding in the central region to allow the dust and molecules to survive.  Gull et al. (2016) have mapped [Fe {\sc{ii}}] wind-wind structures, which may protect the region close to the central source where we suspect much of the molecular gas resides, absorbing UV radiation at $\lambda <$ 2200 {\rm{\AA}} and greatly reducing the radiation field to around 3000 {\rm{\AA}}.  The geometry is roughly that of a hemisphere with an opening angle to the east, and we would expect to see a shell of molecules slightly offset to the west with both positive and negative velocities extending from south through west to north.  The spatial morphology of these structures changes dramatically with orbit phase as the companion approaches from $\approx$30 au  at apastron to $\approx$1.6 au, plunging deep into the primary star's wind at periastron (see Figure 1 in Gull et al. 2016).


\subsection{Implications of the Dust Mass Estimates} 

The presence of iron grains as a major component of the dust in the Homunculus has a strong impact on estimation of the total mass.  As a consequence of the low probability of Fe grain formation, most Fe in the envelope is left in the gas phase (e.g., Gail et al. 2005), perhaps as much as $\sim$80-85\%, accounting for minority inclusions in silicates.   In this case, the gas-to-dust ratio $f_{gd}$ is much higher than 100.   If we adopt $f_{gd}$ = 200 as a  {\em{very}} conservative lower limit for the Fe dust and 100 for all other species, then the total mass of the Homunculus is $M_{\rm{tot}} \approx$ 45 $M_\odot$ in both models.  This mass value is more of a lower limit in Model A, due to the stronger contribution of cool Fe dust.  The predicted dust masses are comparable to those in the surrounding outer ejecta as estimated and proposed by Gomez et al. (2006) to originate in an even earlier eruption, leading to some dust formation by collision of ejected material with the local ISM.  

What can lead to such large mass loss and not  annihilate the system?   There are at least some global consistencies between our results and the stellar merger model by Portegies Zwart \& van den Heuvel (2016).  In their model, the Great Eruption is the result of one of the stars in the triple system plunging into the dense envelope of the primary star, a few decades {\it{after}} formation of the Homunculus bipolar nebula of 20 M$_\odot$ that is the result of massive single-star evolution at the Eddington limit (cf. Owocki et al. 2004).  The triple-star system orbits become partially re-aligned into the equatorial plane of the primary star, leading to the merger event over two orbital passages that forms the massive loops and torus-like structures in the mechanical dumping of angular momentum.   The observed size of that set of structures is consistent with our inferred constraints on the principle dust volume, while the mass is predicted to be $\approx$22 M$_\odot$,  with $\approx$10\% of that in a nitrogen-enriched inner ring of material.  Thus, their predictions for the total mass of the Homunculus and the merger are consistent with our dust model results, the largest difference being in the distribution of the mass which is inferred in our data to be concentrated in the central structures, and far less in the lobes.  Portegies Zwart \& van den Heuvel do not address observational constraints on the age of the Homunculus with respect to the Great Eruption, which has a literature range of about one orbital period, not several decades.\footnote{Most proper-motion studies place the Homunculus age between 1841.2 and 1847.4  (e.g., Currie et al. 1996; Smith \& Gehrz 1998; Morse et al. 2001).}  Also, to what extent the model has assumed that there must be $\approx$20 M$_\odot$ lost into the Homunculus from the primary several decades prior to the merger is not clear, i.e., whether based on observational claims or an  outcome of the modeling. Nonetheless, the general attributes are encouraging, and the parameter space for this modeling may possibly allow for some closer agreement of the mass predictions, or reveal additional paths to significant mass loss. 

\subsection{The carbon isotope ratio and the question of $C^+$ in the Homunculus}\label{carbonisotope}

We have emphasized that the chemistry of our dust models is consistent with observed abundances in the CNO-processed nebula, supported by analysis of $^{12}$CO and $^{13}$CO lines that indicates a subsolar abundance ratio of CO/H$_2$ and an enhancement of  $^{13}$C.  Some caution is needed about the isotopic abundance ratio, however, since chemical fractionation of carbon driven by the reaction $^{13}$C$^+$ + CO $\rightleftharpoons$  $^{12}$C$^+$ + $^{13}$CO + 34.8 K can occur in dense photodissociation regions (PDRs; Langer et al. 1984; Gerin et al. 1998; R\"ollig \& Ossenkopf 2013), and would decrease the $^{12}$C/$^{13}$C abundance ratio.  To check for this most directly requires observations of C~{\sc{i}} or [C~{\sc{ii}}] and their isotopes.  Neutral C at 492.2 GHz ($^3$P$_1 - ^3$P$_0$) and 809.3 GHz ($^3$P$_2 - ^3$P$_1$) has been searched but is not detected in our {\em{Herschel}} observations.  

\begin{figure}
 \begin{center}
 \includegraphics[width=8cm]{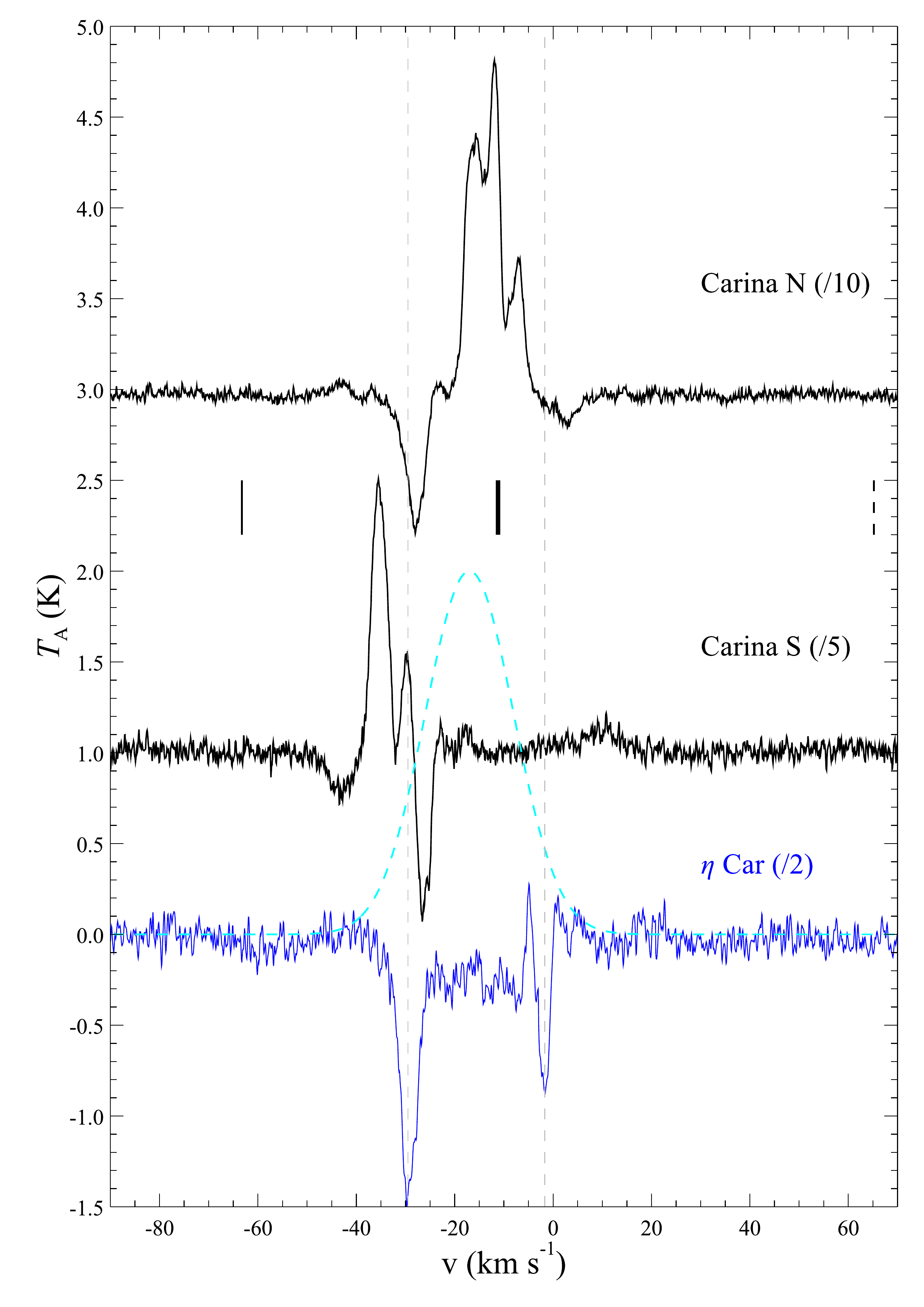}
 \end{center}
 \caption{HIFI observations of  [C~{\sc{ii}}] $^2$P$_{3/2}-^2$P$_{1/2}$ toward $\eta$~Carinae and two positions in the Carina nebula (see Fig.~\ref{spireoffdets}).  Each spectrum has been scaled as labeled.  The Carina N and Carina S spectra have also been offset by $+$1.0 K and $+$3.0 K, respectively.  Continuum emission is not detected in either spectrum. The $\eta$~Carinae spectrum has been offset by $-$1.9~K, which corresponds to the continuum at 1900 GHz (158 $\mu$m), accounting for the native DSB data.  Vertical dashed lines gray lines are centered on the main absorption features in the $\eta$~Carinae spectrum at $-$29.5 km s$^{-1}$ and $-$1.7 km s$^{-1}$.   Short vertical lines below the Carina N spectrum indicate the positions of the [$^{13}$C~{\sc{ii}}] hyperfine lines $F=2-1$ (thick), $1-1$ (thin), and $1-0$ (dashed).  The light-blue dashed Gaussian profile shows an approximation for the emission toward $\eta$~Carinae that would be needed in combination with the observed absorption features, including the broad trough centered near $-$15 km~s$^{-1}$.  In the composite model, $\tau_0$ = 3.9 for the absorption line at $-$29.5 km s$^{-1}$.
   \label{carinaCplus}}
\end{figure}

The [$^{12}$C~{\sc{ii}}] $^2$P$_{3/2} - ^2$P$_{1/2}$ line has been detected with PACS (T. Gull et al., in preparation) and HIFI.  The HIFI  spectrum shown in Figure~\ref{carinaCplus} contains two obvious absorption lines at $v_{\rm{lsr}}$ =  $-$1.7 and $-$29.3 km s$^{-1}$, with widths of 2.1 and 4.5 km s$^{-1}$, respectively.  Emission may be present at velocities near the weaker of the two absorption lines, but this is clearly a complex profile also containing a broad apparent ``trough'' of absorption centered between the two main features.   In fact, the entire baseline around these features is composed of 1900 GHz (158 $\mu$m) continuum emission and much broader [C~{\sc{ii}}] emission detected with PACS.  The spectral resolution of the PACS grating is 230 km s$^{-1}$ at this wavelength, and the main emission line may just be barely resolved, so that the HIFI [$^{12}$C~{\sc{ii}}] spectrum is only showing the structure at the profile peak.  The narrow absorption lines are unresolved to PACS.  Absorption by C$^+$ is a signature of very low excitation temperatures in a foreground cloud, $T_{\rm{ex}}$(C$^+$) $\simeq$ $25-40$ K (Goldsmith et al. 2012), along the line of sight to $\eta$~Carinae's strong continuum and the line emission, which provide disparate sources of background radiation.  {\em{ None of the [$^{13}$C~{\sc{ii}}] hyperfine lines are detected}}, contrary to what we would expect from [$^{12}$C/$^{13}$C] $\approxeq$ 3 in the Homunculus.  We have two possible explanations for this.

{\underline{Scenario 1}}: {\em{The inferred C$^+$ emission is not part of the Homunculus. }}   [$^{12}$C~{\sc{ii}}] 158 $\mu$m emission is widespread across the Carina Nebula as observed with ISO, arising primarily in the PDR (Mizutani et al. 2004) formed at the interface between the H~{\sc{ii}} region and molecular cloud mapped out in CO (Brooks et al. 1998; Zhang et al. 2001; see also Cox 1995;  Smith et al. 2000; Brooks et al. 2003, for reviews of the large-scale structure of the Carina cloud). Our PACS spectroscopy shows significant emission in the vicinity of the Homunculus, the strongest being in the direct sightline.   HIFI spectra at two positions, Carina North and Carina South (Ossenkopf et al. 2013) marked in Figure~\ref{spireoffdets}, are shown  in Figure~\ref{carinaCplus} in a comparison with $\eta$~Carinae for the line excitation and kinematics.\footnote{The processing of these spectra has skipped the OFF reference subtraction to avoid self-chopping from extended line emission; see Sec.~\ref{HifiObsSec}} 

The strongest absorption feature at all three positions is consistent with H~{\sc{i}} absorption at $-$29.4 km s$^{-1}$, interpreted to originate from neutral hydrogen associated with molecular concentrations and absorbing dust in an expanding shell (Dickel 1974).   Additional absorption is present at $-$43.0 and $-$31.6 km s$^{-1}$ in Carina S and at $+$2.8 km s$^{-1}$ in Carina N, which may be similar to the $-$1.7 km~s$^{-1}$ feature in $\eta$~Carinae (given the similar shift of the main absorption feature).  The emission in Carina N may be composed of three or four main velocity components, or it is being self-absorbed at velocities similar to the absorption trough in the $\eta$~Carinae spectrum.  We show a Gaussian line profile in the $\eta$~Carinae spectrum for the emission that is hinted at in the wings, at a strength that would be needed in combination with the observed absorption features to match the total spectrum.  In this composite model, the optical depth at line center $\tau_0$ = 3.9 for the absorption line at $-$29.5 km s$^{-1}$.  The absorption is expected when a significant column of low-density gas is observed toward a strong source of background emission
at temperature $T_{\rm{bg}}$, and the excitation temperature $T_{\rm{ex}}$ is less than the brightness temperature of the background, i.e., 

\begin{equation}\label{eq:TexCplus}
T_{\rm{ex}}({\rm{C}}^+) < \frac{91.2}{{\rm{ln}}(1 + 91.2/T_{\rm{c}}(1900.5))}  \rm{K}
\end{equation}

\noindent where the continuum brightness temperature at 1900.5 GHz $T_{\rm{c}}$(K)$ = 91.2/(e^{91.2/T_{\rm{bg}}} - 1)$.  The value 91.2 K is the equivalent temperature $T_*  \equiv h \nu/k$ at 1900.5 GHz.  At the position of $\eta$~Carinae, where we observe absorption, the net continuum levels peak at around 4.1 K on a main-beam temperature scale (SSB continuum plus inferred background emission), and if we include 2.7 K for the cosmic-ray background temperature, giving $T_{\rm{bg}}$ = 6.8~K, then $T_c$ is very small (negligible)  and $T_{\rm{ex}}^{\rm{abs}} \lesssim$  7~K for the absorption component.  This is very low compared to most sightlines to dense molecular clouds in star forming regions, and in fact we require the background emission component of the composite model to reach this temperature or it would be even lower, less than 5~K.  These thermal excitation conditions on the absorption are independent of the source of the inferred emission, whether in the extended background or part of the Homunculus. 

We can also estimate the excitation temperature for the emission component when it arises in the background and fills the HIFI beam.  Following Goldsmith \& Langer (1999), $T_{\rm{ex}}$ is related to the measured radiation temperature $T_{\rm{R}}$ through the equation

\begin{equation}\label{eq:TRad}
T_{\rm{R}} = \left( \frac{T_*}{\exp (T_*/T_{\rm{{ex}}}) - 1} - \frac{T_*}{\exp (T_*/T_{\rm{{bg}}}) - 1} \right) \left[ 1 - e^{-\tau} \right] (\rm{K}).
\end{equation}

\noindent For optically thick C$^+$ line emission that fills the HIFI beam, 

\begin{equation}\label{eq:CplusTex}
T_{\rm{ex}}({\rm{C}}^+) = \frac{91.2}{ {\rm{ln}} \left[ 1 + 91.2/ \left( T_{\rm{MB}}(^{12}{\rm{C}}^+) + T_{\rm{c}}(1900.5)  \right) \right] } \; ({\rm{K}})
\end{equation}
\noindent where $T_{\rm{MB}}(^{12}{\rm{C}}^+)$ $\approx$ 2.7~K is the line peak on a main-beam temperature scale, yielding $T_{\rm{ex}}^{\rm{em}}({\rm{C}}^+) \simeq$ 27~K.  To match the inferred emission profile, the C$^+$-emitting gas has a column density $N$(C$^+$) of several times 10$^{18}$ cm$^{-2}$, which is not atypical of dense PDR clouds. 

Given the similarity of this empirically derived profile to the observed Carina N emission spectrum, and to some extent the more blue-shifted Carina S emission, it is likely that the C$^+$ observed with HIFI (and PACS) originates in foreground clouds. In this case,  [$^{12}$C/$^{13}$C] $\approx$ 70, and thus the $^{13}$C$^+$ is too weak to detect with HIFI.   Caution with this interpretation is warranted, however, since the emission toward $\eta$~Carinae could include more than one simple component, as well as higher-velocity components as indicated in the wings of the PACS spectra (T. Gull et al., in preparation), outside the IF range of HIFI.  In this case, we should consider a second possibility.

{\underline{Scenario 2}}:  {\em{C$^+$ emission is being produced locally in or near the central region.}}  We consider this scenario since  [$^{12}$C~{\sc{ii}}] is observed to be strongest toward the Homunculus in our PACS spectral mapping (T. Gull et al., in preparation).   While the absorption line at $-$29.5 km s$^{-1}$ is almost certainly associated with the expanding H~{\sc{i}} shell as noted above, the blue-shifted velocities of the other observed absorption lines and the inferred emission are consistent with those of the partially ionized Weigelt structures located in the central 1$''$.  Teodoro et al. (2017) show in an analysis of HST/STIS observations of the two most optically prominent condensations, Weigelt C and D,  that the former is brighter in emission lines originating in intermediate- or high-ionization potential ions, particularly near apastron, which happens to be when our HIFI observations of C$^+$ were obtained.  Because of the large HIFI beam, however, we cannot make a spatial distinction between structures in the central region.   The main problem that we have with this scenario is the lack of detection of nebular [$^{13}$C~{\sc{ii}}], if we expect the Weigelt structures to have the same [$^{12}$C/$^{13}$C] abundance ratio as the rest of the Homunculus.  If [$^{12}$C~{\sc{ii}}] emission does originate from the Weigelt blobs, then the [$^{12}$C/$^{13}$C]  ratio must instead be more ISM-like in order for the strongest $F=2-1$ hyperfine transition of $^{13}$C$^+$ to be undetected by HIFI (or else we have grossly overestimated the strength of the $^{12}$C$^+$ emission).  From proper-motion analysis, Teodoro et al. (2017) estimate that the Weigelt structures are located at a distance of 1240 au from the central source, on the near side of the Homunculus close to the equatorial plane. 
 
The nature of the C$^+$ detected toward the Homunculus favors an origin in the foreground clouds with low $T_{\rm{ex}}$(C$^+$) in the Carina PDR, but it cannot be decisively confirmed without further mapping of the region at high enough spectral resolution to trace the absorption lines detected with HIFI.   

\section{Summary and Conclusions}

Following up the publication by MWB99 of first results for the 1996/1997 ISO spectrum of $\eta$~Carinae, which revealed a much more massive Homunculus Nebula than previously estimated, we have combined new {\em{Herschel}} observations with the legacy ISO data to give us continuous coverage of the 2.4 - 670 $\mu$m 
SED and carry out a full radiative transfer analysis of dust.  A better understanding of the dust chemistry, mass, and mass budget of the Homunculus and surrounding outer ejecta (Gomez et al. 2006) is essential for constraining the mass-loss history and mechanisms responsible for $\eta$~Carinae's enormous eruption(s).

Our results are summarized as follows:

\begin{enumerate}
\item The observations have been taken through large beams compared to the size of the Homunculus; therefore, the continuum SED must be treated as spatially integrated.  However, aperture and beam constraints strongly indicate that the principle source of thermal emission is no more extended than $\approx$ 5$''$ around the central source, while there is a separate feature of source size closer to 2$''$ measured at 350 $\mu$m, and probably related to a small source of free-free emission detected with ALMA by Abraham et al. (2014).  

\item The full SED gives $L_{\rm{IR-submm}}$ = 2.96 $\times 10^6 L_\odot$,  a 25\% decline deduced from an average of photometric levels observed in 1971-1977 in the $10-20$ $\mu$m range, indicating a reduction in circumstellar extinction in conjunction with an increase in visual brightness, allowing 25-40\% of optical and UV radiation to escape from the central source.

\item The SPIRE and ALMA continuum levels appear consistent with further brightening above levels published by Brooks et al. (2005) and Gomez et al. (2010) at submillimeter wavelengths, in phase with the radio cycle in which emission at 3 and 6~cm peaks near apastron (cf. Duncan \& White 2003). A low level of continuum variability at an effective wavelength of 350 $\mu$m consistent with brightening near apastron is tentatively identified in HIFI measurements taken over $\Phi$ = $12.10-12.71$.
\item Dust models fit to the full mid-IR SED indicate compositions that are consistent with a CNO-processed nebula.  Two best-fit models have pure iron, pyroxene and other metal-rich silicates (Mg, Fe, Ca, Al inclusions), corundum, and magnesium-iron sulfide in common.    Our preferred model also has nitrides (AlN and Si$_3$N$_4$ in low abundances, less than 0.5\% of the total dust mass, but makes up the bulk of the emission in the broad 10 $\mu$m band and are therefore strong contributors to the total luminosity in this model.  Our second model contains O-poor silicates and olivine to make up the 10 $\mu$m emission.  Corundum in both models fits the very narrow 20.2 $\mu$m feature when spherical grains are used.  Both crystalline and amorphous species are present in both models, indicating that while some dust has condensed under rapid cooling and expansion conditions (not surprisingly), higher-density material that has cooled more slowly and has allowed some dust to crystallize during condensation. 

\item While the dust masses in our two models differ by almost a factor of 2 owing to the varying amounts of pure Fe dust,  0.25 and 0.44 $M_\odot$, estimates of the total mass both suggest $M_{\rm{tot}} \ge$ 45 $M_\odot$.  This uses the assumption of $f_{gd}$(Fe) = 200, which is most important in Model A, although the ratio is very likely to be much greater given the high-abundance Fe in the gas phase and the experimental and theoretical results on the low sticking probability for the formation of Fe grains applying to this environment.  Similarly, $f_{gd}$ for the silicates and metal oxides more important in Model B may be different from the assumed canonical value of 100.  Furthermore, not all dust species may be optically thin, which we assumed for making conservative lower limit estimates of the dust masses, due to the lack of geometric constraints for each species.   All of these potential systematics lead  to a conservative lower limit on the total mass of the Homunculus.

\item We have no direct observational constraints on the location of each thermal or chemical component of the dust; however, the photometric calibrations of ISO data in conjunction with the aperture and beam properties infers the bulk of the emission to be within 5$''$ (equatorial) $\times$ 7$''$ (polar).  A volume of this size is consistent with mid-IR imaging of the extended structures in the equatorial region, with earlier studies using similar beam size constraints at wavelengths in the mid-IR (e.g. Gehrz et al. 1973) and far-IR where the cool dust is sampled (e.g., Harvey et al. 1978), and with recent modeling by Portegies Zwart \& van den Heuvel (2016) of the present-day primary star as a stellar merger which created the massive equatorial structures, far exceeding the mass of the polar lobes.      
   
\item The C- and O-deficient abundance pattern expected of CNO-processed material has been supported by analysis of $^{12}$CO and $^{13}$CO lines observed with HIFI over the range of energies $E_u$ = 83 - 238 K.  The results indicate that [CO/H$_2$] is some $10-20$ times lower than cosmic levels, and enhancement of $^{13}$C for that reduces the [$^{12}$C/$^{13}$C] ratio to a value that is consistent with evolutionary model prediction for rotating massive stars prior to entering the WC phase of W-R stars.  The CO column densities are much lower than quoted from the analysis of ground-based observations by Loinard et al. (2012), covering CO upper-level energies of $E_u$ = $33-116$ K in which the smaller adopted source size  ($\theta_s$ = 1$''$.0 versus 3$''$.0) drives up the CO column densities and thus the  [CO/H$_2$] ratio to cosmic levels.  The lower column densities and higher range of CO excitation temperatures in our study may be related to differences in the treatment of the radiative interactions between spatially overlapping velocity components. 
  
\item Absorption lines of $^{12}$C$^+$ have been detected toward $\eta$~Carinae, indicating the presence of disparate radiation fields in the line of sight and low excitation temperatures in the absorbing gas. $^{13}$C$^+$ is not detected.  The inferred emission is arising either in the extended PDR or from the Weigelt blobs closer to the core region.  The latter scenario would require the Weigelt condensations to have a more ISM-like [$^{12}$C/$^{13}$C] ratio (at least 40) in order for $^{13}$C$^+$ not to be detected.

\end{enumerate} 

\section{Future Work}

The evidence for a weakening of the dust opacity around the central source should be followed up with further mid-IR observations, with particular attention to the balance between the destruction and creation of dust during the 2020 periastron passage. This will be important to establish changes in the fraction of IR energy that can be assumed to constrain the mass of the primary star of this enigmatic system.   Imaging and spectroscopic observations taken at sufficient spatial resolution will be needed to map secular changes in dust opacities and the distribution  of the coolest dust, which dominates the dust masses.  Such observations can be partly accomplished with the Faint Object Infrared Camera (FORCAST; Herter et al. 2012) and the Far Infrared Field-Imaging Line Spectrometer (FIFI-LS; Colditz et al. 2012) for the SOFIA Telescope.  Additional mapping of the C$^+$ 1900.54 GHz line with the SOFIA German REceiver for Astronomy at Terahertz Frequencies (GREAT; G\"usten et al. 2000) may also elucidate the nature of the absorption and emission toward and around the Homunculus. Our modeling of the molecular gas and dust can be further refined with these observations, including the use of more sophisticated 2D/3D dust codes. 
  
\begin{center}
{\sc{Acknowledgments}} \\ 
\end{center}
\vspace{-0.5em}
We appreciate the comments from our anonymous referee, leading to several important clarifications, and for discussions with A. Mehner, P. Goldsmith, V. Ossenkopf,  and D. Teyssier.  We also extend thanks to L.  Loinard for helpful comments on the manuscript, K. Pitman for pointers on laboratory nitride reflectance spectra that aided in our interpretations, and to M. Teodoro for kindly granting figure adaptation permission. PM also thanks E. Caux for advice on the CASSIS package.  P.M., T.R.G., and K.N.  acknowledge partial financial support from NASA grant SCEX22012D for the {\em{Herschel}} program OT1\_tgull\_3.

\end{document}